\documentclass[aip,numerical,preprint]{revtex4-1}

\usepackage{epsfig,amsmath,amssymb,txfonts,hyperref}
\usepackage{xcolor}
\bibpunct{[}{]}{,}{n}{}{}
\renewcommand{\bibnumfmt}[1]{[#1]\quad}

\newcommand*{\et}[0]{\textit{et~al.}}

\DeclareMathOperator{\erf}{erf}
\DeclareMathOperator{\erfc}{erfc}

\newcommand*{\eps}[0]{\varepsilon}
\newcommand*{\x}[0]{\times}
\newcommand*{\ox}[0]{\otimes}
\newcommand*{\EE}[1]{\times 10^{{#1}}}

\newcommand*{\VECIT}[1]{\mathbf{#1}}
\newcommand*{\VECITS}[1]{\boldsymbol{#1}}  
\newcommand*{\kB}[0]{k_{\text{B}}}

\newcommand*{\qv}[0]{\VECIT{q}}
\newcommand*{\pv}[0]{\VECIT{p}}

\newcommand*{\Rv}[0]{\VECIT{R}}

\newcommand*{\av}[0]{\VECIT{a}}
\newcommand*{\vv}[0]{\VECIT{v}}
\newcommand*{\xv}[0]{\VECIT{x}}
\newcommand*{\xvp}[0]{\VECIT{x}'}
\newcommand*{\xsv}[0]{\VECIT{x}_\text{s}}
\newcommand*{\xvv}[0]{\VECIT{x}_\text{v}}
\newcommand*{\xspv}[0]{\VECIT{x}'_\text{s}}
\newcommand*{\xvpv}[0]{\VECIT{x}'_\text{v}}
\newcommand*{\zv}[0]{\VECIT{0}}  
\newcommand*{\is}[0]{i_\text{s}}
\newcommand*{\iv}[0]{i_\text{v}}
\newcommand*{\isp}[0]{i'_\text{s}}
\newcommand*{\ivp}[0]{i'_\text{v}}
\newcommand*{\stategen}[0]{\xsv\is\xvv\iv}  
\newcommand*{\stategenX}[0]{\overline{\stategen}}  
\newcommand*{\stategenXexplicit}[0]{(\xsv+\xvv)\iv-\xvv\is}  
\newcommand*{\staterep}[0]{\zv\is\xvv\iv}  
\newcommand*{\staterepX}[0]{\overline{\staterep}}  
\newcommand*{\statenew}[0]{\xspv\isp\xvpv\ivp}  
\newcommand*{\statenewrep}[0]{\zv\isp\xvpv\ivp}  
\newcommand*{\statenewrepS}[0]{\zv\is\xvpv\ivp}  
\newcommand*{\Srep}{S} 
\newcommand*{\Sbar}{\bar S} 
\newcommand*{\bdelta}[0]{\bar\delta}  
\newcommand*{\bs}[0]{\overline{cs}}  
\newcommand*{\yv}[0]{\VECIT{y}}
\newcommand*{\uv}[0]{\VECIT{u}}
\newcommand*{\bv}[0]{\VECIT{b}}
\newcommand*{\Gv}[0]{\VECIT{G}}

\newcommand*{\fv}[0]{\VECIT{f}}
\newcommand*{\gv}[0]{\VECIT{g}}
\newcommand*{\gav}[0]{\VECITS{\gamma}}
\newcommand*{\tv}[0]{\VECIT{t}}

\newcommand*{\cv}[0]{c_\text{v}} 
\newcommand*{\cs}[0]{c_\text{s}} 
\newcommand*{\muv}[0]{\mu^\text{v}} 
\newcommand*{\mus}[0]{\mu^\text{s}} 

\newcommand*{\flux}[1]{\VECIT{J}^{#1}}
\newcommand*{\fluxt}[1]{\VECIT{J}^\text{#1}}
\newcommand*{\Onsager}[1]{\underline L^{(#1)}}
\newcommand*{\Onsagert}[1]{\Onsager{\text{#1}}}
\newcommand*{\Dvv}[0]{\underline D^\text{v}}
\newcommand*{\DD}[0]{\underline D^\text{(4)}}
\newcommand*{\Lvv}[0]{\Onsagert{vv}}
\newcommand*{\Lsv}[0]{\Onsagert{sv}}

\newcommand*{\Lss}[0]{\Onsagert{ss}}
\newcommand*{\nconf}[0]{\VECIT{\underline n}}
\newcommand*{\hW}[0]{\hat W}
\newcommand*{\hw}[0]{\hat\omega}
\newcommand*{\dtomega}[0]{\widetilde{\delta\omega}}
\newcommand*{\dhw}[0]{\delta\hat\omega}
\newcommand*{\hP}[0]{\hat P}
\newcommand*{\hH}[0]{\hat H}
\newcommand*{\hEt}[0]{\hat E^{\text{trans}}}
\newcommand*{\etav}[0]{\VECITS{\hat\eta}}
\newcommand*{\dxvh}[0]{\hat{\VECITS{\delta}\VECIT{x}}}
\newcommand*{\dxv}[0]{\VECITS{\delta}\VECIT{x}}
\newcommand*{\bxv}[0]{\VECIT{\hat{wx}}}
\newcommand*{\hbv}[0]{\VECIT{\hat{b}}}
\newcommand*{\hgv}[0]{\VECITS{\hat{\gamma}}}
\newcommand*{\hg}[0]{\hat{g}}
\newcommand*{\Esv}[0]{E^\text{s-v}}
\newcommand*{\di}[0]{d}
\newcommand*{\ei}[0]{\VECIT{e}}
\newcommand*{\pmax}[0]{p_\text{max}}

\newcommand*{\GBZ}[0]{G_\text{BZ}}
\newcommand*{\Ng}[0]{N_\text{group}}
\newcommand*{\VV}[0]{\underline{VV}}
\newcommand*{\R}[0]{R}
\newcommand*{\Rtv}[0]{\{\R,\tv\}}
\newcommand*{\Rtvinv}[0]{\{\R,\tv\}^{-1}}
\newcommand*{\Rtvinvexplicit}[0]{\{\R^{-1},-\R^{-1}\tv\}}
\newcommand*{\Nsites}[0]{N_{\text{sites}}} 
\newcommand*{\ProbV}[1]{P^\text{0,v}_{#1}} 
\newcommand*{\ProbS}[1]{P^\text{0,s}_{#1}} 
\newcommand*{\Lmax}[0]{L_{\text{max}}} 
\newcommand*{\threshold}[0]{\eps_{\text{threshold}}} 
\newcommand*{\Nkpt}[0]{N_{\text{kpt}}} 
\newcommand*{\Nsymmkpt}[0]{N_{\text{symm-kpt}}} 

\newcommand*{\be}[0]{\begin{equation}}
\newcommand*{\ee}[0]{\end{equation}}
\newcommand*{\beu}[0]{\begin{equation*}}
\newcommand*{\eeu}[0]{\end{equation*}}
\newcommand*{\bme}[0]{\begin{multline}}
\newcommand*{\eme}[0]{\end{multline}}
\newcommand*{\bmeu}[0]{\begin{multline*}}
\newcommand*{\emeu}[0]{\end{multline*}}
\newcommand*{\ba}[0]{\begin{array}}
\newcommand*{\ea}[0]{\end{array}}
\newcommand*{\bfig}[0]{\begin{figure}[t]}
\newcommand*{\efig}[0]{\end{figure}}
\newcommand*{\bfigwide}[0]{\begin{figure*}[ht]}
\newcommand*{\efigwide}[0]{\end{figure*}}

\newlength{\wholefigwidth}
\newlength{\smallfigwidth}
\newlength{\halfsmallfigwidth}
\newlength{\figwidth}
\newcommand{\Fig}[1]{Fig.~\ref{fig:#1}}
\newcommand{\Tab}[1]{Table~\ref{tab:#1}}
\newcommand{\Sec}[1]{Section~\ref{sec:#1}}
\newcommand{\App}[1]{Appendix~\ref{sec:#1}}
\newcommand{\Eqn}[1]{Eqn.~\ref{eqn:#1}}
\newcommand{\Step}[1]{Step~\ref{step:#1}}
\newcommand{\rcite}[1]{Ref.~\onlinecite{#1}}

\begin{document}

\title{Automatic numerical evaluation of vacancy-mediated transport for arbitrary crystals: Onsager coefficients in the dilute limit using a Green function approach}

\author{Dallas R. Trinkle}
\email{dtrinkle@illinois.edu}
\affiliation{Department of Materials Science and Engineering, University of Illinois, Urbana-Champaign, Illinois 61801, USA}

\date{\today}
\begin{abstract}
  A general solution for vacancy-mediated diffusion in the dilute-vacancy/dilute-solute limit for arbitrary crystal structures is derived from the master equation. A general numerical approach to the vacancy lattice Green function reduces to the sum of a few analytic functions and numerical integration of a smooth function over the Brillouin zone for arbitrary crystals. The Dyson equation solves for the Green function in the presence of a solute with arbitrary but finite interaction range to compute the transport coefficients accurately, efficiently and automatically, including cases with very large differences in solute-vacancy exchange rates. The methodology takes advantage of the space group symmetry of a crystal to reduce the complexity of the matrix inversion in the Dyson equation. An open-source implementation of the algorithm is available, and numerical results are presented for the convergence of the integration error of the bare vacancy Green function, and tracer correlation factors for a variety of crystals including wurtzite (hexagonal diamond) and garnet.
\end{abstract}

\maketitle

\section{Introduction}
\label{sec:intro}
Mass transport in crystals controls a variety of different phenomena in materials from formation and growth of precipitates to aging to ionic conductivity to irradiation-induced degradation of material properties.\cite{Balluffi-Kinetics} The increasing ability to computationally evaluate rates for atomic scale mechanisms for diffusion using sophisticated first-principles methods\cite{Janotti2004,Mantina2009,Garnier2014c} has increased the need for accurate and extensible theory that can use jump rates as inputs to produce transport coefficients without the introduction of additional approximations. Generally, solute transport can be divided into a few general mechanisms: interstitialcy-based, vacancy-mediated, or self-interstitial-mediated, where the first two are the most common in materials of technological interest. Of these, interstitial diffusion currently has a full mathematical framework for the evaluation of diffusivity in the dilute solute limit for arbitrary crystal structures and interstitial sites, including derivatives of diffusivity with respect to temperature (activation energy tensor) and strain (elastodiffusion tensor).\cite{TrinkleElastodiffusivity2016} However, the state of quantitative evaluation of transport coefficients for vacancy-mediated diffusion in \textit{arbitrary} crystal structures is not as well developed. Vacancy-mediated diffusion occurs via a correlated random walk, which is the source of the complexity and crystal-structure dependence, as has been recognized for decades.\cite{Manning1968,Allnatt1993} This has been followed with specific solutions for particular structures, such as the five-frequency model for face-centered cubic (FCC) crystals,\cite{Leclaire1956,Bocquet1974} the four-frequency model for body-centered cubic (BCC) crystals,\cite{Jones1972,Barbu1996} the eight-frequency model in hexagonal close-packed (HCP) crystals,\cite{Ghate1964,Batra1967} and most recently a thirteen-frequency model in HCP.\cite{Allnatt2014} In all of these models, not only are interactions assumed to end beyond first-neighbor, they also introduce additional constrains on the rates that do not arise from crystal symmetry.

Outside of these specific crystal structures with particular constraints on the form of rates, the current approaches for treating vacancy-mediated diffusivity for arbitrary crystal structures is either a stochastic approach like kinetic Monte Carlo\cite{Murch1984,Belova2000,Belova2001,Belova2003a,Belova2003b} or a master equation method like the self-consistent mean-field method\cite{Nastar2000,Nastar2005} and kinetic mean-field approximation.\cite{Belashchenko1998,Vaks2014,Vaks2016} Kinetic Monte Carlo's appeal lies mainly in being a mathematically ``light-weight'' approach that requires no approximations in the form of the rates: once enumerated, pathways are generated and as trajectories become longer, the stochastic averages converge. This simplicity, however, can limit the practical use: in cases where there are large differences in rates, very large numbers of steps may be needed to accurately sample all states and transitions. For example, cases where the vacancy-solute exchange rate becomes very large or very small require more clever treatments.\cite{Novotny1995,Puchala2010,Nandipati2010,Chatterjee2010,Fichthorn2013} Furthermore, approaching the ``dilute'' limit requires larger cells and more trajectories to converge. Finally, using finite differences to evaluate numerical derivatives is significantly more difficult, with severe limitations on the ability to reduce the numerical and stochastic error.\cite{Li2016} It remains a useful methodology, especially for non-dilute diffusivities. An alternative is master equation approaches which were developed for non-dilute concentrations but are formally exact in the dilute limit: self-consistent mean field (SCMF) and kinetic mean-field approximation. The SCMF method has been powerful for investigating the effects of correlation in the presence of non-uniform chemical potential gradients.\cite{Nastar2014} These methods start from a cluster variation approach, and develop analytic expressions which can be evaluated numerically but are not fully automated for arbitrary crystal structures and symmetries,\cite{Barbe2007a,Barbe2007b,Garnier2013a,Garnier2013c,Garnier2014a,Garnier2014c} nor are they operationally without approximation even for the dilute limit. For example, in the SCMF, the range of correlations is cutoff at a finite distance, which is an approximation. The error can be reduced by increasing the cutoff distance\cite{Garnier2014a} which requires considering more and more so-called ``kinetic interactions.'' Furthermore, the introduction of the chemical potential gradient is taken to explicitly break crystal symmetry which makes symmetry analysis less effective. Finally, a related approach is the path probability method for irreversible thermodynamics\cite{Kikuchi1966} which has also been applied to diffusion\cite{Sato1983}, including to non-dilute tracer concentrations;\cite{Sato1985} however, there are difficulties in producing exact tracer correlation factors with the method.

While Green function methods were developed more than three decades ago for diffusion,\cite{Montroll1965,Koiwa1983} the need to compute the lattice Green function accurately for each new crystal structure became a serious roadblock. Hence, extending to new structures---or relaxing constraints on the form of rates---requires a new start with each case to be considered. Watson\cite{Watson1939} recognized the complexity in computing the lattice Green function at $\Rv=\zv$, much less for arbitrary sites in a crystal. To compute diffusion, we require the Green function at a series of sites, which is significantly more difficult to evaluate analytically, though a few cases have been solved, such as FCC\cite{Inoue1974,Mano1974,Morita1975} and some other lattices like HCP and tetrahedrally coordinated crystals using matrix methods.\cite{Koiwa1978,Koiwa1979} In this work, we develop a generally applicable Green function based method for vacancy-mediated transport in the dilute-vacancy/dilute-solute limit in arbitrary crystal structures and with arbitrary finite-ranged solute-vacancy interactions; the methodology is implemented in a fully tested numeric open-source software.

Our goal is to evaluate the Onsager transport coefficients in a dilute alloy of solute ``s'' in solvent ``A'' for the case of vacancy-mediated diffusion. In particular, we are interested in the three tensor transport coefficients, $\Lvv$, $\Lsv$, and $\Lss$ where
\be
\begin{split}
\fluxt{v} &= -\Lvv\nabla\muv - \Lsv\nabla\mus\\
\fluxt{s} &= -\Lsv\nabla\muv - \Lss\nabla\mus
\end{split}
\label{eqn:Onsager}
\ee
for vacancies ``v'' and solute ``s.'' The fluxes $\fluxt{v}$ and $\fluxt{s}$ combine to produce the solvent flux $\fluxt{A} = -\fluxt{v}-\fluxt{s}$ due to site conservation. Moreover, the chemical potentials $\muv$ and $\mus$ are defined relative to the solvent chemical potential $\mu^\text{A}$. For our purposes, we are interested in the motion of vacancies and solute in response to their chemical potential gradients; the response of solvent atoms, or the response of vacancies and solute to solvent chemical potential gradients, can be derived from the other transport coefficients.\cite{Allnatt1993} We work from a master equation framework for diffusivity\cite{Allnatt1993} and develop the Green function method\cite{Koiwa1983} to include the presence of a solute atom that can only move in the presence of a vacancy. The full algorithm is implemented in the code \textsc{Onsager}\cite{OnsagerCalc}; c.f. \App{implementation}. We conclude with numerical results for multiple systems, and discuss future extensions of the method.

\subsection{Master equation}
\label{sec:masterequation}
To model this system, we work with a lattice gas model similar to Nastar\cite{Nastar2000,Nastar2005} and Vaks\cite{Belashchenko1998,Vaks2014,Vaks2016}
containing the three species of interest: solvent ``A,'' solute ``s,'' and vacancies ``v.'' If we have a three dimensional crystal with unit cell vectors $\av_1,\av_2,\av_3$ and $\Nsites$ per unit cell at positions $\uv_i$, $i=1\ldots \Nsites$, then we can consider the set of all crystal sites, defined by vectors $\Rv = \xv + \uv_i$ for $\xv$ a lattice vector (a linear combination of unit cell vectors with integer coefficients).\cite{Ashcroft76} The choice of unit cell vectors and sites in the unit cell is such that every crystal site $\Rv$ is represented by exactly one lattice vector $\xv$ plus unit cell position $\uv$. We will work with Born-von~Karman boundary conditions in the thermodynamic limit---large number of sites and system volume $V_0$---to eliminate the introduction of any surfaces; the only defects present will be a vacancy and a solute, which corresponds to the dilute limit (interactions between several vacancies and/or several solutes are neglected); moreover, the crystalline sites need only to be those sites that can be visited by a vacancy or a solute, and hence may represent a sublattice for some systems. A configuration is a vector $\nconf$ where each element $n^\alpha_\Rv$ determines the site occupancy by species $\alpha$ at site $\Rv$; the occupancies are either 0 or 1, and $\sum_\alpha n^\alpha_\Rv = 1$ for all sites $\Rv$. We will identify all configuration dependent quantities with a hat. The system admits possible transitions from configuration $\nconf$ to $\nconf'$ defined by the transition rate matrix $\hW(\nconf\to\nconf')$. Furthermore, this rate matrix gives us the master equation for the evolution of the system probability $\hP(\nconf, t)$ with time $t$ as
\be
\frac{d\hP(\nconf, t)}{dt} =
\sum_{\nconf'} \hW(\nconf'\to\nconf) \hP(\nconf', t)
- \hW(\nconf\to\nconf')\hP(\nconf, t)
\ee
This expression is primarily useful for us to define \textit{steady-state} and \textit{equilibrium} in terms of balance and detailed balance, respectively, where $d\hP(\nconf, t)/dt = 0$. A probability distribution $\hP(\nconf)$ will be in \textit{balance} if it obeys
\be
\sum_{\nconf'} \hW(\nconf'\to\nconf) \hP(\nconf')
= \sum_{\nconf'} \hW(\nconf\to\nconf')\hP(\nconf)
\label{eqn:balance}
\ee
for all configurations $\nconf$. A probability distribution $\hP_0(\nconf)$ will be in \textit{detailed balance} if it obeys
\be
\hW(\nconf'\to\nconf) \hP_0(\nconf')
= \hW(\nconf\to\nconf')\hP_0(\nconf)
\label{eqn:detailedbalance}
\ee
for all configurations $\nconf$ and $\nconf'$. Clearly, detailed balance is sufficient for balance, but not necessary. In particular, we note that a system satisfying detailed balance will be in equilibrium (zero flux), while a system satisfying balance will be in steady-state, and may admit non-zero fluxes. Our approach to determining transport coefficients is to use near-equilibrium thermodynamics: we will find steady-state solutions that are the equilibrium distribution plus a small perturbation in response to a chemical potential gradient $\nabla\mu$ along an arbitrary direction. We will then determine the fluxes and solve directly for the transport coefficients in \Eqn{Onsager}.

We assume that our transition matrix corresponds to a physical system with a Hamiltonian $\hH(\nconf)$ and equilibrium probability distribution $\hP_0(\nconf)$; in particular, the equilibrium distribution for chemical potentials $\mu^\alpha$ at temperature $\kB T$ is
\be
\hP_0(\nconf) = \exp\left[\frac{1}{\kB T}\left(
\Phi_0 + \sum_\alpha \mu^\alpha\sum_\Rv n^\alpha_\Rv - \hH(\nconf)
\right)\right]
\label{eqn:equilibrium}
\ee
where $\Phi_0$ is a normalization constant---the grand potential---such that $\sum_\nconf \hP_0(\nconf) = 1$. We will assume that $\hP_0(\nconf)$ obeys detailed balance (\Eqn{detailedbalance}), which relates $\hW$ and $\hH$. Note also that $\hH$ is a lattice function, and as such will obey symmetry relations of the underlying lattice; i.e., it will remain invariant with respect to all space-group operations applied to $\nconf$. Those symmetries also necessarily translate to $\hW$. Moreover, we will assume that \textit{all} non-zero transition rates $\hW(\nconf\to\nconf')$ conserve mass: $\sum_\Rv n^\alpha_\Rv = \sum_\Rv n^{\prime\alpha}_\Rv$ for all species $\alpha$. Hence, even though we work in the grand-canonical ensemble, our particle numbers will remain conserved.

\subsection{Transport coefficients}
\label{sec:transport}
In order to introduce a steady-state solution with chemical potential gradients, we will consider a site-based chemical potential perturbation $\delta\mu^\alpha_\Rv$; these perturbations are such that for any two sites $\Rv$ and $\Rv'$,
\be
\delta\mu^\alpha_\Rv - \delta\mu^\alpha_{\Rv'} = (\Rv-\Rv')\cdot\nabla\mu^\alpha
\label{eqn:deltamudiff}
\ee
where $\nabla\mu^\alpha$ will be considered a homogeneous constant corresponding to an infinitesimally small chemical potential gradient vector along an arbitrary direction. Note that we do not require an explicit form for the perturbation $\delta\mu$, as \textit{only} differences of the form \Eqn{deltamudiff} appear in our equations. Next, we work with an ansatz steady-state solution
\be
\hP_\text{ss}(\nconf) := \hP_0(\nconf)\exp\left[\frac{1}{\kB T}\left(
\delta\Phi_0 + \sum_\alpha\sum_\Rv \delta\mu^\alpha_\Rv n^\alpha_\Rv 
-\sum_\alpha \etav^\alpha(\nconf)\cdot\nabla\mu^\alpha
\right)\right]
\label{eqn:steadystate}
\ee
where $\delta\Phi_0$ is a normalization constant, and $\etav^\alpha(\nconf)$ is a vector lattice function, with the same lattice symmetries as $\hH$, albeit as a \textit{vector}, so that rotations also rotate $\etav$ (while $\hH$ is a scalar). The combination $\etav^\alpha(\nconf)\cdot\nabla\mu^\alpha$ acts as the effective Hamiltonian in the self-consistent mean-field notation.\cite{Nastar2000,Nastar2005} In order to solve for the steady-state, and determine the fluxes, we introduce the mass-transport vector
\be
\dxvh^\alpha(\nconf\to\nconf') := \sum_\Rv n^{\prime\alpha}_\Rv \Rv - \sum_\Rv n^\alpha_\Rv \Rv
= \sum_\Rv \left(n^{\prime\alpha}_\Rv - n^{\alpha}_\Rv\right)\Rv
\label{eqn:deltax}
\ee
which is the total transport of species $\alpha$ in the transition $\nconf\to\nconf'$. Given mass-conservation, and as we work in the laboratory frame, for any non-zero $\hW(\nconf\to\nconf')$, we have $\sum_\alpha \dxvh^\alpha(\nconf\to\nconf') = 0$. The additional symmetries are that $\dxvh^\alpha(\nconf\to\nconf')=-\dxvh^\alpha(\nconf'\to\nconf)$, which requires that $\dxvh^\alpha(\nconf\to\nconf)=0$. Finally, $\dxvh^\alpha$ is a lattice function and as such obeys symmetry operations of the lattice. Then, the flux of species $\alpha$ can be expressed in a system with total volume $V_0$; for convenience, we multiply through by the volume and temperature $\kB T$ to get
\be
V_0\kB T \flux{\alpha} = \sum_{\nconf,\nconf'}
\dxvh^\alpha(\nconf\to\nconf')
\hW(\nconf\to\nconf')
\left(\kB T \hP_\text{ss}(\nconf)\right).
\label{eqn:fluxdef}
\ee
If the equilibrium probability $\hP_0(\nconf)$ is used instead, detailed balance and antisymmetry of $\dxvh^\alpha$ is sufficient to make $\flux{\alpha}=0$ for all species $\alpha$. In the limit of small gradients $\nabla\mu^\alpha$, we can expand our steady-state solution $\kB T \hP_\text{ss}$,
\be
\kB T\hP_\text{ss}(\nconf) = 
\hP_0(\nconf)\left[\kB T + \delta\Phi_0 
+ \sum_\beta\sum_\Rv \delta\mu^\beta_\Rv n^\beta_\Rv 
-\sum_\beta \etav^\beta(\nconf)\cdot\nabla\mu^\beta
\right] + O\left(|\nabla\mu|^2\right)
\ee
which substitutes into \Eqn{fluxdef} to get, to first order in $\nabla\mu^\alpha$,
\be
\begin{split}
V_0\kB T \flux{\alpha} &= \sum_{\nconf,\nconf'}
\dxvh^\alpha(\nconf\to\nconf')
\hW(\nconf\to\nconf')
\hP_0(\nconf)\left[\kB T + \delta\Phi_0 
+ \sum_\beta\sum_\Rv \delta\mu^\beta_\Rv n^\beta_\Rv 
-\sum_\beta \etav^\beta(\nconf)\cdot\nabla\mu^\beta
\right]\\
&= \frac12\sum_{\nconf,\nconf'}
\Bigg(
\dxvh^\alpha(\nconf\to\nconf')\hW(\nconf\to\nconf')
\hP_0(\nconf)\left[\kB T + \delta\Phi_0 + \sum_\beta\sum_\Rv \delta\mu^\beta_\Rv n^\beta_\Rv \right]\\
&\qquad+ 
\dxvh^\alpha(\nconf'\to\nconf)\hW(\nconf'\to\nconf)
\hP_0(\nconf')\left[\kB T + \delta\Phi_0 + \sum_\beta\sum_\Rv \delta\mu^\beta_\Rv n^{\prime\beta}_\Rv \right]
\Bigg)\\
&\quad-
\sum_{\nconf,\nconf'}
\hW(\nconf\to\nconf')\hP_0(\nconf)
\dxvh^\alpha(\nconf\to\nconf')\sum_\beta \etav^\beta(\nconf)\cdot\nabla\mu^\beta
\end{split}
\ee
where the second expression comes from symmetrizing the double summation. This expression can be simplified in a few quick steps. First, we note by detailed balance and antisymmetry of $\dxvh$ that $\dxvh^\alpha(\nconf'\to\nconf)\hW(\nconf'\to\nconf)\hP_0(\nconf')
=-\dxvh^\alpha(\nconf\to\nconf')\hW(\nconf\to\nconf')\hP_0(\nconf)$. Next, we note that $\sum_{\Rv}\delta\mu^\beta_\Rv(n^\beta_\Rv-n^{\prime\beta}_\Rv) = -\dxvh^\beta(\nconf\to\nconf')\cdot\nabla\mu^\beta$. Then,
\be
\begin{split}
V_0\kB T \flux{\alpha} &= 
-\sum_\beta\Bigg[
\frac12\sum_{\nconf,\nconf'}
\hP_0(\nconf)\hW(\nconf\to\nconf')\dxvh^\alpha(\nconf\to\nconf')\dxvh^\beta(\nconf\to\nconf')
\\
&\qquad+\sum_{\nconf,\nconf'}
\hW(\nconf\to\nconf')\hP_0(\nconf)
\dxvh^\alpha(\nconf\to\nconf')\etav^\beta(\nconf)\Bigg]\cdot\nabla\mu^\beta
\\
\end{split}
\ee
and thus our transport coefficients are
\be
\Onsager{\alpha\beta} = 
\frac1{\kB T V_0}\sum_{\nconf,\nconf'}
\hP_0(\nconf)\hW(\nconf\to\nconf')\left[\frac12 \dxvh^\alpha(\nconf\to\nconf')\ox\dxvh^\beta(\nconf\to\nconf')
+\dxvh^\alpha(\nconf\to\nconf')\ox\etav^\beta(\nconf)\right],
\label{eqn:transportcoeff}
\ee
where $\ox$ is the outer (or dyad) product of two vectors. The first term is the ``bare'' (uncorrelated) mobility, and the second term contains correlations.\cite{Allnatt1993}

Two brief notes about the second term in the right hand side of \Eqn{transportcoeff}. First, only differences in $\etav$ are important. This can be shown by symmetrizing with respect to $\nconf$ and $\nconf'$, in a similar fashion to the first term of \Eqn{transportcoeff}. This gives
\be
\sum_{\nconf,\nconf'}
\hP_0(\nconf)\hW(\nconf\to\nconf')
\dxvh^\alpha(\nconf\to\nconf')\ox\etav^\beta(\nconf)=
\frac12\sum_{\nconf,\nconf'}
\hP_0(\nconf)\hW(\nconf\to\nconf')
\dxvh^\alpha(\nconf\to\nconf')\ox(\etav^\beta(\nconf)-\etav^\beta(\nconf')).
\ee
Secondly, we identify the \textit{velocity vector},
\be
\bxv^\alpha(\nconf) := \sum_{\nconf'}
\hW(\nconf\to\nconf')\dxvh^\alpha(\nconf\to\nconf')
\label{eqn:ratebias}
\ee
which is a non-zero vector when the jumps in one direction occur with a different rate for the opposite direction. Then, the transport coefficients are 
\be
\Onsager{\alpha\beta} = 
\frac1{\kB T V_0}\sum_{\nconf}\hP_0(\nconf)\left\{
\sum_{\nconf'}\left[\frac12\hW(\nconf\to\nconf') \dxvh^\alpha(\nconf\to\nconf')\ox\dxvh^\beta(\nconf\to\nconf')\right]
+\bxv^\alpha(\nconf)\ox\etav^\beta(\nconf)\right\}.
\label{eqn:transportcoeff-eta}
\ee

\subsection{Balance equation}
\label{sec:balance}
Finally, we solve for the deviation from the equilibrium probability distribution, $\etav$, using balance. If we take \Eqn{balance}, to linear order in $\nabla\mu^\beta$, we have
\be
\begin{split}
&\sum_{\nconf'}\hW(\nconf'\to\nconf)
\hP_0(\nconf')\left[\kB T + \delta\Phi_0 
+ \sum_\beta\sum_\Rv \delta\mu^\beta_\Rv n^{\prime\beta}_\Rv 
-\sum_\beta \etav^\beta(\nconf')\cdot\nabla\mu^\beta
\right]
=\\
&\sum_{\nconf'}\hW(\nconf\to\nconf')
\hP_0(\nconf)\left[\kB T + \delta\Phi_0 
+ \sum_\beta\sum_\Rv \delta\mu^\beta_\Rv n^\beta_\Rv 
-\sum_\beta \etav^\beta(\nconf)\cdot\nabla\mu^\beta
\right].
\end{split}
\label{eqn:balance-eta1}
\ee
We apply detailed balance, $\hW(\nconf'\to\nconf)\hP_0(\nconf')=\hW(\nconf\to\nconf')\hP_0(\nconf)$ to the left-hand side of \Eqn{balance-eta1} to cancel out the first two terms of each side, and rearrange the remaining terms,
\be
\sum_\beta
\sum_{\nconf'}\hP_0(\nconf)\hW(\nconf\to\nconf')
\dxvh^\beta(\nconf\to\nconf')\cdot\nabla\mu^\beta
=
\sum_\beta
\sum_{\nconf'}\hP_0(\nconf)\hW(\nconf\to\nconf')
\left[
\etav^\beta(\nconf')-\etav^\beta(\nconf)\right]\cdot\nabla\mu^\beta.
\label{eqn:balance-eta2}
\ee
This must hold for any arbitrary direction of $\nabla\mu^\beta$. We define a matrix representation of $\hW$, where
\be
\hW_{\nconf\nconf'} =
\begin{cases}
\hW(\nconf\to\nconf') &: \nconf' \ne \nconf\\
-\sum_{\nconf'} \hW(\nconf\to\nconf') &: \nconf' = \nconf
\end{cases}
\ee
and, divide out $\hP_0(\nconf)$ from \Eqn{balance-eta2} to produce
\be
\begin{split}
\sum_\beta
\bxv^\beta(\nconf)
\cdot\nabla\mu^\beta &=
\sum_\beta\sum_{\nconf'} \hW_{\nconf\nconf'}\etav^\beta(\nconf')
\cdot\nabla\mu^\beta\\
\bxv^\beta(\nconf) &= \sum_{\nconf'} \hW_{\nconf\nconf'}\etav^\beta(\nconf').
\end{split}
\label{eqn:balance-eta}
\ee
Thus, the diffusion problem involves solving \Eqn{balance-eta} for $\etav$ to evaluate \Eqn{transportcoeff-eta}.

There are a few approaches to solve \Eqn{balance-eta}. The equation as written does not rely on assumptions about the dilute limit, and is valid for interstitial diffusion;\cite{TrinkleElastodiffusivity2016} we will consider, going forward, the case of vacancy-mediated diffusion in the dilute limit. One approach is the self-consistent mean-field method (SCMF).\cite{Nastar2000,Nastar2005} The SCMF approach solves \Eqn{balance-eta} by (a) selecting a particular direction for diffusion, (b) multiplying by $\hP_0(\nconf)$ and summing over $\nconf$ to convert the equation into thermodynamic averages, (c) writing out $\etav(\nconf)$ in terms of pair interactions that are invariant along the diffusion direction, and (d) are cutoff after a fixed distance (setting $\etav(\nconf)=\zv$ for vacancy-solute distance greater than a cutoff). This is an approximate solution for the effective Hamiltonian, which becomes more accurate as the cutoff distance is increased. Note that truncating $\etav(\nconf)$ is different than truncating the interaction energy between a solute and a vacancy: even for the case of a radioisotopic tracer where the interaction \textit{energy} is exactly zero, truncating the \textit{correlation} effects at a finite distance introduces error, as the Green function has infinite range. The second approach---laid out here---is a Green-function approach, which is fairly straightforward for the dilute-vacancy/solute limit for vacancy-mediated diffusion, and is exact. The Green function approach to the problem seeks to solve \Eqn{balance-eta} by constructing the \textit{exact} pseudo-inverse of $\hW_{\nconf\nconf'}$ for the dilute-vacancy/dilute-solute limit; we treat a single vacancy and single solute in the total volume $V_0$, while we take the thermodynamic limit of $V_0\to\infty$. We do this by (a) breaking $\hW_{\nconf\nconf'}$ into three contributions---the bare vacancy, vacancy near a solute, and vacancy-solute exchange---and (b) taking advantage of translational invariance for our lattice functions. Moreover, we will also take advantage of space-group symmetry operations to maximally reduce the rank of the linear problem to be solved. Note that this is similar in approach to Koiwa and Ishioka;\cite{Koiwa1983} we automate the computation of the Green function for the vacancy and the vacancy-solute complex for an arbitrary crystal, where we can take advantage of automated crystal symmetry analysis.\cite{OnsagerCalc}

\subsection{Matrix symmetrization}
\label{sec:symmetry}
Before we reduce to the dilute-vacancy/dilute-solute limit, we rewrite \Eqn{balance-eta} in terms of the pseudo-inverse of a \textit{symmetric} matrix. Define the components of the matrix $\hw_{\nconf\nconf'}$,
\be
\hw_{\nconf\nconf'}:= \hP_0^{1/2}(\nconf)\hW_{\nconf\nconf'}\hP_0^{-1/2}(\nconf')
\label{eqn:omega-def}
\ee
which is symmetric by detailed balance,
\be
\begin{split}
\hw_{\nconf\nconf'}&=\hP_0^{1/2}(\nconf)\hW_{\nconf\nconf'}\hP_0^{-1/2}(\nconf')
=\hP_0^{-1/2}(\nconf)\hP_0(\nconf)\hW_{\nconf\nconf'}\hP_0^{-1/2}(\nconf')\\
&=\hP_0^{-1/2}(\nconf)\hW_{\nconf'\nconf}\hP_0(\nconf')\hP_0^{-1/2}(\nconf')
=\hP_0^{-1/2}(\nconf)\hW_{\nconf'\nconf}\hP_0^{1/2}(\nconf')
=\hw_{\nconf'\nconf}.
\end{split}
\ee
This form of the matrix can be related to the linear-interpolated migration barrier (LIMB) approximation;\cite{Senhaji1992,Vanderven2001} from \Eqn{equilibrium},
\be
\hP_0^{1/2}(\nconf)\hP_0^{-1/2}(\nconf')=
\exp\left[\frac{\hH(\nconf') - \hH(\nconf)}{2\kB T}\right]
\ee
as we only consider transitions that conserve particle number. If we have a transition state energy $\hEt(\nconf-\nconf')=\hEt(\nconf'-\nconf)$ between $\nconf$ and $\nconf'$ so that $\hW(\nconf\to\nconf')\propto\exp(-(\hEt(\nconf-\nconf')-\hH(\nconf))/(\kB T))$, then for $\nconf\ne\nconf'$,
\be
\hw_{\nconf\nconf'}\propto
\exp\left[-\frac{\hEt(\nconf-\nconf') - (\hH(\nconf') + \hH(\nconf))/2}{\kB T}\right]
\ee
which is \textit{constant} for allowed jumps in the LIMB approximation.

Next we define, the \textit{bias vector}, in terms of the velocity vector
\be
\hbv^\alpha(\nconf) := \hP_0^{1/2}(\nconf)\bxv^\alpha(\nconf)
\ee
and the symmetrized correction vector,
\be
\hgv^\alpha(\nconf) := \hP_0^{1/2}(\nconf)\etav^\alpha(\nconf).
\ee
which then, by \Eqn{balance-eta}, gives
\be
\begin{split}
\hP_0^{-1/2}(\nconf)\hbv^\alpha(\nconf)
&= \sum_{\nconf'} W_{\nconf\nconf'}\hP_0^{-1/2}(\nconf')\hgv^\alpha(\nconf')\\
\hbv^\alpha(\nconf)
&= \sum_{\nconf'} \hP_0^{1/2}(\nconf)W_{\nconf\nconf'}\hP_0^{-1/2}(\nconf')\hgv^\alpha(\nconf')\\
&= \sum_{\nconf'} \hw_{\nconf\nconf'}\hgv^\alpha(\nconf').
\end{split}
\ee
Let the pseudo-inverse of $\hw$ be $\hg$, the Green function. Then,
\be
\hgv^\alpha(\nconf) = \sum_{\nconf'} \hg_{\nconf\nconf'}\hbv^\alpha(\nconf')
\ee
and
\be
\begin{split}
\sum_{\nconf} \hP_0(\nconf)\bxv^\alpha(\nconf)\ox\etav^\beta(\nconf)
&= 
\sum_{\nconf} \hP_0(\nconf)\hP_0^{-1/2}(\nconf)
\hbv^\alpha(\nconf)\ox\hP_0^{-1/2}(\nconf)\hgv^\beta(\nconf)\\
&= 
\sum_{\nconf,\nconf'} \hbv^\alpha(\nconf)\ox\hg_{\nconf\nconf'}\hbv^\beta(\nconf').
\end{split}
\ee
This shows that $\Onsager{\alpha\beta}=\Onsager{\beta\alpha}$ as $\hg_{\nconf\nconf'}=\hg_{\nconf'\nconf}$. It also means that we only need to find $\hg_{\nconf\nconf'}$ for those configurations where $\hbv^\alpha(\nconf)\ne0$. Thus, our transport coefficients are
\be
\Onsager{\alpha\beta} = 
\frac1{\kB T V_0}\sum_{\nconf,\nconf'}
\frac12\hP_0^{1/2}(\nconf)\hw_{\nconf\nconf'}\hP_0^{1/2}(\nconf')
\dxvh^\alpha(\nconf\to\nconf')\ox\dxvh^\beta(\nconf\to\nconf')
+\hbv^\alpha(\nconf)\ox\hg_{\nconf\nconf'}\hbv^\beta(\nconf').
\label{eqn:Onsager-symmetric}
\ee

\subsection{Dilute-vacancy/dilute-solute limit}
\label{sec:dilutelimit}
For the dilute-vacancy/dilute-solute limit, our state $\nconf$ simplifies to the position of the solute and the vacancy. With only a single solute and a single vacancy, the solute concentration $\cs=1/V_0$ and the vacancy concentration $\cv=1/V_0$. We will take advantage of translational invariance; moreover, we will specify the position of the vacancy relative to the unit cell of the solute: $\xsv+\uv_{\is}$ will be the position of the solute in the lattice, and $\xvv+\xsv+\uv_{\iv}$ the position of the vacancy---thus, the state of our system is captured by $\stategen$. We assume that the vacancy and solute have a finite interaction range, so that for large enough $\xvv$, the site probability $\hP^0(\stategen)\propto\exp(-\Esv_{\is\iv}(\xv)/\kB T)$ is independent of $\xvv$, and instead is given by the product of solute and vacancy probabilities. We construct the transition rate matrix by adding three contributions: vacancy without solute, vacancy near solute, and vacancy-solute exchange. We use the historical nomenclature, and identify these rates with superscripts ``0,'' ``1,'' and ``2.''

First, we consider the migration of the vacancy without solute. For this case, we track the solute state only in anticipation of later contributions; otherwise, the solute is ignored. The vacancy can occupy the same state as the solute (which will be corrected with a subsequent contribution), and so we ensure translational invariance for the transition rate matrix. A vacancy at the site $\xvv\iv$ transitions to another size $\xvpv\ivp$ with rate $w^{0,\iv\iv'}_{\xvv-\xvpv}$, so
\be
\hW^0_{\stategen,\statenew} := \delta(\xsv-\xspv)\delta_{\is\isp}\biggl\{
w^{0,\iv\iv'}_{\xvv-\xvpv}-\delta(\xvv-\xvpv)\delta_{\iv\ivp}\sum_{\xv i}w^{0,\iv i}_{\xv}\biggr\}
\label{eqn:omega0definition}
\ee
where $\delta$ is the Kronecker delta function, and the second term gives the correct value for $\hW_{\nconf\nconf}$. Because this contribution to the transition matrix ignores any solute-vacancy interaction,
\be
\hw^0_{\stategen,\statenew} = \hW^0_{\stategen,\statenew} (\ProbV{\iv}/\ProbV{\ivp})^{1/2}
\label{eqn:omega0ratesymmetrization}
\ee
where $\ProbV{i}$ is the probability for a vacancy to occupy the unit cell site $i$. In the special case of a Bravais lattice, the ratio of probabilities is 1. The transition matrix $\hW^0$ (and $\hw^0$) has the space group symmetry of the lattice; the corresponding Green function for the vacancy without a solute is derived in \Sec{vacancy}.

Next, we consider the contribution from solute-vacancy exchange. For a solute at $\xsv+\uv_{\is}$ and a vacancy at $\xsv+\xvv+\uv_{\iv}$, exchange will place the solute at $(\xsv+\xvv)+\uv_{\iv}$, and the vacancy at $(\xsv+\xvv)-\xvv+\uv_{\is}$. In the dilute limit, there is only one solute and one vacancy, and so if exchange is possible for a state, there is only one endpoint state where the solute changes to a new position with a non-zero rate. As a shorthand, define $\stategenX:=\stategenXexplicit$, the final state in solute-vacancy exchange. Then, we identify the set of all states $\{\stategen\}$ where $\hW_{\stategen,\stategenX}\ne 0$ as the \textit{exchange space}. Note that if $\stategen$ is in the exchange space, so is $\stategenX$; and $\overline{\stategenX}=\stategen$. Then, we define
\be
  \hw^2_{\stategen,\statenew} :=
  \begin{cases}
    \hP_0^{1/2}(\stategen)\hW_{\stategen,\stategenX}\hP_0^{-1/2}(\stategenX)&:
    \statenew=\stategenX \in \text{exchange space}\\
    -\hW_{\stategen,\stategenX} &:
    \statenew=\stategen \in \text{exchange space}\\
    0 &:
    \text{otherwise}
  \end{cases}
\label{eqn:omega2exchange}
\ee
as the symmetrized transition rate matrix for solute exchange. Note that we do not explicitly require that the solute-vacancy exchange occur with the same jump vectors as the vacancy jumps, but only that the same crystalline lattice is used for solute and vacancy migration. The diagonal components of $\hw^2$ correct the escape rate for complexes that are in the exchange space.

The final contribution corresponds to the vacancy jumping around the solute and contains all the remaining changes in rates of the vacancy where the solute does not change position. This includes the changes in site probability (solute-vacancy interaction), ``site exclusion,'' where solutes and vacancies cannot occupy the same site, and the subsequent escape rate changes for a complex. We define
\be
\begin{split}
  \hw^1_{\stategen,\statenew} &:= \delta(\xsv-\xspv)\delta_{\is\isp}\biggl\{(1-\delta(\xvv)\delta_{\is\iv})(1-\delta(\xvpv)\delta_{\isp\ivp})\hP_0^{1/2}(\stategen)\hW_{\stategen,\xsv\is\xvpv\ivp}\hP_0^{-1/2}(\xsv\is\xvpv\ivp) \\
  &\quad -\hw^0_{\stategen,\xsv\is\xvpv\ivp}
  -\delta(\xvv-\xvpv)\delta_{\iv\ivp}
  \sum_{\xv i}\hW_{\xsv\is\xvv\iv,\xsv\is\xv i}\biggr\}
\end{split}
\label{eqn:omega1ratesymmetrization}
\ee
which will be zero as $\xvv$ gets far away from the solute. The solute position remains unchanged, and we first consider the changes in rates when the vacancy does not jump into or originate from the solute site. The next term is the replacement of the simple vacancy rates: removing the jump where the vacancy would occupy the position of the solute, and the change in the escape rate (for $\statenew=\stategen$). The final summation corrects the escape rate for a complex corresponding to the changed rates \textit{but excluding solute-vacancy exchange} already included in $\hw^2$. This also includes so-called ``association'' (vacancy coming into the interaction range) and ``dissociation'' (vacancy leaving the interaction range) jumps that correspond to the formation/dissolution of the vacancy-solute complex.

Thus, we have $\hw = \hw^0 + \hw^1 + \hw^2$. This breakdown partly follows the labeling of the five-frequency model, though both ``3'' (dissociation) and ``4'' (association) jumps are subsumed as part of $\hw^1$, and produce non-zero bias vectors for sites at the edge of association for a solute-vacancy complex. We assume vacancy-solute interactions to be zero outside some finite range, which makes both $\hw^1$ and $\hw^2$ local.

\subsection{Green function solution}
\label{sec:Dyson}
The separation of jumps allows for the solution of the Green function first for $\hw^0$, which will be $\hg^0$, followed by the corrections due to $\dhw:=\hw^1+\hw^2$. This is particularly useful as $\dhw$ is strictly zero beyond a finite range; hence, the full Green function can be found exactly using
\be
\hg = ((\hg^0)^{-1} + \dhw)^{-1} = (\mathbf{1} + \hg^0\dhw)^{-1}\hg^0,
\label{eqn:GFcorrection}
\ee
which can be done for any subspace of states where $\dhw=0$ for all states \textit{not} in the subspace. This Dyson equation solution is exact for any value of $\dhw$; it does not rely on $\dhw$ being ``small'' in any sense, only that the full $\hw=\hw^0+\dhw$ is not a pathological transition matrix. We briefly outline the approach that takes advantage of translational invariance: First, we solve for $\hg^0$ by transforming $\hw^0$ to reciprocal space; the inverse is written as the sum of a pole, a discontinuity, and a smooth periodic function, which are transformed back to real space analytically for the first two terms, and numerically for the last. Next, we consider the subspace of states with non-zero bias vectors, and express our bias and correction vectors in a fully symmetrized representation, called \textit{vector-stars}; the Green function, and the changes in rates $\dhw$ can be written as matrices in this representation. We also note that, due to translational invariance in the solute position for the bias vector, we only need to consider $\hw^2$ in reciprocal space at $\qv_\text{s}=\zv$ (i.e., summed over all combinations of $\xsv-\xspv$). Finally, we can write \Eqn{GFcorrection} as a finite-dimensional matrix inversion problem, which can be solved numerically and used in \Eqn{Onsager-symmetric} to construct the transport coefficients. The use of a symmetrized representation---stars, and introducing vector-stars---also dictates the minimum information required for the computation of site probabilities (energies) and rates (energy barriers), providing for an automated computation of transport coefficients that is also efficient.

\section{Vacancy Green function}
\label{sec:vacancy}
We solve for the Green function $\hg^0$, the pseudoinverse of $\hw^0$. We note first that $\hw^0$ is diagonal and invariant in $\xsv\is,\xspv\isp$, so we will simplify by writing everything in terms of $\xvv\iv,\xvpv\ivp$ only, and we will drop the unnecessary ``v'' subscript throughout this section; we will also use $\alpha$ and $\beta$ to index eigenvalues rather than specify chemical species. Next, $\hw^0$ only depends on the difference of vacancy unit cells $\xv-\xvp$, so it will be block-diagonal in reciprocal space. The approach that follows is similar that Yasi and Trinkle\cite{Yasi2012b} for the lattice Green function; however, the treatment of the pole is different in this formulation, leading to a simpler solution without the need for truncated spherical harmonic expansions. The first step is a change of basis to reciprocal space and with the eigenvectors of $\hw^0$ at $\qv=\zv$; for a crystal with $\Nsites$ per unit cell, identify the non-positive eigenvalues $r^\alpha$, $\alpha=0\ldots\Nsites-1$ and normalized real eigenvectors $s^\alpha_a$ such that
\be
\sum_{\xvp,b}\left[\sum_{\xv} \hw^0_{\xv a,\xvp b}\right] s^\alpha_b = r^\alpha s^\alpha_a
\label{eqn:relaxivedefinition}
\ee
for all $\alpha$ and $a$. As $\hw^0$ is singular and negative-definite, there is one zero eigenvalue which we identify as $\alpha=0$; all other $r^\alpha<0$ for $\alpha>0$. The corresponding eigenvector is $s^0_a = \sqrt{\ProbV{a}}$, the square root of the probability for a vacancy to occupy the unit cell site $a$. Note that the eigenvalues $r^\alpha$ and vectors $s^\alpha$ depend on the rates, and hence the temperature $\kB T$. We construct basis vectors $\phi^\alpha_{\qv,\xv a}:=s^\alpha_a\exp(i \qv\cdot(\xv+\uv_a))/\sqrt{N}$ for a periodic system with $N$ unit cells (Born-von~Karman boundary conditions), so that
\be
\begin{split}
\hw^{\alpha\beta}(\qv,\qv') &:= \frac{1}{N}\sum_{\xv a,\xvp b} e^{i\qv\cdot(\xv+\uv_a)} s^\alpha_a\hw^0_{\xv a,\xvp b} s^\beta_b e^{-i\qv'\cdot(\xvp+\uv_b)}\\
&=\delta(\qv-\qv')\sum_{\xv,ab} s^\alpha_a\hw^0_{\zv a,\xv b} s^\beta_b e^{i\qv\cdot(\uv_a-(\xv+\uv_b))}
\end{split}
\ee
is the Fourier transform of $\hw^0$; as it is only nonzero for $\qv=\qv'$, we will use the shorthand $\hw^{\alpha\beta}(\qv,\qv')=\delta(\qv-\qv')\omega^{\alpha\beta}(\qv)$. Note that the vector $(\xv+\uv_b)-\uv_a$ corresponds to $\dxvh_{\zv a,\xv b}$ for the vacancy jump. Then, our inverse Fourier transform is given by
\be
\begin{split}
\hw^0_{\xv a,\xvp b} &= \frac{1}{N}\sum_{\qv\alpha,\qv'\beta} e^{-i\qv\cdot(\xv+\uv_a)}s^\alpha_a\hw^{\alpha\beta}(\qv,\qv')s^\beta_b e^{i\qv'\cdot(\xvp+\uv_b)}\\
  &=V\int_\text{BZ}\frac{d^3q}{(2\pi)^3} e^{-i\qv\cdot(\xv+\uv_a-\xvp-\uv_b)}
  s^\alpha_a\omega^{\alpha\beta}(\qv)s^\beta_b,
\end{split}
\ee
where $V=V_0/N$ is the volume per unit cell in the lattice and we replace the sum with an integral in the thermodynamic limit. We take advantage of similar definitions for the Fourier transform of $\hg^0_{\xv a,\xvp b}$ to get $g^{\alpha\beta}(\qv)$, and find that for all $\qv$, $g(\qv) = (\omega(\qv))^{+}$ the pseudoinverse of $\omega(\qv)$. For all $\qv\ne\zv$ in the Brillouin zone, $\omega(\qv)$ is non-singular and so the pseudoinverse is the inverse; at $\qv=\zv$, $\omega(\qv)$ is diagonal with one zero entry corresponding to $r^0=0$. Thus,
\be
\hg^0_{\xv a,\xvp b} = V\int_\text{BZ}\frac{d^3q}{(2\pi)^3} e^{-i\qv\cdot(\xv+\uv_a-\xvp-\uv_b)} \sum_{\alpha\beta}s^\alpha_a(\omega(\qv))^{-1}_{\alpha\beta}s^\beta_b,
\label{eqn:Green-function-integral}
\ee
and we need only to evaluate the integral in \Eqn{Green-function-integral} to find the Green function.

\subsection{Block inversion and Taylor expansion}
\label{sec:blockinversion}
To integrate \Eqn{Green-function-integral}, we first block $\omega(\qv)$ in the $\{s^\alpha\}$ basis, and Taylor expand the blocks in series from $\qv$ at the origin. The $\alpha=0$ eigenvector corresponds to the equilibrium distribution, which we call a \textit{diffusive} mode; the $\alpha>0$ eigenvectors correspond to distributions that decay with time (as $r^\alpha<0$), which we call \textit{relaxive} modes.%
\footnote{Diffusive and relaxive modes are analogous to acoustic and optical modes for phonons in a crystal, and a similar separation is used by Yasi and Trinkle to solve for the force-constant lattice Green function.\cite{Yasi2012b}}
Note that relaxive modes are only possible for $\Nsites>1$. We write our block matrix,
\be
\omega^{\alpha\beta}(\qv) = \left(
\begin{array}{c|c}
  DD(\qv)&DR(\qv)\\
  \hline
  RD(\qv)&RR(\qv)
  \end{array}
\right)
\label{eqn:blockomega}
\ee
where $DD(\qv)$ is a real $1\x1$ matrix corresponding to $\alpha=\beta=0$, $DR(\qv)=RD^\dagger(\qv)$ is a $1\x(\Nsites-1)$ matrix corresponding to $\alpha=0$ and $\beta>0$, and $RR(\qv)$ is a Hermitian $(\Nsites-1)\x(\Nsites-1)$ matrix corresponding to $\alpha>0$ and $\beta>0$. For small magnitude $q$, the leading order of $DD$ is $q^2$, of $DR$ and $RD$ is $iq$, and of $RR$ is $q^0$. For $DD(\qv)$, we have
\be
DD(\qv) := \sum_{\xv,ab} \Bigl(\ProbV{a}\ProbV{b}\Bigr)^{1/2}\hw^0_{\zv a,\xv b} e^{i\qv\cdot(\uv_a-(\xv+\uv_b))}
\ee
which can be simplified by noting that, from \Eqn{omega0ratesymmetrization}, $\Bigl(\ProbV{a}\ProbV{b}\Bigr)^{1/2}\hw^0_{\zv a,\xv b} = \hW^0_{\zv a,\xv b}\ProbV{a}$, which is the rate for the vacancy to transition from $\zv a$ to $\xv b$ times the probability to be at unit cell site $a$. The combination $\xv+\uv_b-\uv_a$ is the displacement vector for that same transition, $\dxvh_{\zv a,\xv b}$. Then, we can Taylor expand $\exp(-i\qv\cdot\dxvh_{\zv a,\xv b})$ up to fourth order in $q$ to get
\be
\begin{split}
DD(\qv) &= \sum_{\xv,ab} \hW^0_{\zv a,\xv b}\ProbV{a}\Biggl\{1-i\qv\cdot\dxvh_{\zv a,\xv b} - \frac12
(\qv\cdot\dxvh_{\zv a,\xv b})^2 - \frac{i}{6}(\qv\cdot\dxvh_{\zv a,\xv b})^3
+\frac{1}{24}(\qv\cdot\dxvh_{\zv a,\xv b})^4 + O(q^5)\Biggr\}\\
&=-\qv\cdot\Biggl[\frac12\sum_{\xv,ab} \dxvh_{\zv a,\xv b}\ox\dxvh_{\zv a,\xv b}\hW^0_{\zv a,\xv b}\ProbV{a}\Biggr]\cdot\qv + \frac{1}{24}\sum_{\xv,ab}(\qv\cdot\dxvh_{\zv a,\xv b})^4\hW^0_{\zv a,\xv b}\ProbV{a} + O(q^6)
\label{eqn:DDexpansion}
\end{split}
\ee
where the $q^0$ term is zero as $\sum_{\xv,b}\hW^0_{\zv a,\xv b}=0$ for all $a$, and all terms containing odd powers of $\dxvh_{\zv a,\xv b}$ sum to zero because of detailed balance and antisymmetry of $\dxvh$ for the reverse jumps. The first term in braces is the same as the first term in \Eqn{transportcoeff-eta} (after multiplying by a factor of $\kB T/\cv$), or the ``bare'' (uncorrelated) contribution to the vacancy diffusion.  For $RD(\qv)$, we have
\be
RD^{\alpha0}(\qv) := \sum_{\xv,ab} s^\alpha_a \Bigl(\ProbV{b}\Bigr)^{1/2}\hw^0_{\zv a,\xv b} e^{i\qv\cdot(\uv_a-(\xv+\uv_b))}
\ee
which can be simplified by noting that, from \Eqn{omega0ratesymmetrization}, $\left(\ProbV{b}\right)^{1/2}\hw^0_{\zv a,\xv b} = \hW^0_{\zv a,\xv b}\left(\ProbV{a}\right)^{1/2}$. Then, we can Taylor expand $\exp(-i\qv\cdot\dxvh_{\zv a,\xv b})$ up to third order in $q$ to get
\be
\begin{split}
RD^{\alpha0}(\qv) &= \sum_{\xv,ab} s^\alpha_a\hW^0_{\zv a,\xv b}\Bigl(\ProbV{a}\Bigr)^{1/2}\biggl\{1-i\qv\cdot\dxvh_{\zv a,\xv b} - \frac12 (\qv\cdot\dxvh_{\zv a,\xv b})^2 - \frac{i}{6}(\qv\cdot\dxvh_{\zv a,\xv b})^3 + O(q^4)\biggr\}\\
&=-i\qv\cdot\Biggl[\sum_a s^\alpha_a\sum_{\xv,b} \dxvh_{\zv a,\xv b}\hW^0_{\zv a,\xv b}\Bigl(\ProbV{a}\Bigr)^{1/2}\Biggr]\\
&\quad - \sum_{\xv,ab}s^\alpha_a\Biggl[\frac12(\qv\cdot\dxvh_{\zv a,\xv b})^2 + \frac{i}{6}(\qv\cdot\dxvh_{\zv a,\xv b})^3\Biggr]\hW^0_{\zv a,\xv b}\Bigl(\ProbV{a}\Bigr)^{1/2} + O(q^4)
\end{split}
\label{eqn:RDexpansion}
\ee
where the $q^0$ term is zero as $\sum_{\xv,b}\hW^0_{\zv a,\xv b}=0$ for all $a$. The first term in brackets is the bias vector (c.f. \Eqn{ratebias}) for the vacancy at sites $a$, rotated into the relaxive basis $s^\alpha$. Similarly, $DR^{0\alpha}(\qv) = (RD^{\alpha0}(\qv))^*$. Finally, for $RR(\qv)$, we have
\be
RR^{\alpha\beta}(\qv) := \sum_{\xv,ab} s^\alpha_as^\beta_b \hw^0_{\zv a,\xv b} e^{i\qv\cdot(\uv_a-(\xv+\uv_b))}
\ee
which does not admit any obvious simplification. Then, we can Taylor expand $\exp(-i\qv\cdot\dxvh_{\zv a,\xv b})$ up to second order in $q$ to get
\be
\begin{split}
RR^{\alpha\beta}(\qv) &= \sum_{\xv,ab} s^\alpha_a s^\beta_b \hw^0_{\zv a,\xv b}\Biggl\{1-i\qv\cdot\dxvh_{\zv a,\xv b} - \frac12
(\qv\cdot\dxvh_{\zv a,\xv b})^2 + O(q^3)\Biggr\}\\
&= r^\alpha\delta_{\alpha\beta} - \sum_{\xv,ab}s^\alpha_as^\beta_b\biggl[i\qv\cdot\dxvh_{\zv a,\xv b} + \frac12(\qv\cdot\dxvh_{\zv a,\xv b})^2\biggr]\hw^0_{\zv a,\xv b} + O(q^3)
\end{split}
\label{eqn:RRexpansion}
\ee
where the $q^0$ term is diagonal, as $s^\alpha$ are the eigenvectors of $\sum_{\xv} \hw^0_{\zv a,\xv b}$ in \Eqn{relaxivedefinition}. As $RR(\qv)$ is diagonal for $\qv=\zv$, we can also Taylor expand $RR(\qv)^{-1}$ up to second order in $q$ to get
\be
\begin{split}
  (RR(\qv)^{-1})^{\alpha\beta} &= (r^\alpha)^{-1}\delta_{\alpha\beta}
  + \sum_{\xv,ab}(r^\alpha)^{-1}s^\alpha_as^\beta_b(r^\beta)^{-1}\biggl[i\qv\cdot\dxvh_{\zv a,\xv b} + \frac12(\qv\cdot\dxvh_{\zv a,\xv b})^2\biggr]\hw^0_{\zv a,\xv b}\\
  & - \sum_{cc'}\sum_{\xv,a}(r^\alpha)^{-1}s^\alpha_a\bigl[\qv\cdot\dxvh_{\zv a,\xv c}\bigr]\hw^0_{\zv a,\xv c}
  \biggl(\sum_{\gamma>0} s^\gamma_c(r^\gamma)^{-1}s^\gamma_{c'}\biggr)
  \sum_{\xv,b}s^\beta_b(r^\beta)^{-1}\bigl[\qv\cdot\dxvh_{\zv c',\xv b}\bigr]\hw^0_{\zv c',\xv b}
  + O(q^3)
\end{split}
\label{eqn:RRinvexpansion}
\ee

As an intermediate step to the block inversion of $\omega^{\alpha\beta}(\qv)$, we scale reciprocal and real space vectors based on the eigenvalues of the vacancy diffusivity. We construct the $1\times1$ matrix
\be
D(\qv) := DD(\qv) - DR(\qv)\,(RR(\qv)^{-1})\,RD(\qv)
\label{eqn:Dq}
\ee
which is the Schur complement of $RR(\qv)$, and using \Eqn{DDexpansion}, \Eqn{RDexpansion}, and \Eqn{RRinvexpansion}, we can Taylor expand
\be
\begin{split}
  D(\qv) &= -\qv\cdot\Biggl[\frac12\sum_{\xv,ab} \dxvh_{\zv a,\xv b}\ox\dxvh_{\zv a,\xv b}\hW^0_{\zv a,\xv b}\ProbV{a}
    +\sum_{aa'}\biggl(\sum_{\xv,b} \dxvh_{\zv a,\xv b}\hW^0_{\zv a,\xv b}\Bigl(\ProbV{a}\Bigr)^{1/2}\biggr)\\
    &\qquad \biggl(\sum_{\alpha>0}s^\alpha_a(r^\alpha)^{-1}s^\alpha_{a'}\biggr)
    \biggl(\sum_{\xv,b} \dxvh_{\zv a',\xv b}\hW^0_{\zv a',\xv b}\Bigl(\ProbV{a'}\Bigr)^{1/2}\biggr)
    \Biggr]\cdot\qv + O(q^4).
\end{split}
\label{eqn:Dexpansion}
\ee
The terms cubic in $q$ must vanish, as they are purely imaginary, and $D(\qv)$ is real by virtue of being a Hermitian matrix. The $q^4$ term can be expressed as a fourth-order homogeneous polynomial in the components of $\qv$. Then, as $r^\alpha$ are the nonzero eigenvalues of $\sum_{\xv} \hw^0_{\zv a,\xv b}$ for $\alpha>0$, the matrix $\sum_{\alpha>0} s^\alpha_a(r^\alpha)^{-1}s^\alpha_{a'}$ is the pseudoinverse. The two vectors dotted into that matrix are the bias vectors, as identified earlier, and so by \Eqn{Onsager-symmetric}, we can write
\be
D(\qv) = -\qv\cdot \Dvv\cdot\qv + O(q^4)
\label{eqn:Dexpansion2}
\ee
where\cite{TrinkleElastodiffusivity2016}
\be
\Dvv := \frac{\kB T}{\cv} \Lvv =
\frac12\sum_{\xv,ab} \dxvh_{\zv a,\xv b}\ox\dxvh_{\zv a,\xv b}\hW^0_{\zv a,\xv b}\ProbV{a}
    +\sum_{aa'} \bv^\text{v}_a\ox \biggl(\sum_{\alpha>0}s^\alpha_a(r^\alpha)^{-1}s^\alpha_{a'}\biggr)\;\bv^\text{v}_{a'}
\label{eqn:vacancy-diffusivity}
\ee
for the bare vacancy bias vector
\be
\bv^\text{v}_a:=\sum_{\xv,b} \dxvh_{\zv a,\xv b}\hW^0_{\zv a,\xv b}\Bigl(\ProbV{a}\Bigr)^{1/2}.
\label{eqn:bareratebias}
\ee
Note that the second-rank tensor $\Dvv$ is symmetric and positive-definite; therefore, it has three real, positive eigenvalues $\di_i$ with corresponding orthonormal eigenvectors $\ei_i$. Note that if $\Dvv$ is isotropic (e.g., a cubic system), $\di_1=\di_2=\di_3$. We define the following coordinate transforms to ``scaled'' reciprocal and real space coordinates,
\be
p_i := \di_i^{1/2}(\ei_i\cdot\qv),
\quad
y_i := \di_i^{-1/2}(\ei_i\cdot\xv)
\label{eqn:scaled-components}
\ee
and then
\be
\qv = \sum_i \di_i^{-1/2} p_i\ei_i,
\quad
\xv = \sum_i \di_i^{1/2} y_i\ei_i
\ee
so that $-\qv\cdot\Dvv\cdot\qv = -|\pv|^2$ and $\exp(-i\qv\cdot\xv)=\exp(-i\pv\cdot\yv)$. In this scaled coordinate system,
\be
D(\pv) = -p^2 + \sum_{n_1+n_2+n_3 = 4} \DD_{[n_1n_2n_3]}p_1^{n_1} p_2^{n_2} p_3^{n_3} + O(p^6),
\label{Dpexpansion}
\ee
where we have explicitly introduced the components of the fourth order expansion, $\DD_{[n_1n_2n_3]}$, and the expansions in \Eqn{DDexpansion}, \Eqn{RDexpansion}, and \Eqn{RRinvexpansion} retain the same form in $p$. Finally, we Taylor expand $D(\pv)^{-1}$ up to order $p^0$,
\be
D(\pv)^{-1} = -\frac{1}{p^2} - \sum_{n_1+n_2+n_3 = 4} \DD_{[n_1n_2n_3]}\frac{p_1^{n_1} p_2^{n_2} p_3^{n_3}}{p^4} + O(p^2),
\label{eqn:Dinvpexpansion}
\ee
Because all of the Taylor expansions involve products of $(\qv\cdot\dxvh)$, all terms of order $q^n$ (or $p^n$) are strictly homogeneous polynomials of order $n$ in components of $\qv$ (or $\pv$). Note that the eigenvalues $\di_i$ and vectors $\ei_i$ depend on the diffusivity, and hence the temperature $\kB T$.

We can block invert $\omega(\qv)$ in the $\{s^\alpha\}$ basis with scaled coordinates $\pv$, and analytically treat terms that behave as $p^{-2}$, $p^{-1}$, and $p^0$ at the origin. The block inverse of \Eqn{blockomega} is most easily written in terms of the inverses of $D(\pv)$ (\Eqn{Dinvpexpansion}) and $RR(\pv)$ (\Eqn{RRinvexpansion}),
\be
g^{\alpha\beta}(\pv) = \left(
\begin{array}{c|c}
  D(\pv)^{-1} & -(D(\pv)^{-1})\,DR(\pv)\,(RR(\pv)^{-1})\\
  \hline
  -(RR(\pv)^{-1})\,RD(\pv)\,(D(\pv)^{-1}) & RR(\pv)^{-1} + (RR(\pv)^{-1})\, RD(\pv)\, (D(\pv)^{-1})\, DR(\pv)\, (RR(\pv)^{-1})\\
  \end{array}
\right)
\label{eqn:blockGF}
\ee
This form is chosen as the two matrix inverses---$D(\pv)^{-1}$ and $RR(\pv)^{-1}$---admit straightforward Taylor series expansions by virtue of their isotropic (in $\pv$) leading order terms: $p^2$ for $D(\pv)$ and $r^\alpha\delta_{\alpha\beta}$ for $RR(\pv)$. The leading order of the diffusive-diffusive block is $-p^{-2}$; the leading order of the relaxive-diffusive (and diffusive-relaxive) blocks is $\sim i\pv/p^2 \sim p^{-1}$; the leading order of the relaxive-relaxive block is $\sim \pv\pv/p^2 \sim p^0$, which can appear as a discontinuity at the origin (approaching the origin from different directions will give different values\cite{Yasi2012b}). In the subsequent series expansions, all terms up to $p^0$ can be written as homogenous polynomials of order four or lower divided by a power of $p$. We follow a semicontinuum approach\cite{Tewary1971, TrinkleLGF2008, Yasi2012b} and expand each block as a sum of terms of order $p^{-2}$, $ip^{-1}$, and $p^0$ multiplied by a Gaussian, and a finite, smooth, periodic function; for example,
\be
g^{00}(\pv) = -\frac{\exp(-p^2/\pmax^2)}{p^2} - 
\exp(-p^2/\pmax^2) \sum_{n_1+n_2+n_3 = 4} \DD_{[n_1n_2n_3]}\frac{p_1^{n_1} p_2^{n_2} p_3^{n_3}}{p^4}
+ g_\text{sc}^{00}(\qv),
\ee
for a width $\pmax$ (described below) and where the (smooth) semicontinuum piece $g_\text{sc}^{00}(\qv)$ is the difference between the first two terms and $D(\pv)^{-1}$. The first term is a second-order pole in $\pv$, while the second term is a discontinuity at $\pv=\zv$; it has different values in the limit as $\pv\to0$ depending on the direction for approaching the origin. Note also that as $p\to 0$, $g_\text{sc}^{00}(\pv=\zv)=-1/\pmax^2$. The first two terms needs to be inverse Fourier transformed analytically, while the last term can be evaluated numerically on a finite grid of $\qv$ in the Brillouin zone.\cite{TrinkleLGF2008,Ghazisaeidi2009} We will evaluate the analytic inverse Fourier transforms by expanding the integral in \Eqn{Green-function-integral} to all space. This requires that $\exp(-p^2/\pmax^2)$ be sufficiently small at the Brillouin zone edge; hence, the width $\pmax$ is chosen so that
\be
\pmax\le\left(\frac{\inf\{\qv\cdot\Dvv\cdot\qv: \qv\in\text{BZ boundary}\}}{-\ln\threshold}\right)^{1/2}
\label{eqn:pmax}
\ee
for a threshold $\threshold$; then $\exp(-p^2/\pmax^2)\le\threshold$ everywhere on the boundary of the Brillouin zone. Note that smaller values of $\pmax$ require more grid points for the inverse Fourier transform of $g_\text{sc}^{00}(\qv)$. Note also that large anisotropy in $\Dvv$ may necessitate an associated anisotropy in the grid of $\qv$; see \Sec{semicontinuum}.

\subsection{Inverse transform of second-order pole}
\label{sec:pole}
When we inverse fourier transform the $p^{-2}$ pole, we recover the large $x$ diffusive behavior where the Green function is inversely proportional to $x$. First,
\be
-V\int_\text{BZ}\frac{d^3q}{(2\pi)^3} e^{-i\qv\cdot(\xv+\uv_a-\xvp-\uv_b)} 
\frac{\exp(-p^2/\pmax^2)}{p^2} = 
-\frac{V}{(\di_1 \di_2 \di_3)^{1/2}}\int \frac{d^3p}{(2\pi)^3} \frac{e^{-i\pv\cdot\yv}\exp(-p^2/\pmax^2)}{p^2},
\label{eqn:G2-FT}
\ee
where we have taken the limits of integration from the Brillouin zone out to all space; see below for an estimate of the error induced. The function to inverse Fourier transform is spherically symmetric, and so is the solution in $\yv$,
\be
-\frac{V}{4\pi(\di_1 \di_2 \di_3)^{1/2} y}\erf\left(\frac{y\pmax}{2}\right)
=
-\frac{V}{4\pi}\frac{\erf\biggl(\frac12\Bigl(\xv\cdot(\Dvv)^{-1}\cdot\xv)\Bigr)^{1/2}\pmax
\biggr)}{\Bigl(\det\Dvv(\xv\cdot(\Dvv)^{-1}\cdot\xv)\Bigr)^{1/2}}
\label{eqn:GF-pole}
\ee
by noting that
\be
y = \left(\sum_i \di_i^{-1} (\ei_i\cdot\xv)^2\right)^{1/2} =
\left(\xv\cdot(\Dvv)^{-1}\cdot\xv\right)^{1/2}
\label{eqn:ymagn}
\ee
where $\xv$ is the vector connecting any two vacancy sites, and that $\di_1 \di_2 \di_3 = \det\Dvv$. The inverse Fourier transform value at $\xv=\zv$ is
\be
\frac{V\pmax}{4\sqrt{\pi^3 \di_1 \di_2 \di_3}},
\ee
which is finite, and the leading term in large $\xv$ is
\be
-\frac{V}{4\pi}\Bigl(\det\Dvv(\xv\cdot(\Dvv)^{-1}\cdot\xv)\Bigr)^{-1/2},
\ee
which is the solution to $-\nabla\cdot(\Dvv\nabla g) = \delta(\xv)$. To estimate the error, we integrate instead from a sphere in $\pv$-space inscribed in the Brillouin zone out to infinity; this is given by a radius $p_0 = \pmax \sqrt{-\ln\threshold}\approx \pi D^{1/2}/a_0$ for lattice constant $a_0$. The largest error occurs for $\xv=\zv$, where
\be
\begin{split}
  \frac{V}{(\di_1 \di_2 \di_3)^{1/2}}\int_{p_0}^\infty \frac{d^3p}{(2\pi)^3} \frac{\exp(-p^2/\pmax^2)}{p^2} &= \frac{V}{(\di_1 \di_2 \di_3)^{1/2}} \frac{\pmax\sqrt\pi}{4\pi^2} \erfc(p_0/\pmax)\\
  &\approx \frac{V/a_0}{4\pi D}\threshold,
\end{split}
\label{eqn:pmax-error}
\ee
which, as $g^0(\zv)\sim V/(Da_0)\sim \hw_0^{-1}$, shows that the truncation error is approximately $g^0(\zv)\threshold$.

\subsection{Inverse transform of first-order poles and discontinuities}
\label{sec:disc}
The remaining analytic terms are inverse Fourier transformed by grouping homogeneous polynomials with common orbital angular momenta $\ell$. The first-order poles and discontinuities can each be written as a sum of up to fourth order polynomials in the normalized components $\pv/p$. Moreover, as shown in \App{orientation}, a homogeneous polynomial expansion of order $L$ can be expanded into a series of homogeneous polynomial expansions of order from $\ell=0$ to $\ell=L$ such that each expansion \textit{only} contains contributions from spherical harmonics of a single orbital angular momentum. The inverse Fourier transform of the separable contribution $p^n \exp(-p^2/\pmax^2) Y_\ell^m(\hat\pv)$ for power $n=-1,0$ and spherical harmonic $Y_\ell^m$ is
\be
\begin{split}
  &V\int_\text{BZ}\frac{d^3q}{(2\pi)^3} e^{-i\qv\cdot(\xv+\uv_a-\xvp-\uv_b)} 
  p^n\exp(-p^2/\pmax^2) Y_\ell^m(\hat\pv)\\
  &\quad= 
  \frac{V}{(\di_1 \di_2 \di_3)^{1/2}}\int \frac{d^3p}{(2\pi)^3} p^n\exp(-p^2/\pmax^2) e^{-i\pv\cdot\yv} p^nY_\ell^m(\hat\pv) \exp(-p^2/\pmax^2)\\
  &\quad= f_{n\ell}(y) Y_\ell^m(\hat\yv)
\end{split}
\label{eqn:power-IFT}
\ee
where
\be
\begin{split}
f_{n\ell}(y) &:= \frac{V}{(\di_1 \di_2 \di_3)^{1/2}}(-i)^\ell \int_0^\infty
\frac{dp}{2\pi^2} p^{2+n} j_\ell(py) \exp(-p^2/\pmax^2)\\
&= \frac{(-i)^\ell V}{\sqrt{\pi^3\di_1 \di_2 \di_3}} \frac{1}{y^{3+n}}\cdot \frac{\Gamma\Bigl(\frac{3+n+\ell}{2}\Bigr)}{2^{-n}\Gamma\Bigl(\frac32 + \ell\Bigr)}
\left(\frac{y\pmax}{2}\right)^{3+n+\ell}
{}_1F_1\biggl(\frac{3+n+\ell}{2}, \frac32 + \ell; -\left(\frac{y\pmax}{2}\right)^2\biggr),
\end{split}
\label{eqn:fnl-definition}
\ee
for spherical Bessel function $j_\ell$ and confluent hypergeometric function $_1F_1$ (c.f. 9.210 in Gradshteyn and Ryzhik\cite{IntegralTables}). As the inverse Fourier transform transforms $Y_\ell^m(\hat\pv)$ into $Y_\ell^m(\hat\yv)$, the homogeneous polynomial expansions in $\hat\pv$ of order $\ell$ corresponding to a single orbital angular momentum transforms to the same homogeneous polynomial expansion in $\hat\yv$, where the radial contribution $p^n\exp(-p^2/\pmax^2)$ transforms to $f_{n\ell}(y)$. For small $y$,
\be
f_{n\ell}(y)= y^\ell
\frac{(-i)^\ell V}{\sqrt{\pi^3\di_1 \di_2 \di_3}} \frac{\Gamma\Bigl(\frac{3+n+\ell}{2}\Bigr)}{2^{-n}\Gamma\Bigl(\frac32 + \ell\Bigr)}
\left(\frac{\pmax}{2}\right)^{3+n}
+ O\bigl(y^{2+\ell}\bigr),
\label{eqn:fnl-smally}
\ee
and for large $y$,
\be
f_{n\ell}(y)= 
y^{-(3+n)}\frac{(-i)^\ell V}{\sqrt{\pi^3\di_1 \di_2 \di_3}} \frac{\Gamma\Bigl(\frac{3+n+\ell}{2}\Bigr)}{2^{-n}\Gamma\Bigl(\frac{\ell-n}{2}\Bigr)}
+ O\bigl(y^{-(5+n)}\bigr).
\label{eqn:fnl-largey}
\ee
Note also that $f_{-2\,0}(y)$ is a special case, captured in the \Sec{pole}. The relationship between $\xv$ and $y$ is given by \Eqn{ymagn} and the components by \Eqn{scaled-components}.

\subsection{Inverse transform of semicontinuum piece}
\label{sec:semicontinuum}
The final contribution to the inverse Fourier transform of $g^{\alpha\beta}$ is $g^{\alpha\beta}_\text{sc}$, which requires numerical integration on a regularly spaced grid in the Brillouin zone. This function is smooth (after subtracting off the poles and discontinuity) and periodic, so it converges quickly with the number of grid points.\cite{Ghazisaeidi2009} We use a regular, gamma-centered $N_1\x N_2\x N_3$ mesh (each $N_i$ is even) in terms of the reciprocal lattice vectors $\bv_1, \bv_2, \bv_3$ as
\be
\qv = \frac{m_1}{N_1}\bv_1 +  \frac{m_2}{N_2}\bv_2 +  \frac{m_3}{N_3}\bv_3.
\ee
Note that alternative meshes, like Monkhorst-Pack\cite{Monkhorst76} are possible as well. We initially generate the mesh of $\qv$ using $m_i = -(N_i/2)+1\ldots(N_i/2)$, but then we translate $\qv$ so that they remain entirely within the Brillouin zone. Our Brillouin zone is defined by a set of reciprocal lattice vectors $\GBZ:=\{\Gv\}$ where $\qv$ is in the Brillouin zone if and only if $\qv\cdot\Gv \le \Gv^2/2$ for all $\Gv\in\GBZ$. So, once we generate our initial set of $\qv$, we check that each lies inside the Brillouin zone; if we find a $\Gv\in\GBZ$ such that $\qv\cdot\Gv > \Gv^2/2$, we replace it with $\qv-\Gv$. At this stage, all of our $\qv$ are equally weighted, and so we approximate our integral $V\int_\text{BZ} d^3q/(2\pi)^3$ as the average value over our $\qv$.

Next, we take advantage of space group symmetry (c.f. \Sec{symmetry2}) to reduce the number of unique $\qv$ we need to consider, and replace our average with a weighted average. We group our $\qv$ points in \textit{stars}; that is, a set of points that are all related to one another by rotation operations $\R$. As the group is closed, we can select a single $\qv$ representative from each star, and compute $g^{\alpha\beta}_\text{sc}$ for that $\qv$; the weight $w_\qv$ in the average will be the number of $q$-points in that star divided by the total number of $q$-points. Furthermore, we can rotate back to the original site indices $a,b$ from our eigenvectors $s^\alpha_a$ and $s^\beta_b$.
When a group operation $\Rtv$ is applied to the initial position $\xv+\uv_a$ and final position $\xvp+\uv_b$, it transforms the vector $\dxv = \xvp+\uv_b-\xv+\uv_a$ to $\R\,\dxv$ but also changes the corresponding site indices to $a' = \Rtv(a)$ and $b' = \Rtv(b)$, which are of the same Wyckoff position as $a$ and $b$.
To perform the inverse Fourier transform, the contribution to the inverse Fourier transform for a star $\qv$ is given by
\be
\sum_{\qv} w_\qv
\frac1{\Ng}\sum_R e^{i\qv\cdot(\R\,\dxv)} g^{\Rtv(a),\Rtv(b)}_\text{sc}(\qv)
\label{eqn:symmqpoint}
\ee
where there are $\Ng$ group operations $\R$. As we can apply $\R$ either to $\dxv$ or $\qv$, it is computationally more efficient to apply to $\dxv$. Note that the use of a regular grid to inverse Fourier transform requires that we include sufficient density to avoid aliasing errors; that is, for the largest $\dxv$, the smallest non-zero value of $\qv\cdot\dxv$ must be smaller than $\pi$. As the number of $q$-points increases, the error scales no slower than $N_1^{-4} + N_2^{-4} + N_3^{-4}$ (c.f., \rcite{Ghazisaeidi2009}).

\subsection{Algorithm summary}
\label{sec:GFsummary}
We summarize the computational steps in the calculation of the vacancy Green function for an arbitrary crystal, given a network of jumps between crystalline sites with known occupation probabilities $\ProbV{}$ and the corresponding rates $\hW^0$:
\begin{enumerate}
\item Construct the symmetric rate matrix $\hw^0$ for the vacancy (\Eqn{omega0ratesymmetrization});
\item Find the eigenvalues $r^\alpha$ and eigenvectors $s^\alpha$ (\Eqn{relaxivedefinition});
\item Rotate $\omega$ to identify diffusive-relaxive block matrices; construct Taylor expansions of corresponding blocks (\Eqn{DDexpansion}, \Eqn{RDexpansion}, \Eqn{RRinvexpansion}, \Eqn{Dexpansion});
\item Find $\Dvv$ from expansion of $D(\qv)$ (\Eqn{vacancy-diffusivity}), and diagonalize to find the coordinate transformation from the eigenvalues $\di_i$ and eigenvectors $\ei_i$ (\Eqn{scaled-components}), and $\pmax$ (\Eqn{pmax});
\item Transform all Taylor expansions to $\pv$, and compute Taylor expansions of blocks of $g(\pv)$ up to $p^0$ (\Eqn{blockGF}); separate the Taylor expansions by powers of $p^n$ and homogeneous polynomials of constant orbital angular momentum $\ell$ (\App{orientation});
\item For each block, and for each $\qv$ grid point (\Sec{semicontinuum}), find the semicontinuum $g^{\alpha\beta}_\text{sc}$ by subtracting the Taylor expansions multiplied by $\exp(-p^2/\pmax^2)$ from the inverse of $\omega(\qv)$, and then rotate back to $g^{ab}_\text{sc}(\qv)$;
\item Rotate the block Taylor expansion back to the original site basis.
\end{enumerate}
This completes all preparatory work necessary to efficiently compute the Green function for a series of different initial and final vacancy positions by expressing the Fourier transform as a series of analytic terms plus a smooth function represented on a finite grid. For initial vacancy position  $\xv+\uv_a$ and final vacancy position $\xvp+\uv_b$, we
\begin{enumerate}
\item Compute $\dxv = \xvp+\uv_b-\xv+\uv_a$, and find the $y_i$ components (\Eqn{scaled-components}), $y$ magnitude (\Eqn{ymagn}), and normalized components $\hat y_i = y_i/y$;
\item Inverse Fourier transform all Taylor expansion pieces (\Eqn{GF-pole}, \Eqn{power-IFT}, and \Eqn{fnl-definition});
\item Add the semicontinuum discrete inverse Fourier transform contribution (\Eqn{symmqpoint}).
\end{enumerate}
This approach is optimal when the Green function is needed for a large number of $\xvv\iv$-$\xvpv\ivp$ pairs for a given set of rates (at a single temperature). For each new set of rates, the first seven steps must be repeated.

\section{Dilute-vacancy/dilute-solute transport}
\label{sec:final}
Given our vacancy Green function without a solute, we construct the Green function solution for our system with a solute using \Eqn{GFcorrection}, and to evaluate transport using \Eqn{Onsager-symmetric}. This requires a few considerations: (1) translational invariance for the solute; (2) crystalline symmetry and the expansion of our lattice functions; (3) systems where vacancy states have non-zero bias before the introduction of the solute; and (4) avoiding catastrophic roundoff error in systems with rates differing by more than ten orders of magnitude. We conclude with expressions for the Onsager coefficients in the dilute-vacancy/dilute-solute limit.

\subsection{Translational invariance of the Green function solution}
\label{sec:translate}
We need to consider all states $\stategen$ that have a non-zero bias vector. Due to translational invariance, the bias vector is independent of $\xsv$. This simplifies our problem, as we are not interested in $\hg$ itself, but rather  $\sum_{\nconf,\nconf'}\bv^\alpha(\nconf)\hg_{\nconf\nconf'}\bv^\beta(\nconf')$. We note that $\bv^\alpha(\stategen)$ is independent of $\xsv$, so
\begin{multline}
\sum_{\xsv\is,\xspv\isp}\sum_{\xvv\iv,\xvpv\ivp}\bv^\alpha(\stategen)\ox\hg_{\stategen,\statenew}\bv^\beta(\statenew) \\
= \sum_{\is,\isp}\sum_{\xvv\iv,\xvpv\ivp}\bv^\alpha(\staterep)\ox\left(N\sum_{\xsv}\hg_{\staterep,\xsv\isp\xvpv\ivp}\right)\bv^\beta(\zv\isp\xvpv\ivp).
\end{multline}
The most straightforward way to evaluate the quantity in parenthesis is to note that, as $\hg_{\stategen,\statenew}$ depends only on $\xsv-\xspv$, this is equal to the $\qv_\text{s}=\zv$ term of its Fourier transform (which is diagonal). Then, if we return to \Eqn{GFcorrection}, we note that both $\hg$ and $\hg^{-1}$ have the same translational symmetry with respect to $\xsv-\xspv$, so that
\be
\sum_{\xspv}\hg_{\staterep,\statenew} = 
\left((\hg^0_{\xvv\iv,\xvpv\ivp})^{-1}\delta_{\is\isp} +  \hw^1_{\staterep,\statenewrep}
+ \sum_{\xspv}\hw^2_{\staterep,\statenew}
\right)^{-1}.
\label{eqn:GF-translation}
\ee
That is: we can replace $\hw^2$ in \Eqn{GFcorrection} with the sum over all solute positions, and work entirely with the positions of the vacancy $\xvv\iv,\xvpv\ivp$ and solute indices $\is,\isp$. Then our matrix inverse is strictly in the space of the kinetic shells (see below). The first two terms in the right-hand side of \Eqn{GF-translation} zero if $\is\ne\isp$; only $\hw^2$ produces translation between sites for the solute when $\Nsites>1$.

The reduction due to translational invariance dictates a finite set of states to consider for any vacancy-mediated diffusion problem, following Nastar~\et\cite{Nastar2000,Nastar2005,Garnier2013a,Garnier2013c,Garnier2014a,Garnier2014c}: the thermodynamic and kinetic ``shells.'' Restricting to $\xsv=\zv$, there are a finite set of states where $\staterep$ has a different (free) energy than an isolated solute at $\is$ and vacancy at $\iv$; these can be thought of as ``shells'' of vacancy states surrounding a solute with a non-zero interaction, called the \textit{thermodynamic shell}. The change in energy can also change the rates for states that can transition to and from those states; we define the \textit{kinetic shell} as any state $\staterep$ for which $\dhw_{\staterep,\statenew}\ne0$ for some $\statenew$.\footnote{%
  It should be noted that our use of ``kinetic shell'' is different than Nastar~\et\ Our kinetic shell is defined for states with changes in rates only---making it a finite space for a finite interaction range---while Nastar~\et\ uses the kinetic shell to identify the set of states for which the effective Hamiltonian will be non-zero, and hence it acts as a convergence parameter in SCMF calculations.}
This finite set of states dictates the minimum subspace necessary to compute $\hg$, for these are also the states with changes to $\hbv$. For simplicity, we will identify the size of the subspace by how the minimum number of transitions necessary for the vacancy in any of the states to transition onto the solute site\cite{Garnier2013a,Garnier2013c,Garnier2014a,Garnier2014c}: a 1nn ``first neighbor'' thermodynamic shell will require a (1nn)$^2$ kinetic shell---first neightbors of first neighbors---which might include more than the second nearest neighbors. For example, in FCC, the (1nn)$^2$ kinetic shell includes up to the fourth nearest neighbor of the solute.

\subsection{Symmetry and state functions}
\label{sec:symmetry2}
The space group symmetry of the crystal allows for significant reduction in the complexity of the diffusion problem. Below, we explicitly develop the linear basis to represent our configuration-dependent quantities for the dilute limit, and construct fully symmetrized basis functions. This requires an expansion of the notion of a star (a set of vectors related by point group operations) to that of a ``crystal star'' which correspond with symmetry-equivalent solute-vacancy complexes under space group operations. We then consider vectors for a given vector quantities ascribed to our configurations, which motivates the development of a ``vector star'' to develop the basis functions for the bias and symmetrized correction vectors.

We use the Seitz notation\cite{Glazer2013} for a symmetry operation $\Rtv$, where for a point $\xv$, $\Rtv\xv:=R\,\xv + \tv$. Then, the inverse $\Rtvinv = \Rtvinvexplicit$. We can apply a symmetry operation to a state $\stategen$ by defining $\Rtv(\stategen):=\statenew$ where $\Rtv(\xsv+\uv_{\is}) = \xspv + \uv_{\isp}$ and $\Rtv(\xsv+\xvv+\uv_{\iv}) = \xspv + \xvpv + \uv_{\ivp}$. This is well-defined as each symmetry operation maps a position in the crystal to another position in the crystal, and each position in the crystal has a unique representation. The full set of operations make up the \textit{space group}; for our purposes here, we will be interested in a subset of group operations that map $\xsv=\zv$ to $\xspv=\zv$; after lattice translations are added, this subset generates the entire space group. We are interested in operations on our state space---the full set of states $\staterep$ with a non-zero $\dhw$---and specifically scalars and vectors at each state. A trivial extension to \textit{tensor state functions} is possible, but not described here.\cite{TrinkleElastodiffusivity2016}

A \textit{state scalar function} $f_{\stategen}$ is a function that has a scalar value for each state $\stategen$; we will primarily consider translationally invariant state scalar functions where $f_{\stategen}=f_{\staterep}$ for all $\stategen$. The application of $\Rtv$ to $f_{\stategen}$ produces a new lattice function $\Rtv f := g$ such that $f_{\stategen} = g_{\Rtv(\stategen)}$ for all $\stategen$; or, $g_{\stategen} = f_{\Rtvinv(\stategen)}$. Moreover, if $f$ is translationally invariant then so is $\Rtv f$. This definition is such that, for example, the state scalar delta function, $\delta(\xsv^0\is^0\xvv^0\iv^0)$ where
\be
\delta(\xsv^0\is^0\xvv^0\iv^0)_{\stategen} :=
\begin{cases}
  0 &: \stategen \ne \xsv^0\is^0\xvv^0\iv^0\\
  1 &: \stategen = \xsv^0\is^0\xvv^0\iv^0
\end{cases}
\label{eqn:deltafunctiongeneral}
\ee
gives $\Rtv\delta(\xsv^0\is^0\xvv^0\iv^0) = \delta(\Rtv(\xsv^0\is^0\xvv^0\iv^0))$, as one would expect. A \textit{state vector function} $\fv_{\stategen}$ is a function that has a vector value for each state $\stategen$; we will primarily consider translationally invariant state vector functions where $\fv_{\stategen}=\fv_{\staterep}$ for all $\stategen$. Then, the application of $\Rtv$ to $\fv_{\stategen}$ produces a new state vector function $\Rtv\fv := \gv$ such that $\fv_{\stategen} = R(\gv_{\Rtv(\stategen)})$ for all $\stategen$; or, $\gv_{\stategen} = R^{-1}(\fv_{\Rtvinv(\stategen)})$. Extending this to a tensor state function is straightforward, but not necessary for what follows. Note that, written this way, each $\Rtv$ now also acts as a linear operator on our (vector) space of scalar and vector lattice functions.

Restricted to the space of translationally invariant scalar state functions and vector state functions, we define scalar products to develop our symmetric basis functions. The scalar product of two translationally invariant scalar state functions $f$ and $g$ is defined as
\be
f \cdot g := \sum_{\is\xvv\iv} f_{\staterep} g_{\staterep},
\ee
or the sum of the product of the function values. The scalar product for two translationally invariant vector state functions $\fv$ and $\gv$ is defined as
\be
\fv \cdot \gv := \sum_{\is\xvv\iv} \fv_{\staterep}\cdot \gv_{\staterep},
\ee
or the sum of the dot product of the vector function values. This scalar product allows for construction of orthonormal bases for our scalar and lattice vector functions. One example of such basis functions are the set of translationally invariant delta functions,
$\bdelta(\zv\is^0\xvv^0\iv^0)$ where
\be
\bdelta(\zv\is^0\xvv^0\iv^0)_{\stategen} :=
\begin{cases}
  0 &: \is\xvv\iv \ne \is^0\xvv^0\iv^0\\
  1 &: \is\xvv\iv = \is^0\xvv^0\iv^0
\end{cases},
\label{eqn:deltafunction}
\ee
for all $\is^0,\xvv^0,\iv^0$. For vector lattice functions, the basis would be $\VECIT{e}\bdelta(\zv\is^0\xvv^0\iv^0)$ for different orthonormal 3-vectors $\VECIT{e}$.

Next, we consider a (real) symmetric linear operator $A$ that is closed over translationally invariant state functions. We can represent $A$ with a matrix $A_{\staterep,\statenewrep}$ where
\be
A_{\staterep,\statenewrep} := \bdelta(\staterep) \cdot(A\bdelta(\statenewrep)).
\ee
If $A$ is a symmetric operator, then $f\cdot(Ag)=g\cdot(Af)$ for any two translationally invariant state functions $f$ and $g$. Since $A$ is a real, symmetric linear operator, it has real eigenvalues and eigenvectors that fully span the vector space. Our symmetry operators $\Rtv$ are unitary operators, and so have complex eigenvalues and eigenvectors that fully span the vector space; the eigenvalues are all roots of unity. If we have an operator $A$ that \textit{also} commutes with a symmetry operator $\Rtv$---that is, $A\Rtv = \Rtv A$---then eigenvectors of $A$ are also eigenvectors of $\Rtv$. In particular, if we take all of the eigenvectors of $\Rtv$ that all have the same eigenvalue, then $A$ will remain closed on that set; thus, we can construct block-diagonal matrix versions of $A$.

Using the space group operations, we can define crystal stars and vector stars from our space group, and construct fully symmetrized translationally invariant scalar and vector state basis functions. We define a \textit{crystal star} as a generalization of a star for a point group; here, a crystal star is a set of all states $cs:=\{\stategen\}$ such that for any two $\stategen,\statenew\in cs$, there exists a space group operation $\Rtv$ such that $\Rtv\stategen=\statenew$, and for $\stategen\in cs$, $\Rtv\stategen\in cs$ for all space group operations $\Rtv$. That is: $cs$ is a minimal closed set of states for the space group. We will identify a unique set of stars as the \textit{origin-state stars} where $\zv\is\zv\is$ are members; these correspond to states where the solute and vacancy are superimposed and which are excluded in $\hw$ but included in $\hw^0$. For each crystal star $cs$, we define a lattice function $\bs$ where
\be
\bs := \frac1{N_s}\sum_{\stategen\in cs} \delta(\stategen)
\label{eqn:stardefinition}
\ee
and $N_s$ is the cardinality of the subset of $\{\zv\is\xvv\iv \in cs\}$---the number of representative states in $cs$. This scalar state function is such that $\Rtv \bs=\bs$ for all $\Rtv$, and has translational invariance. We define a \textit{vector star} derived from a crystal star $cs$: a vector star $vs$ is a set of tuples, $vs:=\{(\stategen,\vv)\}$ such that $\{\stategen\text{ for }(\stategen,\vv)\in vs\} = cs$, for any two $(\stategen,\vv), (\statenew,\vv') \in vs$, there exists a space group operation $\Rtv$ such that $(\Rtv\stategen,\R\,\vv)=(\statenew,\vv')$, and for $(\stategen,\vv)\in vs$, $(\Rtv\stategen,\R\,\vv)\in vs$ for all space group operations $\Rtv$. That is: $vs$ is a minimal closed set of states with vectors for the space group. Note that this requires all of the $\vv$ vectors have the same magnitude. For each vector star $vs$, we define a normalized vector state function $\VECIT{vs}$ where
\be
\VECIT{vs} := \frac1{N_s |\vv|}\sum_{(\stategen,\vv)\in vs} \vv\delta(\stategen)
\ee
This vector state function is such that $\Rtv \VECIT{vs}=\VECIT{vs}$ for all $\Rtv$, and has translational invariance.

Each non-origin state star $cs$ has at least one, and as many as three, unique vector stars associated with it. We construct the vector stars to be orthonormal: if $vs$ and $vs'$ are based on the same star $cs$, then for $(\stategen,\vv)\in vs$ and $(\stategen,\vv')\in vs'$, $\vv\cdot\vv'=0$ if $vs\ne vs'$; otherwise $\vv\cdot\vv' = 1/N_s$ where $N_s$ is the number of representative states of the star $cs$, so that the corresponding vector lattice functions are orthonormal. For each star $cs$, there is the \textit{parallel vector star} where for each $(\stategen,\vv)\in vs$, $\vv\propto(\xvv+\uv_{\iv}-\uv_{\is})$, as this trivially satisfies the definition of a vector star. However, there may be one or two possible \textit{perpendicular vectors stars}; these are such that for each $(\stategen,\vv)\in vs$, $\vv\cdot(\xvv+\uv_{\iv}-\uv_{\is}) = 0$. To be vector stars, we require that for each $\Rtv$ such that $\Rtv(\stategen) = \xspv\is\xvv\iv$, the corresponding $\vv$ obeys $\R\,\vv = \vv$; there may be one or two unique solutions in addition to the parallel star. As an example, in a simple cubic system, the star generated by a vacancy at $\langle 100\rangle$ only has the parallel vector star, while the star generated by a vacancy at $\langle 210\rangle$ has both the parallel vector star and one perpendicular vector star, and the state generated by a vacancy at $\langle 321\rangle$ has the parallel vector star and two perpendicular vector stars. In the special case of origin-state stars, some crystals have \textit{no} corresponding vector star, as it is excluded by symmetry. This corresponds to an empty site vector basis\cite{TrinkleElastodiffusivity2016} which occurs for any site whose point group includes inversion or a 3-, 4-, or 6-fold axis combined with a mirror. The exclusion of origin-state vector stars produces no bias for the vacancy without a solute, and provides for significant simplification below; c.f. \Sec{nonzerobias}.

Finally, the symmetric linear operators we consider---transition rate matrices and Green functions---are invariant under all space group operations: $\Rtv A = A\Rtv = A$. Our state bias vectors can be expanded in vector stars, and the Green function and transition rate matrices can be expressed entirely in components of those vector stars: $G_{ab} = \VECIT{vs}_a\cdot (G\,\VECIT{vs}_b)$ for two vector stars $vs_a$ and $vs_b$. Thus, symmetry can significantly reduce the computational complexity: in the case of a face-centered cubic lattice with 1nn ``first neighbor'' thermodynamic shell and a (1nn)$^2$ kinetic shell, the subspace is represented by four stars that generate five vector stars, producing $5\x5$ symmetric matrices.

\subsection{Vacancy diffusion with bias in all states}
\label{sec:nonzerobias}
For a crystal where symmetry does not preclude origin state vector stars, there will be translationally invariant bias in the solute and the vacancy that needs to be handled differently. For the vacancy, it can have non-zero bias even fully separated from the solute; while \Eqn{GFcorrection} could be computed locally for the changes in bias vector, the non-zero bias for vacancy states that are outside the kinetic shell require a specific separation so that only ``local'' changes need to be calculated. For the solute, the projection of the bias vector into the vacancy null space $\sum_{\xvv\iv}\bv_{\staterep}(\ProbV{\iv})^{1/2}$ can be non-zero; this requires a corresponding correction vector for the solute based on the solute transition matrix $\hw^2$ that compensates the ``bare'' solute transport contributions to $\Onsagert{ss}$ and $\Onsagert{sv}$.

A non-zero vacancy bias can be separated into a translationally invariant bias (absent the solute) plus localized changes due to the solute. We write the bias vector for the vacancy as $\bv^\text{v}_{\staterep} = \bv^\text{0,v}_{\staterep} + \delta\bv^\text{v}_{\staterep}$; in this case, $\bv^\text{0,v}_{\staterep}$ is the bare vacancy bias vector, \Eqn{bareratebias}, computed using the terms in $\hW^0$ (\Eqn{omega0definition}) with $\ProbV{\iv}$, and $\delta\bv^\text{v}_{\staterep}$ is the remaining (local) terms from $\hw^1$ and $\hw^2$. Thus, $\bv^0$ has translational invariance and is non-local. Note that $\delta\bv^\text{v}$ includes terms that \textit{remove} the bias at the origin states when they are disconnected from the full jump network. Similarly, we write the Green function as $\hg = \hg^0 + \delta\hg$. Then, the second term in \Eqn{Onsager-symmetric} for the vacancy-vacancy contribution is
\be
\begin{split}
  \bv^\text{v}\ox\hg\,\bv^\text{v} &= (\bv^{0,\text{v}}+\delta\bv^\text{v})\ox(\hg^0 + \delta\hg)(\bv^{0,\text{v}}+\delta\bv^\text{v})\\
  &= \bv^\text{0,v}\ox\hg^0\,\bv^\text{0,v} + \delta\bv^\text{v}\ox\hg\,\delta\bv^\text{v} \\
  &\quad + \Bigl(\delta\bv^\text{v}\ox\hg^0\,\bv^\text{0,v} + \bv^\text{0,v}\ox\hg^0\,\delta\bv^\text{v}\Bigr)
  + \Bigl(\delta\bv^\text{v}\ox\delta\hg\,\bv^\text{0,v} + \bv^\text{0,v}\ox\delta\hg\,\delta\bv^\text{v}\Bigr)\\
  &\quad + \bv^\text{0,v}\ox\delta\hg\,\bv^\text{0,v},
\end{split}
\label{eqn:VVexpansion}
\ee
where the first term contributes to the diffusion of the vacancy \textit{without} any solute, the second term is fully localized and represents the correlation in the absence of any non-solute vacancy bias, and the remaining terms are corrections to be evaluated. We wish to continue working in our kinetic state space (the kinetic shell), and so require a few simplifications. First, $\hg^0\,\bv^\text{0,v} = \gav^\text{0,v}$, the symmetrized correction vector for the vacancy. Next, while the \textit{change} in the Green function $\delta\hg$ is not local, it can be rewritten using the Dyson equation as
\be
\delta\hg = \hg^0\Bigl(-\delta\hw + \delta\hw\,\hg\,\delta\hw \Bigr)\hg^0.
\ee
For convenience, we define
\be
\dtomega := \delta\hw - \delta\hw\,\hg\,\delta\hw
\label{eqn:tildeomega}
\ee
so that $\delta\hg = -\hg^0\,\dtomega\,\hg^0$ and $\dtomega$ is non-zero only for states in our kinetic shell. Then
\be
\begin{split}
  \bv^\text{v}\ox\hg\,\bv^\text{v} &= \bv^\text{0,v}\ox\gav^\text{0,v} + \delta\bv^\text{v}\ox\hg\,\delta\bv^\text{v} 
  + \Bigl(\delta\bv^\text{v}\ox\gav^\text{0,v} + \gav^\text{0,v}\ox\delta\bv^\text{v}\Bigr)\\
  &\quad - \Bigl(\delta\bv^\text{v}\hg^0\ox\dtomega\,\gav^\text{0,v} + \gav^\text{0,v}\dtomega\ox\hg^0\,\delta\bv^\text{v}\Bigr)
  - \gav^\text{0,v}\ox\dtomega\,\gav^\text{0,v},
\end{split}
\label{eqn:VVexpansion-local}
\ee
only contains terms that are fully periodic ($\bv^\text{0,v}$, $\gav^\text{0,v}$) or fully localized ($\delta\bv^\text{v}$, $\dtomega$). The first term is the only one including $\bv^\text{0,v}$, and it contributes to the vacancy diffusivity without solute; all of the remaining terms are leading order $\cs$.

The non-zero solute bias requires a different treatment as it represents a different subspace of the state space: the solute diffusion network. Without the solute, the transition matrix $\hw^0$ has a null space corresponding to the square root of the vacancy probability, $(\ProbV{\iv})^{1/2}$ for any $\xsv\is$. When there are origin state vector stars, the projection of the solute bias vector into that vacancy null space is nonzero. We construct the total solute bias vector
\be
\bar\bv^\text{s}_{\is} := \sum_{\xvv\iv}\bv^\text{s}_{\staterep}(\ProbV{\iv})^{1/2},
\label{eqn:totalsolutebias}
\ee
and the solute transition matrix
\be
\bar\omega^2_{\is\isp} := \sum_{\xvv\iv,\statenew}(\ProbV{\iv})^{1/2}\hw^2_{\staterep,\statenew}(\ProbV{\ivp})^{1/2}
\label{eqn:totalsolutejump}
\ee
which only includes the transitions that move the solute atom. Note that as this is the null space projection, the vacancy probability appears to the half power twice. The solute transition matrix can be mapped into a single species diffusion problem,\cite{TrinkleElastodiffusivity2016} and we define the solute total symmetrized correction vector $\bar\gav^\text{s} := (\bar\omega^2)^+\bar\bv^\text{s}$ for the pseudoinverse $(\bar\omega^2)^+$. We will add to $\Onsagert{ss}$ a term corresponding to $\bar\bv^\text{s}\ox\bar\gav^\text{s}$, and an opposite correction $-\bar\bv^\text{s}\ox\bar\gav^\text{s}$ to $\Onsagert{sv}$. Finally, we construct the projected solute bias vector
\be
\delta\bv^\text{s}_{\staterep}:=\bv^\text{s}_{\staterep}
- \sum_{\statenew}\hw^2_{\staterep,\statenew}(\ProbV{\ivp})^{1/2}\bar\gav^\text{s}_{\isp}
\label{eqn:correctedsolutebias}
\ee
which now has zero projection into the vacancy null space.

\subsection{Modifications to accommodate large exchange rates}
\label{sec:roundoff-error}
The rates in a vacancy-mediated diffusion problem can differ by many orders of magnitude, which can contribute to round-off error when calculating diffusivity numerically. The predominant type of round-off error in vacancy mediated diffusion is caused by large exchange rates ($\hw^2$) compared with all other rates; this causes the correlation to increase so that the correlated contribution to the diffusivity is almost exactly equal (and opposite) to the uncorrelated diffusivity contribution. This leads to catastrophic round-off error; moreover, as our rates are Arrhenius, as long as the activation barrier for vacancy-solute exchange is lower than the other activation barriers in the system, large differences in rates will \textit{always} occur at some finite temperature. Hence, a modified version of the diffusivity is required.

We work in the exchange space (the subspace of states involved in solute-vacancy exchange) to separate the $\hw^2$ contribution to diffusivity from the rest of the state space. The exchange space are the only states that have a non-zero contribution from our ``fast'' exchange rates $\hw^2$. Then, we can apply the Dyson equation in two stages, where $\hg^1 = ((\hg^0)^{-1} + \hw^1)^{-1}$ and $\hg = ((\hg^1)^{-1} + \hw^2)^{-1}$, and the second Dyson equation will be computed in the exchange space. Then, if $(\hg^1\hw^2)\gg 1$, rewriting the Green function as
\be
\begin{split}
  \hg &= \Bigl(\bigl(\hg^1\bigr)^{-1} + \hw^2\Bigr)^{-1}
  = \biggl(\hw^2\Bigl[\mathbf{1} + \bigl(\hg^1\,\hw^2\bigr)^{-1}\Bigr]\biggr)^{-1}\\
  &= \bigl(\hw^2\bigr)^{-1} + \Bigl\{ \Bigl[\mathbf{1} + \bigl(\hg^1\,\hw^2\bigr)^{-1}\Bigr]^{-1} - \mathbf{1}\Bigr\}\bigl(\hw^2\bigr)^{-1}\\
  &=  \bigl(\hw^2\bigr)^{-1} - \Bigl[\hw^2 + \hw^2 \hg^1 \hw^2\Bigr]^{-1}\\
\end{split}
\label{eqn:GFlargeomega2}
\ee
produces a first term that is the inverse of $\hw^2$ and the second term is, to leading order, $O\bigl(\bigl(\hg^1\bigr)^{-1}\bigl(\hw^2\bigr)^{-2}\bigr)$. Hence, when the matrix $\bigl(\hg^1\,\hw^2\bigr)$ is large, this expansion can capture many more significant figures in two separate matrices. The calculation in \Eqn{GFlargeomega2} must be performed in the complement of the null space of $\hw^2$ in the exchange space; see below. While \Eqn{GFlargeomega2} is exact, it only shows particular utility for large $\hw^2$ values evaluated with floating-point arithmetic.

The difference between the uncorrelated and correlated diffusivity can be directly calculated without significant round-off error for large $\hw^2$. The exchange space contains all of the states $\stategen$ with non-zero bias for the solute,  $\bv^\text{s}_{\stategen}$. Moreover, as there is only a single vacancy and a single solute, there is only one transition that contributes to the bias for any state $\stategen$, and \Eqn{omega2exchange} shows that $\hw^2$ has a block diagonal structure consisting of $2\x2$ blocks of $\stategen$ and $\stategenX$. Then, we can consider each pair of states, $\stategen$ and $\stategenX$ individually; call one state $\Srep:=\stategen$ and the other $\Sbar := \stategenX$. Then, let $\rho_{\Srep}:=\hP_0(\stategen)$, $\rho_{\Sbar}:=\hP_0(\stategenX)$, $\nu := \hP_0^{1/2}(\stategen)\hW_{\stategen,\stategenX}\hP_0^{-1/2}(\stategenX)$, and $\dxv:=\dxvh^\text{s}(\stategen\to\stategenX)$. In this shortened notation,
\be
  \bv_{\Srep} := \bv^\text{s}_{\stategen} = \dxv\, \nu \rho^{1/2}_{\Sbar},\quad
  \bv_{\Sbar} := \bv^\text{s}_{\stategenX} = -\dxv\, \nu \rho^{1/2}_{\Srep},\quad
  \omega^2 =
  \begin{pmatrix}
    -\nu (\rho_{\Sbar}/\rho_{\Srep})^{1/2} & \nu\\
    \nu & -\nu (\rho_{\Srep}/\rho_{\Sbar})^{1/2}
  \end{pmatrix}
\ee
and the contribution to $\Onsagert{ss}$ from both states is
\be
\frac1{\kB T V_0}\bigl\{
\dxv\ox\dxv (\rho_{\Srep}\rho_{\Sbar})^{1/2}\nu + \bigl(\bv_{\Srep},\bv_{\Sbar}\bigr) \ox(\omega^2)^+\bigl(\bv_{\Srep},\bv_{\Sbar}\bigr)^\text{T}
- \bigl(\bv_{\Srep},\bv_{\Sbar}\bigr) \ox\bigl(\hw^2 + \hw^2 \hg^1\hw^2\bigr)^{-1}\bigl(\bv_{\Srep},\bv_{\Sbar}\bigr)^\text{T}\bigl\}.
\label{eqn:exchangeLss}
\ee
The first two terms add to zero, so that the second term in \Eqn{GFlargeomega2} contributes to the diffusivity, giving an overall contribution $O\bigl(\bigl(\hg^1\bigr)^{-1}\bigl(\hw^2)^{0}\bigr)$ for large $\omega^2$. To see this, we evaluate the pseudoinverse of the $2\times2$ matrix $\omega^2$,
\be
(\omega^2)^+ = \nu^{-1}\frac{(\rho_{\Srep}\rho_{\Sbar})^{1/2}}{(\rho_{\Srep}+\rho_{\Sbar})^2}
\begin{pmatrix}
  -\rho_{\Sbar} & (\rho_{\Srep}\rho_{\Sbar})^{1/2}\\
  (\rho_{\Srep}\rho_{\Sbar})^{1/2} & -\rho_{\Srep}
\end{pmatrix}
\ee
and so
\be
\begin{split}
  \bigl(\bv_{\Srep},\bv_{\Sbar}\bigr)(\omega^2)^+\bigl(\bv_{\Srep},\bv_{\Sbar}\bigr)^\text{T} &=
  \nu^{-1}\frac{(\rho_{\Srep}\rho_{\Sbar})^{1/2}}{(\rho_{\Srep}+\rho_{\Sbar})^2}
  \begin{pmatrix}
    \dxv \nu\rho_{\Sbar}^{1/2}&-\dxv \nu\rho_{\Srep}^{1/2}\\
  \end{pmatrix}
  \begin{pmatrix}
    -\rho_{\Sbar} & (\rho_{\Srep}\rho_{\Sbar})^{1/2}\\
    (\rho_{\Srep}\rho_{\Sbar})^{1/2} & -\rho_{\Srep}
  \end{pmatrix}
  \begin{pmatrix}
    \dxv \nu\rho_{\Sbar}^{1/2}\\
    -\dxv \nu\rho_{\Srep}^{1/2}\\
  \end{pmatrix}\\
  &= \frac{\nu\dxv\ox\dxv}{(\rho_{\Srep}+\rho_{\Sbar})^2}
  \Bigl[-(\rho_{\Srep}\rho_{\Sbar})^{1/2}\rho_{\Sbar}^2 - 2(\rho_{\Srep}\rho_{\Sbar})^{3/2}
  -(\rho_{\Srep}\rho_{\Sbar})^{1/2}\rho_{\Srep}^2
  \Bigr]\\
  &= -(\rho_{\Srep}\rho_{\Sbar})^{1/2}\nu\dxv\ox\dxv,
\end{split}
\ee
which exactly cancels the bare term in \Eqn{exchangeLss}. This is expected, as a pair of states that only transition between themselves (as they would if $\hw = \hw^2$ only) does not contribute to transport and $\Onsagert{ss}$ is exactly zero. \Eqn{exchangeLss} also shows that while in the limit of $\hw^2\to\infty$, $\Onsagert{ss}\sim\bigl(\hw^2\bigr)^0$, in the limit $\hw^2\to0$,  $\Onsagert{ss}\sim\bigl(\hw^2\bigr)^1$.

\subsection{Algorithm summary}
\label{sec:Lijsummary}
We summarize the computational steps in the calculation of of the transport coefficients for an arbitrary crystal in the dilute-vacancy/dilute-solute limit. Our inputs are the same as \Sec{GFsummary}, with the addition of rates for the vacancy near the some $\hW^1$, exchange with the solute $\hW^2$, and solute-vacancy interaction energies (probabilities) $\Esv_{\is\iv}(\xv)$ ($\hP^0(\stategen)$); c.f, \Sec{dilutelimit}. We assume that the introduction of the solute does not cause any vacancy states to become unstable, or introduce new vacancy states or transitions in the jump network; the extensions of this work to consider those cases is possible (c.f., \Sec{discussion}), but beyond the scope of this work.
\begin{enumerate}
\item For a given thermodynamic shell, define the kinetic shell (c.f., \Sec{translate}), and construct a set of crystal stars corresponding to the thermodynamic shell where $\Esv_{\is\iv}(\xv)\ne 0$, and both crystal stars and vector stars for the kinetic shell; index the vector state functions by $i$ as $\VECIT{vs}_i$. Construct the outer product of vector state functions $\VECIT{vs}_i$ and $\VECIT{vs}_j$,
\be
\VV_{ij}:=\VECIT{vs}_i\ox\VECIT{vs}_j =
\sum_{(\staterep,\vv)\in vs_i}\sum_{(\statenewrep,\vv')\in vs_j} \vv\ox\vv' \delta(\xvv-\xvpv)\delta_{\is\isp}\delta_{\iv\ivp}.
\ee
and when crystal symmetry does not preclude origin state vector stars, construct the projection matrix $\Upsilon$ between vector state function $\VECIT{vs}_i$ and an origin state vector function $\VECIT{vs}_k$,
\be
\Upsilon_{ik} := \sum_{(\staterep,\vv)\in vs_i}\sum_{(\zv\ivp\zv\ivp,\vv')\in vs_k} \vv\cdot\vv'\delta_{\iv\ivp}.
\ee

\item\label{step:prob} Compute the probabilities $\ProbV{\iv}$, $\ProbS{\is}$, and $\hP_0(\staterep)=\ProbV{\iv}\ProbS{\is}\exp(-\Esv_{\is\iv}(\xv)/\kB T)$, with normalization $\sum_i \ProbV{i} = \sum_i \ProbS{i} = \Nsites$.

\item Construct the bare transport contributions
\be
\begin{split}
  d^\text{2,ss}&:=\frac12 \sum_{\staterep}
  (\xvv+\uv_{\iv}-\uv_{\is})\ox(\xvv+\uv_{\iv}-\uv_{\is})\;
  \hw^2_{\staterep,\staterepX}\hP_0^{1/2}(\staterep)\hP_0^{1/2}(\staterepX)\\
  d^\text{2,sv}&:=-d^\text{2,ss}\\
  d^\text{1,vv}&:=\frac12 \sum_{\staterep}\sum_{\xvpv\ivp}
  (\xvpv+\uv_{\ivp}-\xvv-\uv_{\iv})\ox(\xvpv+\uv_{\ivp}-\xvv-\uv_{\iv})\\
  &\qquad \cdot  \Bigl[\bigl(\hw^1_{\staterep,\statenewrepS} + \hw^0_{\staterep,\statenewrepS}\bigr)\hP_0^{1/2}(\staterep)\hP_0^{1/2}(\statenewrepS)
-\hw^0_{\staterep,\statenewrepS}\bigl(\ProbV{\iv}\ProbV{\ivp}\bigr)^{1/2}\ProbS{\is}\Bigr]\\
  d^\text{2,vv}&:=d^\text{2,ss}
  \end{split}
\ee
where the last two terms contribute to the first order change in $\Onsagert{vv}$ with $\cs$. These tensors can be written as linear combinations of the symmetry unique $\omega^0$, $\omega^1$, and $\omega^2$ values with square roots of appropriate probabilities.

\item Construct the bias vectors in the vector state function basis. Separate into the vacancy bias (1) and exchange bias vectors (2),
\be
\begin{split}
  b^1_i&:=\sum_{(\staterep,\vv)\in vs_i}\sum_{\xvpv\ivp}
  \vv\cdot(\xvpv+\uv_{\ivp}-\xvv-\uv_{\iv})\;
  \Bigl[\bigl(\hw^1_{\staterep,\statenewrepS} + \hw^0_{\staterep,\statenewrepS}\bigr)\hP_0^{1/2}(\statenewrepS)\\
    &\qquad-\hw^0_{\staterep,\statenewrepS}\bigl(\ProbV{\ivp}\ProbS{\is}\bigr)^{1/2}\Bigr]\\
  b^2_i&:=\sum_{(\staterep,\vv)\in vs_i}
  \vv\cdot(\xvv+\uv_{\iv}-\uv_{\is})\;
  \hw^2_{\staterep,\staterepX}\hP_0^{1/2}(\staterepX)\\
\end{split}
\ee
so that $\bv^\text{s}$ is given by $b^2_i$ and the localized change in bias vector for the vacancy $\delta\bv^\text{v}$ is $b^1_i-b^2_i$; this is superimposed onto the bare vacancy bias vector $\bv^\text{0,v}$ (c.f., \Sec{nonzerobias}). These vectors can be written as linear combinations of the symmetry unique $\omega^0$, $\omega^1$, and $\omega^2$ values with square roots of appropriate probabilities.

\item Evaluate the vacancy Green function $g^0_{\staterep,\statenewrepS}$, the vacancy diffusivity without a solute $\Dvv$ (where $\Onsager{0,\text{vv}}=\cv\Dvv/\kB T$), and the periodic symmetrized correction vector $\gav^\text{0,v}_{\iv}$, which is projected into origin state vector functions $\VECIT{vs}_k$,
\be
\gamma^\text{0,v}_k := \sum_{(\zv\iv\zv\iv,\vv)\in vs_k} \vv\cdot\gav^\text{0,v}_{\iv}.
\ee

\item Construct the vector state function basis matrices 
\be
\begin{split}
  g^0_{ij}&:=\sum_{(\staterep,\vv)\in vs_i}\sum_{(\statenewrepS,\vv')\in vs_j}(\vv\cdot\vv')
  \hg^0_{\xvv\iv,\xvpv\ivp}\\
  \omega^1_{ij}&:=\sum_{(\staterep,\vv)\in vs_i}\sum_{(\statenewrepS,\vv')\in vs_j}(\vv\cdot\vv')
  \hw^1_{\staterep,\statenewrepS}\\
  \omega^2_{ij}&:=\sum_{(\staterep,\vv)\in vs_i}\sum_{(\staterepX,\vv')\in vs_j}(\vv\cdot\vv')
  \hw^2_{\staterep,\staterepX}
\end{split}
\ee
from which the matrix $g_{ij}$ is constructed in \Step{GF}. These matrices can be written as linear combinations of the symmetry unique $g^0$, $\omega^1$ and $\omega^2$ values.

\item Evaluate the total solute bias vector $\bar\bv^\text{s}_{\is}$ (\Eqn{totalsolutebias}), solute-solute transition matrix $\bar\omega^2_{\is\isp}$ (\Eqn{totalsolutejump}), symmetrized correction vector $\bar\gav^\text{s}_{\is}$. Add $\bar\bv^\text{s}\ox\bar\gav^\text{s}$ to $d^\text{2,ss}$ and add $-\bar\bv^\text{s}\ox\bar\gav^\text{s}$ to $d^\text{2,sv}$. Construct $\delta\bv^\text{s}_i$ from \Eqn{correctedsolutebias}.

\item\label{step:GF} Construct $g^1 = (\mathbf{1}+g^0\omega^1)^{-1}g^0$, $g = (\mathbf{1}+g^1\omega^2)^{-1}g^1$, and $\dtomega$ (\Eqn{tildeomega}) in the vector state function basis. If $g^1\omega^2$ contains ``large'' entries (above $10^8$ for a double floating-point representation), then
  \begin{enumerate}
  \item find the eigenvectors of $\omega^2$ in the exchange space and the pseudoinverse $\omega^{2+}$;
  \item rotate $g$ into the subspace; replace $g$ with $-(\omega^2+\omega^2g^1\omega^2)^{-1}$ in the non-null subspace;
  \item rotate back to exchange space;
  \item set the bare transport contributions to
    \be
    \begin{split}
      d^\text{2,ss} &:= 0\\
      d^\text{2,sv} &:= \sum_{ijk} b^1_i \VV_{ij} \omega^{2+}_{jk} b^2_k\\
      d^\text{1,vv} &:= 0\\
      d^\text{2,vv} &:= \sum_{ijk} (b^1_i-2b^2_i) \VV_{ij} \omega^{2+}_{jk} b^2_k
    \end{split}
    \ee
  \end{enumerate}

\item The transport coefficients are, to linear order in $\cs$ and $\cv$,
\be
\begin{split}
  \Onsagert{ss} &= \frac{\cv\cs}{\Nsites \kB T}\Biggl[ d^\text{2,ss}+
  \sum_{ijk} b^2_i \VV_{ij} g_{jk}b^2_k
  \Biggr]\\
  \Onsagert{sv} &= \frac{\cv\cs}{\Nsites \kB T}\Biggl[ d^\text{2,sv}+
  \sum_{ijk} (b^1_i-b^2_i) \VV_{ij} g_{jk}b^2_k
  \Biggr]\\
  \Onsagert{vv} &= \frac{\cv}{\kB T}\Dvv +
  \frac{\cv\cs}{\Nsites \kB T}\Biggl [d^\text{1,vv}+d^\text{2,vv}+
    \sum_{ijk}(b^1_i-b^2_i) \VV_{ij} g_{jk}(b^1_k-b^2_k)\\
    &\qquad +\sum_{ijk} \gamma_k^\text{0,v}\Upsilon_{ik}\VV_{ij}\biggl(2b^1_j
    - 2\sum_{mn}\dtomega_{jm}g^0_{mn}b^1_n
    - \sum_{mk'} \dtomega_{jm}\Upsilon_{mk'}\gamma^\text{0,v}_{k'}\biggr)\\
    &\qquad -\sum_{ii'kk'} \gamma_k^\text{0,v}\Upsilon_{ik}\VV_{ii'}\Upsilon_{i'k'} b^\text{0,v}_{k'}
  \Biggr]\\
\end{split}
\ee
where $\Nsites$ appears due to the normalization of $\ProbV{\iv}$ and $\ProbS{\is}$ (c.f., \Step{prob}).
\end{enumerate}
The algorithm, with symmetry analysis, is available in an open source implementation, described in \App{implementation}. Note that only crystals with low symmetry have origin state vector stars; otherwise, $\Upsilon$, $\bv^\text{0,v}_{\iv}$, $\gav^\text{0,v}_{\iv}$, $\bar\bv^\text{s}$ and $\bar\gav^\text{s}$ are zero.

\section{Numerical results}
\label{sec:numerics}
To demonstrate the efficacy and accuracy of the algorithm, we consider several cases of interest. First, we demonstrate the scaling of the error of the evaluation of bare vacancy Green function for FCC and HCP lattices. We then show the effect of roundoff error in the large $\hw^2$ limit, and that our corrections successfully circumvent catastrophic error. Next, we compute tracer coefficients for a series of crystal structures, compare with known values where available, and show new results for a few crystals. As a more detailed example, we consider the case of a garnet structure, which was recently studied with a related method, and produce the first accurate tracer correlation coefficients for that structure. Finally, we compare with prior simulations of solute drag of silicon in FCC nickel computed using the self-consistent mean-field method,\cite{Garnier2013a,Garnier2014b,Garnier2014bData} and solute drag of tin and zinc in HCP magnesium\cite{Agarwal2017} computed using kinetic Monte Carlo. The source code for all of the results are available as Jupyter notebooks online;\cite{OnsagerCalc} c.f.~\App{implementation}.

\subsection{Vacancy Green function error scaling without solute}
\label{sec:GFerror}

\begin{figure}[ht]
  \begin{center}
  \includegraphics[width=\figwidth]{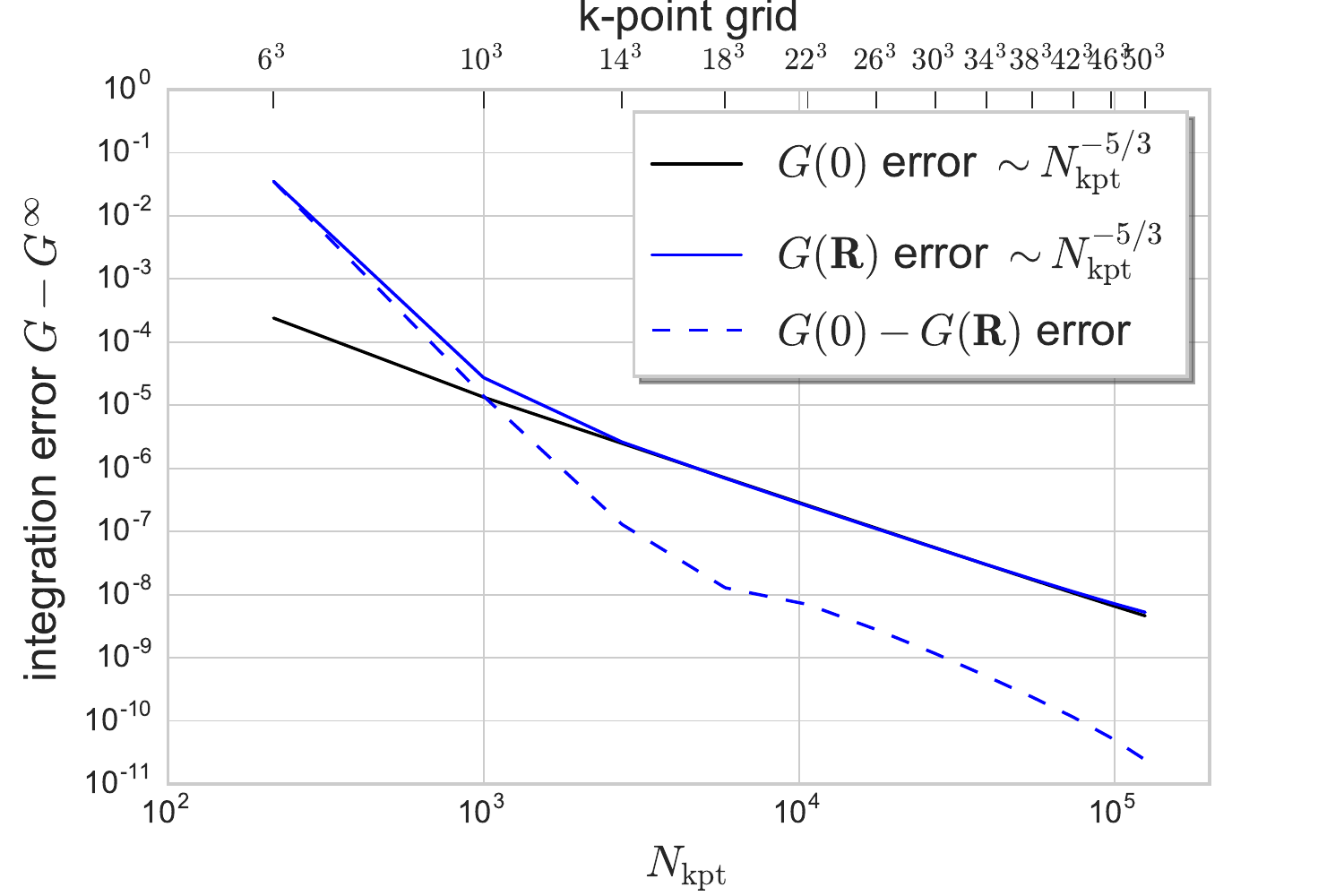}
  \end{center}
\caption{Scaling of error in FCC Green function. The absolute error is due to the integration of the semicontinuum contribution $g^{\alpha\beta}_\text{sc}$ over the Brillouin zone, c.f., \Sec{semicontinuum}. A minimum number of $k$-point divisions is required to avoid aliasing errors for larger $\Rv$; as $\Nkpt$ increases, the error is controlled by the integration error of a smooth, periodic function, which scales\cite{Ghazisaeidi2009} no slower than $\Nkpt^{-4/3}$; empirically, we find a scaling of $\Nkpt^{-5/3}$.}
\label{fig:FCC-GFerror}
\end{figure}

We consider progressively denser $k$-point meshes to integrate the Brillouin zone for a face-centered cubic (FCC) lattice. The transition rate is $\omega^0 = 1/12$ and we start from a $6\x6\x6$ mesh in progression $(4n+2)^3$ up to $34\x34\x34$ with $\threshold = 10^{-8}$ in \Eqn{pmax}. The values of the Green function and differences appears in \Tab{FCC-GFerror}, with the convergence of integration errors in \Fig{FCC-GFerror}. The largest distance to appear in the correlation factor for a first-neighbor thermodynamic interaction is $\Rv=4\mathbf{a}_1 = 2a_0\mathbf{\hat j} + 2a_0\mathbf{\hat k}$, and so we compare the convergence of $g^0(\zv)$, $g^0(\Rv)$, and the difference $g^0(\zv)-g^0(\Rv)$ (as differences in Green function values are important for errors in correlation factors). To separate the integration error from the $\threshold$ error, we fit the large $\Nkpt$ limit of the Green function value, then subtract the limiting value. The default value in our algorithm of $\Nkpt = 18^3$ provides an error in the Green function difference of $\sim 10^{-8}$. Roundoff error becomes an issue as $\Nkpt$ increases, as the error is limited by $\sim 10^{-16} \Nsymmkpt$. Empirically, we observe an error scaling of $\sim\Nkpt^{-5/3}$---faster than the minimum convergence rate of $\Nkpt^{-4/3}$ from \rcite{Ghazisaeidi2009}---in the Green function values, and an even faster convergence for the difference of Green function values.

\begingroup
\squeezetable
\begin{table}[ht]
  \caption{Green function convergence with $k$-point density for FCC. The $k$-point meshes can be reduced by cubic symmetry to a smaller set of unique points ($\Nsymmkpt$), the number of which is listed parenthetically next to $\Nkpt$. The computational effort scales with the number of symmetry unique points (c.f., \Eqn{symmqpoint}) while the error is controlled by $\Nkpt$ (c.f., \Fig{FCC-GFerror}). The $\pmax$ value is chosen so that $\threshold=10^{-8}$ in \Eqn{pmax}, and $\Rv=4\mathbf{a}_1 = 2a_0\mathbf{\hat j} + 2a_0\mathbf{\hat k}$.}
  \label{tab:FCC-GFerror}
  \begin{tabular}{crlccc}
    \hline\hline
    $k$-point mesh&$\Nkpt$&$(\Nsymmkpt)$&$g^0(\zv)$&$g^0(\Rv)$&$g^0(\zv)-g^0(\Rv)$\\
    \hline
    $6\x6\x6$&216&(16) & --1.344\,901\,582\,401 & --0.119\,888\,361\,621 & --1.225\,013\,220\,779\\
    $10\x10\x10$&1000&(48) & --1.344\,674\,624\,975 & --0.084\,566\,077\,531 & --1.260\,108\,547\,444\\
    $14\x14\x14$&2744&(106) & --1.344\,663\,672\,542 & --0.084\,541\,308\,263 & --1.260\,122\,364\,278\\
    $18\x18\x18$&5832&(200) & --1.344\,661\,890\,661 & --0.084\,539\,383\,601 & --1.260\,122\,507\,060\\
    $22\x22\x22$&10\,648&(337) & --1.344\,661\,442\,418 & --0.084\,538\,941\,204 & --1.260\,122\,501\,213\\
    $26\x26\x26$&17\,576&(528) & --1.344\,661\,295\,591 & --0.084\,538\,798\,573 & --1.260\,122\,497\,018\\
    $30\x30\x30$&27\,000&(778) & --1.344\,661\,238\,153 & --0.084\,538\,742\,761 & --1.260\,122\,495\,392\\
    $34\x34\x34$&39\,304&(1095) & --1.344\,661\,212\,587 & --0.084\,538\,717\,850 & --1.260\,122\,494\,737\\
    $38\x38\x38$& 54\,872&(1488) & --1.344\,661\,200\,054 & --0.084\,538\,705\,591 & --1.260\,122\,494\,464\\
    $42\x42\x42$& 74\,088&(1971) & --1.344\,661\,193\,423 & --0.084\,538\,699\,082 & --1.260\,122\,494\,341\\
    $46\x46\x46$& 97\,336&(2547) & --1.344\,661\,189\,691 & --0.084\,538\,695\,410 & --1.260\,122\,494\,281\\
    $50\x50\x50$& 125\,000&(3222) & --1.344\,661\,187\,483 & --0.084\,538\,693\,232 & --1.260\,122\,494\,251\\
    \hline\hline
  \end{tabular}
\end{table}
\endgroup

\begin{figure}[ht]
  \begin{center}
  \includegraphics[width=\figwidth]{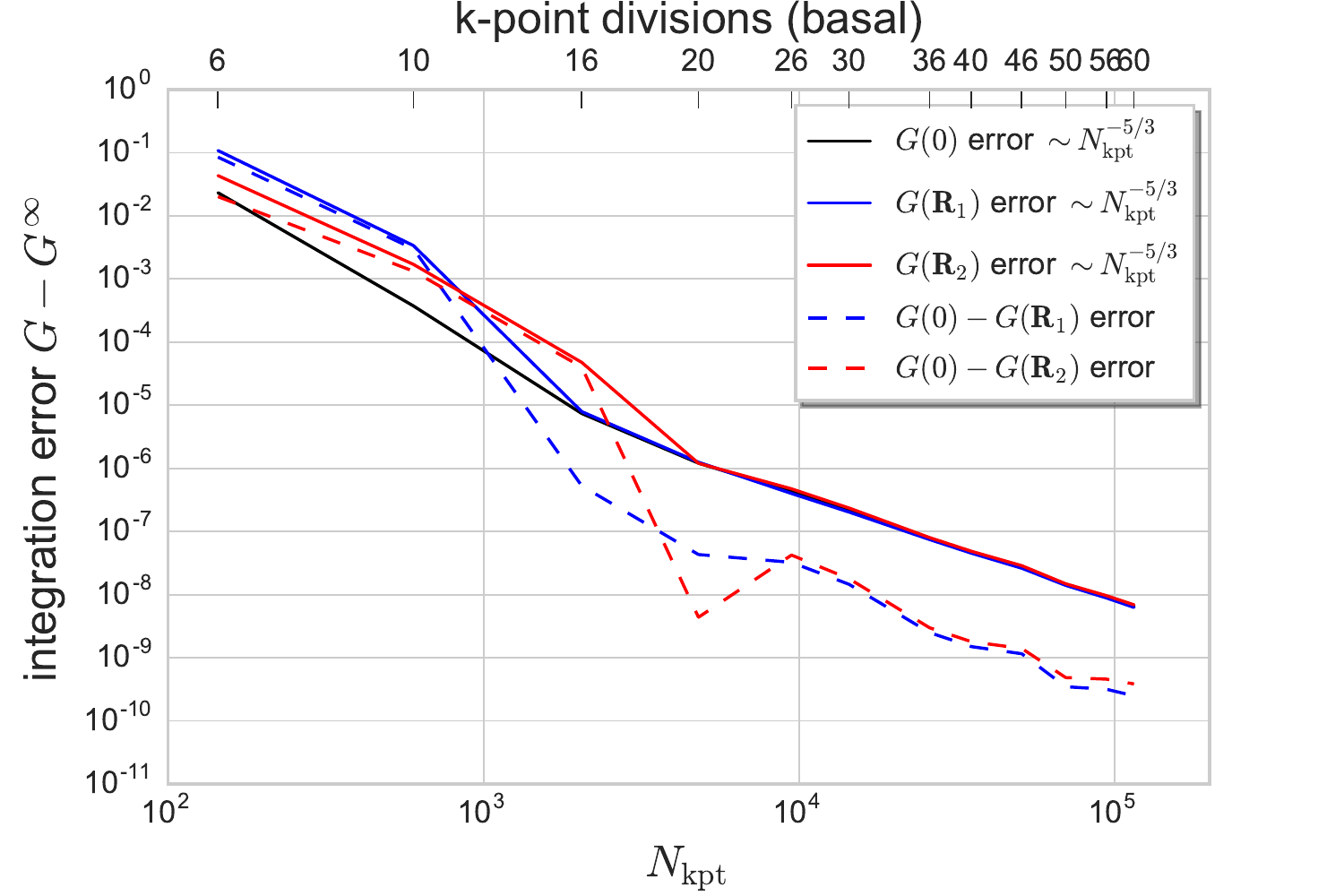}
  \end{center}
\caption{Scaling of error in HCP Green function. The absolute error is due to the integration of the semicontinuum contribution $g^{\alpha\beta}_\text{sc}$ over the Brillouin zone, c.f., \Sec{semicontinuum}. A minimum number of $k$-point divisions is required to avoid aliasing errors for larger $\Rv$; as $\Nkpt$ increases, the error is controlled by the integration error of a smooth, periodic function, which scales\cite{Ghazisaeidi2009} no slower than $\Nkpt^{-4/3}$; empirically, we find a scaling of $\Nkpt^{-5/3}$.} \label{fig:HCP-GFerror}
\end{figure}

We also consider progressively denser $k$-point meshes to integrate the Brillouin zone for a hexagonal close-packed (HCP) lattice. The basal and pyramidal transition rates are $\omega^0 = 1/12$, and we start from a $6\times6\times4$ mesh in progression up to $60\times60\times32$ with $\threshold = 10^{-8}$ in \Eqn{pmax}. The values of the Green function and differences appears in \Tab{HCP-GFerror}, with the convergence of integration errors in \Fig{HCP-GFerror}. The largest distances to appear in the correlation factor for a first-neighbor thermodynamic interaction are $\Rv_1=4\mathbf{a}_1 + 4\mathbf{a}_2= 4a_0\mathbf{\hat i}$ and $\Rv_2=2\mathbf{a}_1+2\mathbf{a}_2+2\mathbf{a}_3 = 2a_0\mathbf{\hat i} + 2c_0\mathbf{\hat k}$, and so we compare the convergence of $g^0(\zv)$, $g^0(\Rv_1)$, $g^0(\Rv_2)$, and the differences $g^0(\zv)-g^0(\Rv_1)$ and $g^0(\zv)-g^0(\Rv_2)$ (as differences in Green function values are important for errors in correlation factors in the basal plane and along the $c$-axis). To separate the integration error from the $\threshold$ error, we fit the large $\Nkpt$ limit of the Green function value, then subtract the limiting value. The default value in our algorithm of $20\x20\x12$ provides an error in the Green function difference of $\sim 10^{-7}$. Roundoff error becomes an issue as $\Nkpt$ increases, as the error is limited by $\sim 10^{-16}\Nsymmkpt$. Empirically, we observe an error scaling of $\sim\Nkpt^{-5/3}$---faster than the minimum convergence rate of $\Nkpt^{-4/3}$ from \rcite{Ghazisaeidi2009}---in the Green function values, and a similar convergence for the difference of Green function values with an order-of-magnitude smaller prefactor.

\begingroup
\squeezetable
\begin{table}[ht]
  \caption{Green function convergence with $k$-point density for HCP. The $k$-point meshes can be reduced by hexagonal symmetry to a smaller set of unique points ($\Nsymmkpt$), the number of which is listed parenthetically next to $\Nkpt$. The computational effort scales with the number of symmetry unique points (c.f., \Eqn{symmqpoint}) while the error is controlled by $\Nkpt$ (c.f., \Fig{HCP-GFerror}). The $\pmax$ value is chosen so that $\threshold=10^{-8}$ in \Eqn{pmax}, and $\Rv_1=4\mathbf{a}_1 + 4\mathbf{a}_2= 4a_0\mathbf{\hat i}$ and $\Rv_2=2\mathbf{a}_1+2\mathbf{a}_2+2\mathbf{a}_3 = 2a_0\mathbf{\hat i} + 2c_0\mathbf{\hat k}$.}
  \label{tab:HCP-GFerror}
  \begin{tabular}{crlccccc}
    \hline\hline
    $k$-point mesh&$\Nkpt$&$(\Nsymmkpt)$&$g^0(\zv)$&$g^0(\Rv_1)$&$g^0(\Rv_2)$&$g^0(\zv)-g^0(\Rv_1)$&$g^0(\zv)-g^0(\Rv_2)$\\
    \hline
    $6\x6\x4$& 144&(21) & --1.367\,909\,503\,563 & --0.192\,892\,722\,514 & --0.131\,552\,967\,388 & --1.175\,016\,781\,049 & --1.236\,356\,536\,175\\
      $10\x10\x6$& 600&(56) & --1.345\,034\,474\,341 & --0.087\,913\,619\,020 & --0.089\,866\,654\,871 & --1.257\,120\,855\,321 & --1.255\,167\,819\,470\\
      $16\x16\x8$&2048&(150) & --1.344\,668\,575\,390 & --0.084\,546\,609\,595 & --0.088\,212\,957\,806 & --1.260\,121\,965\,795 & --1.256\,455\,617\,584\\
      $20\x20\x12$&4800&(308) & --1.344\,662\,392\,185 & --0.084\,539\,941\,251 & --0.088\,166\,498\,574 & --1.260\,122\,450\,934 & --1.256\,495\,893\,611\\
      $26\x26\x14$&9464&(560) & --1.344\,661\,615\,456 & --0.084\,539\,088\,966 & --0.088\,165\,768\,509 & --1.260\,122\,526\,490 & --1.256\,495\,846\,946\\
      $30\x30\x16$&14\,400&(819) & --1.344\,661\,401\,027 & --0.084\,538\,892\,419 & --0.088\,165\,529\,659 & --1.260\,122\,508\,608 & --1.256\,495\,871\,368\\
      $36\x36\x20$&25\,920&(1397) & --1.344\,661\,260\,564 & --0.084\,538\,764\,009 & --0.088\,165\,374\,312 & --1.260\,122\,496\,555 & --1.256\,495\,886\,252\\
      $40\x40\x22$&35\,200&(1848) & --1.344\,661\,230\,214 & --0.084\,538\,734\,661 & --0.088\,165\,342\,770 & --1.260\,122\,495\,553 & --1.256\,495\,887\,444\\
      $46\x46\x24$&50\,784&(2600) & --1.344\,661\,210\,808 & --0.084\,538\,715\,598 & --0.088\,165\,322\,977 & --1.260\,122\,495\,211 & --1.256\,495\,887\,832\\
      $50\x50\x28$&70\,000&(3510) & --1.344\,661\,197\,817 & --0.084\,538\,703\,416 & --0.088\,165\,309\,065 & --1.260\,122\,494\,400 & --1.256\,495\,888\,752\\
      $56\x56\x30$&94\,080&(4640) & --1.344\,661\,192\,649 & --0.084\,538\,698\,279 & --0.088\,165\,303\,871 & --1.260\,122\,494\,370 & --1.256\,495\,888\,778\\
    $60\x60\x32$&115\,200&(5627) & --1.344\,661\,189\,980 & --0.084\,538\,695\,678 & --0.088\,165\,301\,128 & --1.260\,122\,494\,302 & --1.256\,495\,888\,852\\
    \hline\hline
  \end{tabular}
\end{table}
\endgroup

The analytic values for the Green functions are available as Watson integrals, which allows us to compare the error induced from a non-zero value of $\threshold$. In the case of $g^0(\zv)$ for both FCC and HCP lattices, the value\cite{Watson1939,Koiwa1978} is the same:
\begin{equation*}
  g^0(\zv) = -\frac{9\Gamma\left(\frac13\right)^6}{2^{14/3}\pi^4} \approx
  -1.344\,661\,183\,165\,144 \cdots
\end{equation*}
which provides an error, when using $\Nkpt\sim 10^5$ from \Tab{FCC-GFerror} and \Tab{HCP-GFerror} of $\approx 5\EE{-9}$ for $\threshold=10^{-8}$. Just as $g^0(\zv)$ is equal for FCC and HCP, $g^0(\Rv)$ for FCC and $g^0(\Rv_1)$ for HCP are also equal; the closed form for $g^0(\Rv)$ can be evaluated from recursion relations\cite{Inoue1974,Mano1974} as
\begin{equation*}
  g^0(\Rv) \approx -0.084\,538\,688\,992\,554 \cdots
\end{equation*}
which provides similar errors as in $g^0(\zv)$. However, the difference with $g^0(\zv)-g^0(\Rv)$ is
\begin{equation*}
  g^0(\zv)-g^0(\Rv) \approx -1.260\,122\,494\,172\,590 \cdots
\end{equation*}
which provides an error from \Tab{FCC-GFerror} and \Tab{HCP-GFerror} of $\approx 1\EE{-10}$ and $\approx 3\EE{-10}$, respectively, for $\threshold=10^{-8}$; this is expected to be the level of the controlling error in correlation factors.

\subsection{Large $\omega^2$ treatment for FCC five-frequency model}
\label{sec:largeOmega2numerics}

\begin{figure}[ht]
  \begin{center}
  \includegraphics[width=\figwidth]{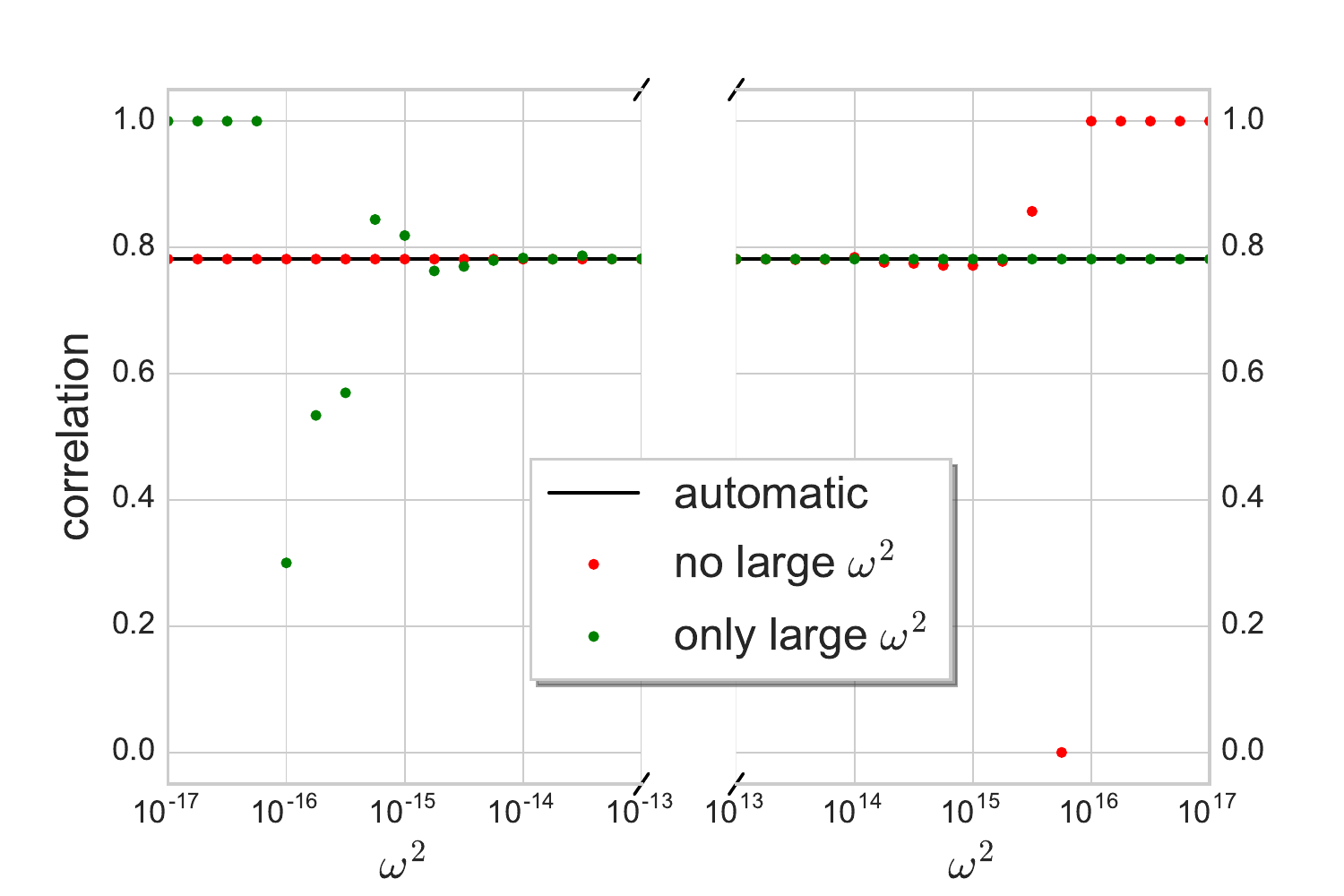}
  \end{center}
\caption{Tracer correlation factor $f=-\Onsagert{ss}/\Onsagert{sv}$ for FCC with different treatment of $\Onsagert{ss}$ and $\Onsagert{sv}$ for $\omega^2$. When the ratio $f$ is numerically meaningless ($\Onsagert{sv}$ is numerically zero, or any quantity is reported as Inf of NaN), $f$ is plotted as having the value 1.}
\label{fig:largeomega2}
\end{figure}

\Sec{roundoff-error} describes the modifications necessary to compute the transport coefficients when $\hw^2$ becomes large, to avoid catastrophic roundoff error; \Fig{largeomega2} shows numerical values of the roundoff error depending on which treatment of the transport coefficients is used. Our implementation uses the algorithm of \Sec{roundoff-error} when $\hg^1\hw^2$ has values greater than $10^8$; this is the ``automatic'' algorithm of \Fig{largeomega2}; alternatively, we can choose to never use the ``large'' $\hw^2$ algorithm, or use it for any value of $\hw^2$. To test the accuracy when changing the algorithm, we consider an FCC five-frequency case where $\omega^1=\omega^3=\omega^4=\omega^0=1$, and we systematically vary $\omega^2$ from $10^{-17}$ to $10^{17}$. The correlation factor $f=-\Onsagert{ss}/\Onsagert{sv}$ is constant and equal to the tracer value $\approx 0.781\,451\,419$ for any nonzero value of $\omega^2$, and our ``automatic'' treatment shows negligible deviation ($<10^{-15}$) in $f$ over the full range. However, not using the large $\hw^2$ treatment shows deviation at the $10^{-4}$ level when $\omega^2=10^{13}$, and the errors increase until at $\omega^2\approx10^{16}$ they are so large that $\Onsagert{sv}$ is reported as numerically 0. Similar ``mirrored'' behavior is shown when using the ``large'' $\hw^2$ algorithm as $\omega^2=10^{-13}$ and as it decreases below $10^{-16}$.

\subsection{Tracer correlation factors for various crystal structures}
\label{sec:tracernumerics}

\Tab{tracer} summarizes tracer correlation factors for nine different crystal structures. The nine crystals cover networks with connectivity ranging from 4 (diamond, wurtzite), 6 (simple cubic), 8 (body-centered cubic, NbO), to 12 (face-centered cubic, hexagonal close-packed), and two mixed networks (hexagonal omega, HCP octahedral-tetrahedral\cite{TrinkleElastodiffusivity2016}). The correlation factors are computed assuming all jumps have the same rate, and all vacancies positions have equal probability; then, the $\hw^2$ rates are also equal, and the correlation factors are temperature independent. The $\threshold=10^{-8}$ and the $k$-point meshes are chosen so that integration errors are $<10^{-8}$; hence, all correlation factors are reported to 8 digits. The cubic structures have isotropic transport coefficients, and hence isotropic correlation factors. The hexagonal crystal structures use the ``ideal'' $c/a$ ratios, which then produces isotropic $\Onsagert{vv}$ transport coefficients for HCP and wurtzite structures, while the hexagonal omega and HCP octahedral-tetrahedral networks have anisotropic $\Onsagert{vv}$ transport. Despite HCP having isotropic $\Onsagert{sv}$, the solute-solute $\Onsagert{ss}$ for the tracer is \textit{anisotropic} at $\approx2.5\EE{-4}$. The similarity of the wurtzite structure to the diamond structure produces the same 1/2 correlation coefficient. The case of garnet is discussed in more detail in \Sec{garnet}.

\begingroup
\squeezetable
\begin{table}[ht]
  \caption{Tracer correlation factor $f=-\Onsagert{ss}/\Onsagert{sv}$ for multiple crystal structures. Each crystal is identified by the common crystal name, and its space group with occupied Wyckoff positions.\cite{Wyckoff} For the cases of wurtzite, hexagonal close-packed and HCP octahedral-tetrahedral we use the ideal $c/a = \sqrt{8/3}$, and for hexagonal omega, we use the ideal $c/a = \sqrt{3/8}$. For NbO and garnet, we only consider jumps on a single subnetwork. All sites whose common distance is less than the cutoff distance are connected with a jump, which contributes to the connectivity. Some structures have more than one unique type of jump, and the connectivity is listed as a sum. All of the jumps are taken to have the same rate. The cubic crystals show isotropic tracer correlation factors, while the hexagonal crystals other than wurtzite show anisotropic correlation. Where available, we report reference calculations of the tracer correlation factor.}
  \label{tab:tracer}
  \begin{tabular}{lrlccccc}
    \hline\hline
    crystal&space group&Wyckoff&cutoff&connectivity&$f_{xx}$&$f_{zz}$&reference values\\
    \hline
    simple cubic&Pm$\bar3$m&1a&$a_0$&6&0.653\,108\,84&0.653\,108\,84& 0.653\,109\cite{Koiwa1983}\\
    body-centered cubic&Im$\bar3$m&2a&$0.9a_0$&8&0.727\,194\,14&0.727\,194\,14& 0.727\,194\cite{Manning1964,Koiwa1983}\\
    face-centered cubic&Fm$\bar3$m&4a&$0.75a_0$&12&0.781\,451\,42&0.781\,451\,42& 0.781\,451\,42\cite{Manning1964,Koiwa1983}\\
    diamond&Fd$\bar3$m&8a&$0.45a_0$&4&0.500\,000\,00&0.500\,000\,00& 1/2\cite{Manning1964,Koiwa1978}\\
    wurtzite&P6$_3$/mmc&4f ($z=1/16$)&$0.62a_0$&1+3&0.500\,000\,00&0.500\,000\,00\\
    hexagonal close-packed&P6$_3$/mmc&2c&$a_0$&6+6&0.781\,204\,88&0.781\,451\,42& 0.781\,204\,89, 0.781\,451\,42\cite{Koiwa1978}\\
    NbO&Pm$\bar3$m&3c (Nb) + 3d (O)&$0.8a_0$&8 (3c)&0.688\,916\,12&0.688\,916\,12& 0.688\,916\cite{Koiwa1983b}\\
    hexagonal omega&P6/mmm&1a + 2d&$0.66a_0$&2+12, 2+3&0.781\,226\,49&0.781\,573\,39\\
    HCP octahedral-tetrahedral&P6$_3$/mmc&2a + 4f ($z=5/8$)&$0.71a_0$&6, 1+3&0.630\,523\,07&0.652\,302\,73\\
    garnet (A$_3$M'$_2$(M''X$_4$)$_3$)&Ia$\bar3$d&24c + 16a + 24d + 96h&$0.31a_0$&4 (24c)&0.374\,972\,67&0.374\,972\,67&0.2491\cite{Carlson2016}\\
    \hline\hline
  \end{tabular}
\end{table}
\endgroup

\begin{figure}[ht]
  \begin{center}
  \includegraphics[width=\figwidth]{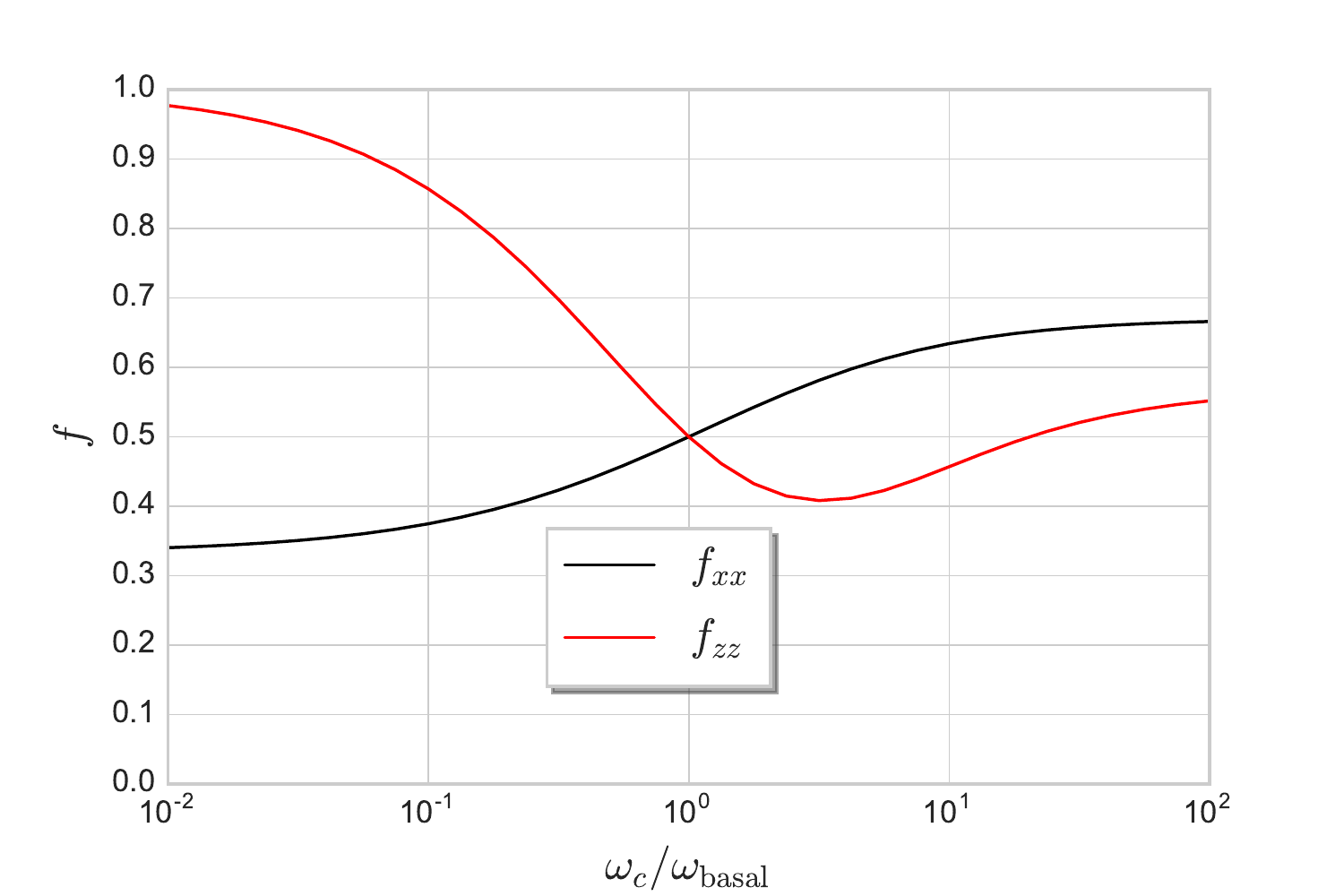}
  \end{center}
\caption{Wurtzite structure tracer correlation factor dependence on $\omega_c$ and $\omega_\text{basal}$. The wurtzite (hexagonal diamond) structure is tetrahedrally coordinated, with one jump along the $c$-axis connecting pairs of 4f sites, and three honeycomb directed jumps primarily in the basal plane.}
\label{fig:wurtzite}
\end{figure}

\Fig{wurtzite} shows the variation in tracer correlation factor with the two unique jump rates for a wurtzite crystal network. The wurtzite (hexagonal diamond) network is tetrahedrally coordinated, where each site has one jump along $c$-axis (jump frequency $\omega_c$), and three jumps that move primarily in basal directions in a honeycomb network. This network requires $c$-axis and ``basal'' jumps to move in the $z$ direction, and ``basal'' jumps to move in the $xy$ plane. As $\omega_c\to 0$, the wurtzite crystal looks like a series of nearly disconnected honeycomb lattices in parallel $xy$ planes; hence, the basal correlation factor $f_{xx}\to 1/3$,\cite{Manning1968} and $f_{zz}\to 1$ as the $c$-axis jumps become uncorrelated. In the wurtzite structure, the $c$-axis jumps do not form one-dimensional chains, but rather connect pairs of 4f sites; hence, as $\omega_c\to\infty$, the basal correlation approaches $\approx 0.6699$, and the $c$-axis correlation approaches $\approx 0.57$. This is different than the the limit of correlation factor for an HCP lattice with zero basal jump rate\cite{Koiwa1983c} (0.644545 and 0.653109); mapping the two sites of a wurtzite lattice onto a single HCP site allows a vacancy and solute to occupy the same HCP ``site,'' which is forbidden in the true HCP case.

\subsection{Green function and tracer correlation factors for garnet}
\label{sec:garnet}
A recent calculation by Carlson and Wilson computed the tracer correlation factors for diffusion on the dodecahedral sites in the garnet structure, using Monte Carlo integration of the lattice Green function.\cite{Carlson2016} The garnet crystal structure appears in many minerals in the earth, and hence predicting diffusion in this structure is important for modeling mass transport in geology. The garnet structure includes pyrope, Mg$_3$Al$_2$Si$_3$O$_{12}$ which is a cubic structure with space group Ia$\bar3$d. The dodecahedral sites are occupied by Mg atoms on the 24c Wyckoff site,\cite{Wyckoff} while Al occupies 16a, Si occupies 24d, and O occupies 96h (0.03284, 0.05014, 0.65330), from Gibbs and Smith.\cite{Gibbs1965} The dodecahedral sites have coordination number 4 with other dodecahedral sites with a cutoff of $0.31a_0$; however, the connectivity is such that there are two fully separated, symmetry related networks that each connect half of the Mg sites.

\Tab{garnet} shows the vacancy Green function entries needed for the first thermodynamic shell, including the agreement and disagreement with the stochastic estimates.\cite{Carlson2016} The stochastic calculation uses a Monte Carlo scheme to compute the lattice Green function values, and the authors' estimated statistical error is $10^{-4}$, which is consistent with the comparison to our computed values. There are two additional discrepancies which are likely due to errors in the analysis of symmetry of the sites: two cases show a difference of a factor of two, while two others conflate two difference vectors that seem superficially equivalent by (cubic) symmetry but actually involve different pathways. This symmetry breaking happens as the fully connected network does \textit{not} maintain cubic symmetry but rather the \textit{pair} of jump networks have cubic symmetry. In addition to these errors, the reported correlation factor (0.2491) differs from our computed value of 0.37497267. This difference could be due to the differences in Green function values used, or an errant factor of 3/2 in the calculation of $\Onsagert{sv}$ used as the denominator in the correlation factor.

\begingroup
\squeezetable
\begin{table}[ht]
  \caption{Green function values for the garnet structure. We use the notation of Carlson and Wilson,\cite{Carlson2016} where $(lmn)$ corresponds to the vector $\dxv = \frac18a_0(l\,\mathbf{\hat i} + m\mathbf{\hat j} + n\mathbf{\hat k})$. The stochastic error in Carlson and Wilson's calculation is $\approx 10^{-4}$; however, there are two other types of error present. In two cases---$(420)$ and $(640)$---the stochastic estimate differs by a factor of 2; and in two more cases---$(444)$ and $(800)$---the stochastic estimate conflated symmetry inequivalent vectors into a single vector, as shown by the reduced error compared with the ``average'' value. Both of these errors are likely due to symmetry analysis errors.}
  \label{tab:garnet}
  \begin{tabular}{cccc}
    \hline\hline
    $(lmn)$&$G(\Rv)$ (this work)&$G(\Rv)$ (\rcite{Carlson2016})&error\\
    \hline
    $(000)$& 2.308\,081\,141\,615 & 2.307\,960\,22 &$1.2092\EE{-4}$\\
    $(211)$& 1.308\,081\,132\,926 & 1.308\,072\,61 &$8.5229\EE{-6}$\\
    $(332)$& 0.806\,767\,995\,595 & 0.806\,695\,36 &$7.2636\EE{-5}$\\
    $(420)$& 0.809\,394\,258\,097 & 0.404\,690\,85 &$\mathbf{4.0470\EE{-1}}$\\
    \hline
    $(\bar444)$& 0.457\,297\,218\,361 & 0.502\,420\,46 &$\mathbf{4.5123\EE{-2}}$\\
    $(\bar44\bar4)$& 0.547\,635\,344\,309 & 0.502\,420\,46 &$\mathbf{4.5215\EE{-2}}$\\
    $\frac12(\bar444)+\frac12(\bar44\bar4)$& 0.502\,466\,281\,335 & 0.502\,420\,46 &$4.5821\EE{-5}$\\
    \hline
    $(532)$& 0.561\,961\,239\,416 & 0.561\,957\,44 &$3.7994\EE{-6}$\\
    $(611)$& 0.560\,766\,700\,022 & 0.560\,710\,92 &$5.5780\EE{-5}$\\
    $(640)$& 0.449\,091\,350\,780 & 0.224\,606\,54 &$\mathbf{2.2448\EE{-1}}$\\
    $(653)$& 0.420\,386\,782\,427 & 0.420\,284\,88 &$1.0190\EE{-4}$\\
    $(655)$& 0.401\,425\,331\,863 & 0.401\,378\,97 &$4.6362\EE{-5}$\\
    $(721)$& 0.444\,350\,262\,895 & 0.444\,378\,78 &$2.8517\EE{-5}$\\
    \hline
    $(800)$& 0.403\,566\,247\,455 & 0.419\,386\,75 &$\mathbf{1.5821\EE{-2}}$\\
    $(00\bar8)$& 0.427\,361\,034\,009 & 0.419\,386\,75 &$\mathbf{7.9743\EE{-3}}$\\
    $\frac13(800)+\frac23(00\bar8)$& 0.419\,429\,438\,491 & 0.419\,386\,75 &$4.2688\EE{-5}$\\
    \hline\hline
  \end{tabular}
\end{table}
\endgroup

\subsection{Solute drag of Si in Ni: comparison with SCMF}
\label{sec:NiSi}

\begin{figure}[ht]
  \begin{center}
  \includegraphics[width=\figwidth]{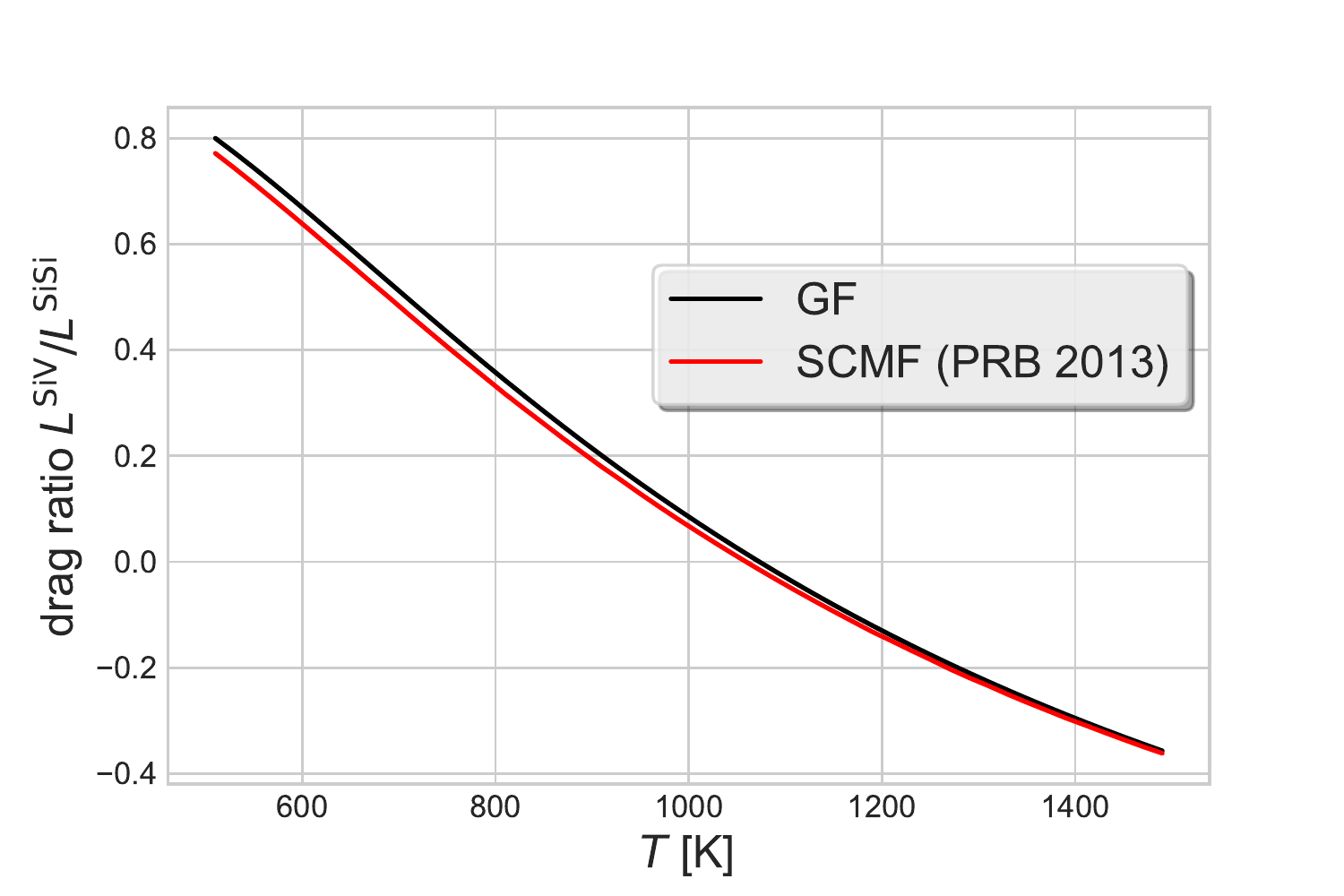}
  \end{center}
\caption{Solute drag ratio of Si in FCC Ni, computed using the Green function method and the self-consistent mean-field method. Identical DFT data from Ref.~\protect\onlinecite{Garnier2014b,Garnier2014bData} is used in both transport coefficient calculations, but the finite range of the Green function in the SCMF method introduces a small underestimation of the correlation. The magnitude of error becomes larger at lower temperatures.}
\label{fig:NiSi}
\end{figure}

To compare with another computational approaches to diffusion, we consider the example of silicon substitutional solutes in nickel, whose drag coefficients were previously calculated using the self-consistent mean-field method.\cite{Garnier2013a} In this case, density-functional theory calculations computed the silicon-vacancy interaction out to the third neighbor: an attraction of 0.1~eV at first neighbor, and a repulsion of 0.045~eV at third neighbor. Using the climbing-image nudged-elastic band method,\cite{Henkelman2000} density-functional theory calculations also determined transition state energies for different transitions around the solute. The computational data, including derivatives with respect to strain, can be found in Ref.~\onlinecite{Garnier2014b,Garnier2014bData}. The SCMF calculations used a ``3nn+3nn'' approach, where the effective Hamiltonian is cutoff at the third-neighbors of the third-neighbors of a vacancy. This truncation of the Green function is an approximation, which produces some error in the transport coefficients. \Fig{NiSi} shows a comparison to the drag coefficients ($\Lsv/\Lss$, also known as the ``vacancy wind'') using the same density-functional theory data. The errors in the drag ratio are due to the cutoff of the Green function in the SCMF method, where the error decreases at higher temperatures as the correlation becomes smaller.

\subsection{Solute drag of Sn and Zn in Mg: comparison with KMC}
\label{sec:MgSnZn}

\begin{figure*}[ht]
  \begin{center}
  \includegraphics[width=0.48\textwidth]{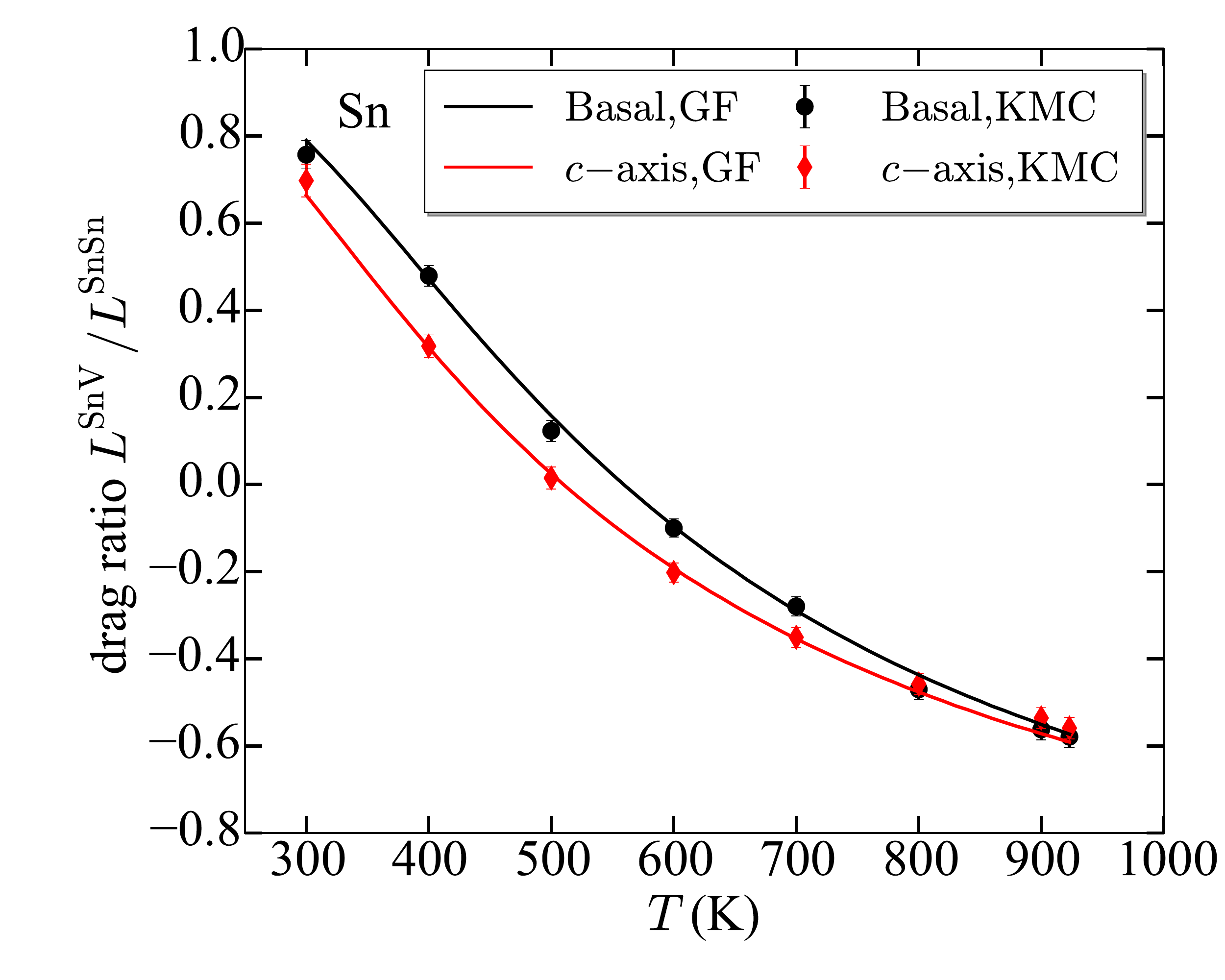}
  \includegraphics[width=0.48\textwidth]{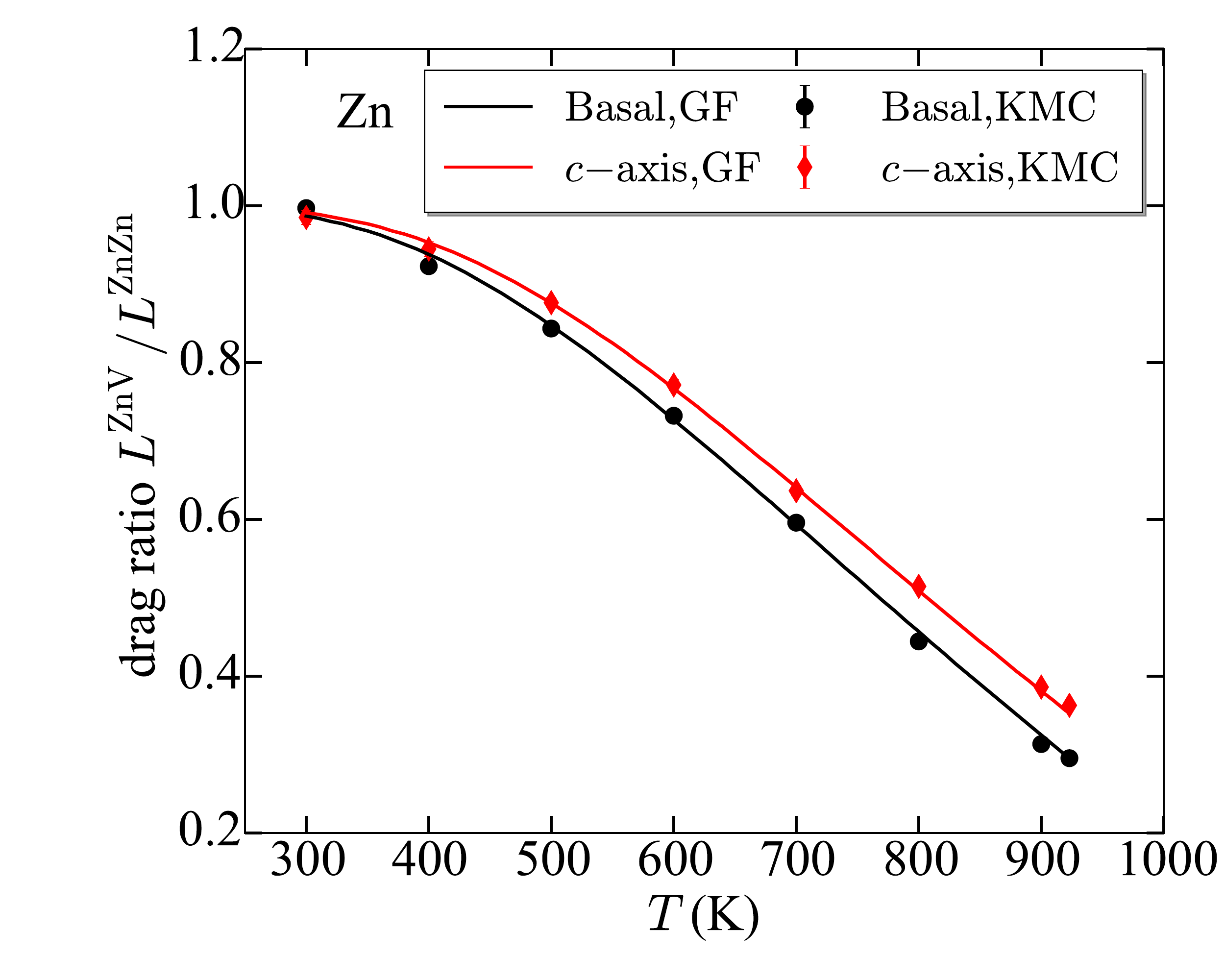}
  \end{center}
\caption{Solute drag ratio of Sn and Zn in HCP Mg, computed using the Green function method and kinetic Monte Carlo. The same density-functional theory data\protect\cite{Agarwal2017} is used for solute-vacancy interaction energy and transition state energies. The deviation between the KMC results and the Green function approach is similar in magnitude to the stochastic error in the KMC results from $12\x10^6$ trajectories.}
\label{fig:Mgdrag}
\end{figure*}

For a comparison with a stochastic approaches to diffusion, we consider the examples of tin and zinc substitutional solutes in magnesium. In this case, the first-principles data are available in the supporting material of Ref.~\onlinecite{Agarwal2017}, where other Green function results are reported for substitutional solutes in HCP magnesium. Here, we compare with kinetic Monte Carlo simulations of the drag coefficient using the same first-principles data. Solute-vacancy interaction energies were computed for nine different neighbors---which correspond to positions that are two jumps from the position of a solute---and all possible transitions of a vacancy from the first neighbor positions, using density-functional theory and the climbing-image nudged-elastic band method.\cite{Henkelman2000} The two cases considered here, Sn and Zn, required careful treatment of the cell sizes, number of jumps, and number of trajectories in order to reduce the stochastic error and to converge the off-diagonal transport coefficients in the most efficient way possible. In particular, $\Lsv$ is very slow to converge. \Tab{KMCparam} contains the simulation choices which vary with temperature, as correlation becomes less important. For all simulations, $12\x10^6$ trajectories were averaged to compute transport coefficients and stochastic errors. \Fig{Mgdrag} shows the comparison of drag coefficients in basal and $c$-axis orientations computed with kinetic Monte Carlo. Good agreement is found over the full temperature range, where the error compared with the KMC is similar to the error predicted from the standard deviation.

\begingroup
\squeezetable
\begin{table}[ht]
  \caption{Kinetic Monte Carlo parameters for Sn and Zn diffusion in Mg. At lower temperatures, correlation is more important, so larger cells with more Monte Carlo steps were used to converge the simulations. At lower temperature, the primary effect of increasing the simulation cell size is to increase the number of steps per trajectory. In all cases, $12\x10^6$ trajectories were used to average the transport coefficients and estimate the stochastic error in \Fig{Mgdrag}.}
  \label{tab:KMCparam}
  \begin{tabular}{c@{\quad}cc@{\quad}cc}
    \hline\hline
    &\multicolumn{2}{c}{Sn}&\multicolumn{2}{c}{Zn}\\
    $T$ [K]&supercell size&number of steps&supercell size&number of steps\\
    \hline
    300&$42\x42\x40$&$1\cdot42\cdot42\cdot40=70560$&
    $22\x22\x20$&$5\cdot22\cdot22\cdot20=48400$\\
    400&$30\x30\x28$&$1\cdot30\cdot30\cdot28=25200$&
    $20\x20\x18$&$5\cdot20\cdot20\cdot18=36000$\\
    500&$28\x28\x26$&$1\cdot28\cdot28\cdot26=20384$&
    $14\x14\x12$&$5\cdot14\cdot14\cdot12=11760$\\
    600&$24\x24\x22$&$1\cdot24\cdot24\cdot22=12672$&
    $14\x14\x12$&$5\cdot14\cdot14\cdot12=11760$\\
    700&$24\x24\x22$&$1\cdot24\cdot24\cdot22=12672$&
    $12\x12\x10$&$5\cdot12\cdot12\cdot10=7200$\\
    800&$24\x24\x22$&$1\cdot24\cdot24\cdot22=12672$&
    $10\x10\x8$&$5\cdot10\cdot10\cdot8=4000$\\
    900&$24\x24\x22$&$1\cdot24\cdot24\cdot22=12672$&
    $10\x10\x8$&$5\cdot10\cdot10\cdot8=4000$\\
    923&$24\x24\x22$&$1\cdot24\cdot24\cdot22=12672$&
    $10\x10\x8$&$5\cdot10\cdot10\cdot8=4000$\\
    \hline\hline
  \end{tabular}
\end{table}
\endgroup

\section{Discussion}
\label{sec:discussion}
The automated numerical approach with controllable errors for arbitrary crystals and arbitrarily large thermodynamic interaction range provides a significant improvement in our ability to compute mass transport coefficients for the dilute-vacancy/dilute-solute limit. It has previously been difficult to make quantitative predictions without approximation from accurate density-functional theory data outside of specific cases, but this new approach enables new predictions. There are also several extensions possible from this approach. First, similar to \rcite{TrinkleElastodiffusivity2016}, we could use a perturbation theory approach to evaluate \textit{derivatives} of the transport coefficients, such as elastodiffusivity, activation energy, and volume directly. Combining that methodology with the current approach would be challenging in the arbitrary crystal symmetry case. In the interim, finite-difference approaches\cite{Garnier2014c} can be used in a straightforward manner with the current implementation. Second, there are crystals---most notably, titanium\cite{Shang2011} with an anomalous vacancy migration jump---where the vacancy diffuses through additional metastable sites in the lattice; these sites are unoccupied unless a vacancy is diffusing. The calculation of the vacancy Green function can be performed identically, but the definition of the $\hw^1$ and $\hw^2$ matrices can become more complex. Third, to consider ordered structures where more than one sublattice is involved in diffusion---including the creation of new defects such as antisites---will require a more complex state space to be defined for a finite number of antisites to treat diffusion in intermetallics such as B2 and L1$_2$. In a related but different problem, more complex diffusion mechanisms where a solute can \textit{both} diffuse via a vacancy-solute complex mechanism \textit{and} as an interstitial introduces new transitions that would need to be considered. Finally, to move beyond the dilute solute (or dilute vacancy) limit, we can rely on other Green-function based solutions such as mean-field approximations, the coherent potential approximation, or diagrammatic series. Just as our new method makes quantitative predictions possible for dilute-vacancy/dilute-solute transport coefficients, new developments will expand where quantitative predictions are possible in the future.

\section*{Acknowledgments}
  The author thanks Maylise Nastar, Thomas Schuler, and Pascal Bellon for helpful conversations and suggestions for the manuscript. Abhinav C.~P.~Jain and Ravi Agarwal contributed kinetic Monte Carlo simulations for hexagonal-close packed systems.

\section*{Disclosure}
  The author has no conflicts of interest.

\section*{Funding}
  This research was supported in part by the U.S. Department of Energy, Office of Basic Energy Sciences, Division of Materials Sciences and Engineering under Award \#DE-FG02-05ER46217, through the Frederick Seitz Materials Research Laboratory, in part by the Office of Naval Research grant N000141210752, and in part by the National Science Foundation Award 1411106. Part of this research was performed while the author was visiting the Institute for Pure and Applied Mathematics (IPAM) at UCLA, which is supported by the National Science Foundation (NSF).

\appendix
\section{Homogeneous polynomials with constant orbital angular momentum}
\label{sec:orientation}
The Taylor expansions that describe analytic terms in the bare vacancy Green function appear as homogeneous polynomials of components $\hat p_i$ up to fourth order; these can be grouped into polynomials of constant orbital angular momentum. While what follows is general, we are interested in the case where our homogeneous polynomials have maximum order $\Lmax = 4$. Then there are $1 + 3 + 5 + 7 + 9 = 25 = (\Lmax+1)^2$ distinct $Y_\ell^m(\hat\pv)$ spherical harmonics and $1 + 3 + 6 + 10 + 15 = 35 = (\Lmax+1)^2 + (\Lmax+1)(\Lmax)(\Lmax-1)/6$ homogeneous polynomials from order 0 to 4. The homogeneous polynomials are a spanning set on the space generated by the spherical harmonics; the homogeneous polynomials of order $\ell$ are a basis for the space generated by the spherical harmonics with orbital angular momentum $\ell$, $\ell-2$, $\ell-4$, \dots. We define the $25\x35$ matrix,
\be
E^{\ell m}_{[n_1n_2n_3]} := \int_0^\pi \int_0^{2\pi}
(-1)^m Y_\ell^{-m}(\theta,\phi)
(\sin\theta\cos\phi)^{n_1}(\sin\theta\sin\phi)^{n_2}(\cos\theta)^{n_3}
\;d\phi\sin\theta\; d\theta
\ee
which is the expansion of $\hat p_1^{n_1} \hat p_2^{n_2} \hat p_3^{n_3}$ in spherical harmonics:
\be
\hat p_1^{n_1} \hat p_2^{n_2} \hat p_3^{n_3} =
\sum_{\ell=0}^{\Lmax}\sum_{m=-\ell}^\ell E^{\ell m}_{[n_1n_2n_3]} Y_\ell^m(\theta,\phi).
\ee
Similarly, we can write the spherical harmonics in terms of the polynomials $\hat p_1^{n_1} \hat p_2^{n_2} \hat p_3^{n_3}$
\be
Y_\ell^m(\theta,\phi) = \sum_{n_1+n_2+n_3\le\ell} C^{\ell m}_{[n_1n_2n_3]} (\sin\theta\cos\phi)^{n_1}(\sin\theta\sin\phi)^{n_2}(\cos\theta)^{n_3}
\ee
for the $25\times35$ coefficient matrix $C^{\ell m}_{[n_1n_2n_3]}$. The coefficient matrix is most easily constructed using recurrence relations for the spherical harmonics.\cite{IntegralTables} These matrices $C$ and $E$ are such that
\be
\sum_{n_1+n_2+n_3\le\Lmax} C^{\ell m}_{[n_1n_2n_3]} E^{\ell'm'}_{[n_1n_2n_3]} = \delta_{\ell\ell'}\delta_{mm'}
\ee
while the matrix
\be
P_{[n_1n_2n_3],[n'_1n'_2n'_3]} :=
\sum_{\ell=0}^{\Lmax}\sum_{m=-\ell}^\ell C^{\ell m}_{[n_1n_2n_3]} E^{\ell m}_{[n'_1n'_2n'_3]}
\label{eqn:power-projection}
\ee
defines a projection from a vector of polynomials into an equivalent (``reduced'') representation consistent with the spherical harmonic basis. The projection $P$ is an idempotent matrix, and
\be
\sum_{n_1+n_2+n_3\le\Lmax}
f_{[n_1n_2n_3]}\hat p_1^{n_1} \hat p_2^{n_2} \hat p_3^{n_3}
=
\sum_{n_1+n_2+n_3\le\Lmax}
\left[\sum_{n'_1+n'_2+n'_3\le\Lmax} P_{[n_1n_2n_3],[n'_1n'_2n'_3]} f_{[n'_1n'_2n'_3]}\right]
\hat p_1^{n_1} \hat p_2^{n_2} \hat p_3^{n_3}
\ee
for any polynomial coefficients $f_{[n_1n_2n_3]}$ and unit vector $\hat\pv$. Similarly, we define
\be
P^\ell_{[n_1n_2n_3],[n'_1n'_2n'_3]} :=
\sum_{m=-\ell}^\ell C^{\ell m}_{[n_1n_2n_3]} E^{\ell m}_{[n'_1n'_2n'_3]}
\label{eqn:l-projection}
\ee
as the projection which only retains polynomials with orbital angular momentum $\ell$. Thus, we can reduce and group a polynomial
\be
\sum_{n_1+n_2+n_3\le\Lmax} f_{[n_1n_2n_3]} \hat p_1^{n_1} \hat p_2^{n_2} \hat p_3^{n_3}
\ee
as
\be
\sum_{\ell=0}^{\Lmax}\left\{
\sum_{n_1+n_2+n_3\le\ell}
\left[\sum_{n'_1+n'_2+n'_3\le\Lmax} P^\ell_{[n_1n_2n_3],[n'_1n'_2n'_3]} f_{n'_1n'_2n'_3}\right]
\hat p_1^{n_1} \hat p_2^{n_2} \hat p_3^{n_3}
\right\}
\ee
where each polynomial expansion in curly braces has a single orbital angular momentum $\ell$.

\section{Implementation}
\label{sec:implementation}
A full numerical implementation of the algorithms in Python described are available on github\cite{OnsagerCalc} under the MIT License. This includes algorithms to analyze a given crystal (lattice and atomic basis), find generators for the space group operations, determine all point group operations for each site, identify Wyckoff positions, generate crystal star sets and vector star bases, construct a jump network for vacancies in the crystal, and identify unique jumps. Once the energies and prefactors are determined for the unique vacancy and solute sites, complexes, and jumps, the numerical implementation can compute the Onsager coefficients for a given temperature. In addition, Jupyter notebooks to compute the numerical results in this paper are available on github. The implementation includes a full test suite of the modules and functions that were used during implementation development.


\begin{thebibliography}{71}%
\makeatletter
\providecommand \@ifxundefined [1]{%
 \@ifx{#1\undefined}
}%
\providecommand \@ifnum [1]{%
 \ifnum #1\expandafter \@firstoftwo
 \else \expandafter \@secondoftwo
 \fi
}%
\providecommand \@ifx [1]{%
 \ifx #1\expandafter \@firstoftwo
 \else \expandafter \@secondoftwo
 \fi
}%
\providecommand \bibnumfmt[1]{[#1]}
\providecommand \natexlab [1]{#1}%
\providecommand \enquote  [1]{\textit{#1}}%
\providecommand \bibnamefont  [1]{#1}%
\providecommand \bibfnamefont [1]{#1}%
\providecommand \citenamefont [1]{#1}%
\providecommand \href@noop [0]{\@secondoftwo}%
\providecommand \href [0]{\begingroup \@sanitize@url \@href}%
\providecommand \@href[1]{\@@startlink{#1}\@@href}%
\providecommand \@@href[1]{\endgroup#1\@@endlink}%
\providecommand \@sanitize@url [0]{\catcode `\\12\catcode `\$12\catcode
  `\&12\catcode `\#12\catcode `\^12\catcode `\_12\catcode `\%12\relax}%
\providecommand \@@startlink[1]{}%
\providecommand \@@endlink[0]{}%
\providecommand \url  [0]{\begingroup\@sanitize@url \@url }%
\providecommand \@url [1]{\endgroup\@href {#1}{\urlprefix }}%
\providecommand \urlprefix  [0]{URL }%
\providecommand \Eprint [0]{\href }%
\providecommand \doibase [0]{http://dx.doi.org/}%
\providecommand \selectlanguage [0]{\@gobble}%
\providecommand \bibinfo  [0]{\@secondoftwo}%
\providecommand \bibfield  [0]{\@secondoftwo}%
\providecommand \translation [1]{[#1]}%
\providecommand \BibitemOpen [0]{}%
\providecommand \bibitemStop [0]{}%
\providecommand \bibitemNoStop [0]{.\EOS\space}%
\providecommand \EOS [0]{\spacefactor3000\relax}%
\providecommand \BibitemShut  [1]{\csname bibitem#1\endcsname}%
\let\auto@bib@innerbib\@empty
\bibitem [{\citenamefont {Balluffi}, \citenamefont {Allen},\ and\ \citenamefont
  {Carter}(2005)}]{Balluffi-Kinetics}%
  \BibitemOpen
  \bibfield  {author} {\bibinfo {author} {\bibfnamefont {R.~W.}\ \bibnamefont
  {Balluffi}}, \bibinfo {author} {\bibfnamefont {S.~M.}\ \bibnamefont {Allen}},
  \ and\ \bibinfo {author} {\bibfnamefont {W.~C.}\ \bibnamefont {Carter}},\
  }\href {\doibase 10.1002/0471749311} {\emph {\bibinfo {title} {Kinetics of
  Materials}}}\ (\bibinfo  {publisher} {John Wiley \& Sons, Inc.},\ \bibinfo
  {year} {2005})\BibitemShut {NoStop}%
\bibitem [{\citenamefont {Janotti}\ \emph {et~al.}(2004)\citenamefont
  {Janotti}, \citenamefont {Kr{\v c}mar}, \citenamefont {Fu},\ and\
  \citenamefont {Reed}}]{Janotti2004}%
  \BibitemOpen
  \bibfield  {author} {\bibinfo {author} {\bibfnamefont {A.}~\bibnamefont
  {Janotti}}, \bibinfo {author} {\bibfnamefont {M.}~\bibnamefont {Kr{\v
  c}mar}}, \bibinfo {author} {\bibfnamefont {C.~L.}\ \bibnamefont {Fu}}, \ and\
  \bibinfo {author} {\bibfnamefont {R.~C.}\ \bibnamefont {Reed}},\ }\bibfield
  {title} {\enquote {\bibinfo {title} {Solute diffusion in metals: Larger atoms
  can move faster},}\ }\href {\doibase 10.1103/PhysRevLett.92.085901}
  {\bibfield  {journal} {\bibinfo  {journal} {Phys. Rev. Lett.}\ }\textbf
  {\bibinfo {volume} {92}},\ \bibinfo {pages} {085901} (\bibinfo {year}
  {2004})}\BibitemShut {NoStop}%
\bibitem [{\citenamefont {Mantina}\ \emph {et~al.}(2009)\citenamefont
  {Mantina}, \citenamefont {Wang}, \citenamefont {Chen}, \citenamefont {Liu},\
  and\ \citenamefont {Wolverton}}]{Mantina2009}%
  \BibitemOpen
  \bibfield  {author} {\bibinfo {author} {\bibfnamefont {M.}~\bibnamefont
  {Mantina}}, \bibinfo {author} {\bibfnamefont {Y.}~\bibnamefont {Wang}},
  \bibinfo {author} {\bibfnamefont {L.~Q.}\ \bibnamefont {Chen}}, \bibinfo
  {author} {\bibfnamefont {Z.~K.}\ \bibnamefont {Liu}}, \ and\ \bibinfo
  {author} {\bibfnamefont {C.}~\bibnamefont {Wolverton}},\ }\bibfield  {title}
  {\enquote {\bibinfo {title} {First principles impurity diffusion
  coefficients},}\ }\href@noop {} {\bibfield  {journal} {\bibinfo  {journal}
  {Acta mater.}\ }\textbf {\bibinfo {volume} {57}},\ \bibinfo {pages}
  {4102--4108} (\bibinfo {year} {2009})}\BibitemShut {NoStop}%
\bibitem [{\citenamefont {Garnier}\ \emph
  {et~al.}(2014{\natexlab{a}})\citenamefont {Garnier}, \citenamefont {Li},
  \citenamefont {Nastar}, \citenamefont {Bellon},\ and\ \citenamefont
  {Trinkle}}]{Garnier2014c}%
  \BibitemOpen
  \bibfield  {author} {\bibinfo {author} {\bibfnamefont {T.}~\bibnamefont
  {Garnier}}, \bibinfo {author} {\bibfnamefont {Z.}~\bibnamefont {Li}},
  \bibinfo {author} {\bibfnamefont {M.}~\bibnamefont {Nastar}}, \bibinfo
  {author} {\bibfnamefont {P.}~\bibnamefont {Bellon}}, \ and\ \bibinfo {author}
  {\bibfnamefont {D.~R.}\ \bibnamefont {Trinkle}},\ }\bibfield  {title}
  {\enquote {\bibinfo {title} {Calculation of strain effects on
  vacancy-mediated diffusion of impurities in {FCC} structures: General
  approach and application to {Ni$_{1-x}$Si}},}\ }\href {\doibase
  10.1103/PhysRevB.90.184301} {\bibfield  {journal} {\bibinfo  {journal} {Phys.
  Rev. B}\ }\textbf {\bibinfo {volume} {90}},\ \bibinfo {pages} {184301}
  (\bibinfo {year} {2014}{\natexlab{a}})}\BibitemShut {NoStop}%
\bibitem [{\citenamefont
  {Trinkle}(2016{\natexlab{a}})}]{TrinkleElastodiffusivity2016}%
  \BibitemOpen
  \bibfield  {author} {\bibinfo {author} {\bibfnamefont {D.~R.}\ \bibnamefont
  {Trinkle}},\ }\bibfield  {title} {\enquote {\bibinfo {title} {Diffusivity and
  derivatives for interstitial solutes: Activation energy, volume, and
  elastodiffusion tensors},}\ }\href {\doibase 10.1080/14786435.2016.1212175}
  {\bibfield  {journal} {\bibinfo  {journal} {Philos. Mag.}\ }\textbf {\bibinfo
  {volume} {96}},\ \bibinfo {pages} {2714--2735} (\bibinfo {year}
  {2016}{\natexlab{a}})}\BibitemShut {NoStop}%
\bibitem [{\citenamefont {Manning}(1968)}]{Manning1968}%
  \BibitemOpen
  \bibfield  {author} {\bibinfo {author} {\bibfnamefont {J.~R.}\ \bibnamefont
  {Manning}},\ }\href@noop {} {\emph {\bibinfo {title} {Diffusion Kinetics for
  Atoms in Crystals}}}\ (\bibinfo  {publisher} {Princeton: van Nostrand},\
  \bibinfo {year} {1968})\BibitemShut {NoStop}%
\bibitem [{\citenamefont {Allnatt}\ and\ \citenamefont
  {Lidiard}(1993)}]{Allnatt1993}%
  \BibitemOpen
  \bibfield  {author} {\bibinfo {author} {\bibfnamefont {A.~R.}\ \bibnamefont
  {Allnatt}}\ and\ \bibinfo {author} {\bibfnamefont {A.~B.}\ \bibnamefont
  {Lidiard}},\ }\enquote {\bibinfo {title} {Atomic transport in solids},}\ \
  (\bibinfo  {publisher} {Cambridge University Press},\ \bibinfo {address}
  {Cambridge},\ \bibinfo {year} {1993})\ Chap.~\bibinfo {chapter} {5}, pp.\
  \bibinfo {pages} {202--203}\BibitemShut {NoStop}%
\bibitem [{\citenamefont {Leclaire}\ and\ \citenamefont
  {Lidiard}(1956)}]{Leclaire1956}%
  \BibitemOpen
  \bibfield  {author} {\bibinfo {author} {\bibfnamefont {A.~D.}\ \bibnamefont
  {Leclaire}}\ and\ \bibinfo {author} {\bibfnamefont {A.~B.}\ \bibnamefont
  {Lidiard}},\ }\bibfield  {title} {\enquote {\bibinfo {title} {{LIII}.
  correlation effects in diffusion in crystals},}\ }\href {\doibase
  10.1080/14786435608238133} {\bibfield  {journal} {\bibinfo  {journal}
  {Philos. Mag.}\ }\textbf {\bibinfo {volume} {1}},\ \bibinfo {pages}
  {518--527} (\bibinfo {year} {1956})}\BibitemShut {NoStop}%
\bibitem [{\citenamefont {Bocquet}(1974)}]{Bocquet1974}%
  \BibitemOpen
  \bibfield  {author} {\bibinfo {author} {\bibfnamefont {J.-L.}\ \bibnamefont
  {Bocquet}},\ }\bibfield  {title} {\enquote {\bibinfo {title} {Expression
  cin{\'e}tique des coefficients ph{\'e}nom{\'e}nologiques de diffusion
  {$L_{AA}$}, {$L_{AB}$}, {$L_{BB}$} dans un alliage dilu{\'e} {A-B}},}\ }\href
  {\doibase 10.1016/0001-6160(74)90119-9} {\bibfield  {journal} {\bibinfo
  {journal} {Acta metall.}\ }\textbf {\bibinfo {volume} {22}},\ \bibinfo
  {pages} {1--5} (\bibinfo {year} {1974})}\BibitemShut {NoStop}%
\bibitem [{\citenamefont {Jones}\ and\ \citenamefont
  {{Le~Claire}}(1972)}]{Jones1972}%
  \BibitemOpen
  \bibfield  {author} {\bibinfo {author} {\bibfnamefont {M.}~\bibnamefont
  {Jones}}\ and\ \bibinfo {author} {\bibfnamefont {A.}~\bibnamefont
  {{Le~Claire}}},\ }\bibfield  {title} {\enquote {\bibinfo {title} {Solvent
  self-diffusion in dilute b.c.c. solid solutions {II.} {Calculation} of
  partial correlation factors},}\ }\href {\doibase 10.1080/14786437208227373}
  {\bibfield  {journal} {\bibinfo  {journal} {Philos. Mag.}\ }\textbf {\bibinfo
  {volume} {26}},\ \bibinfo {pages} {1191} (\bibinfo {year}
  {1972})}\BibitemShut {NoStop}%
\bibitem [{\citenamefont {Barbu}\ and\ \citenamefont
  {Lidiard}(1996)}]{Barbu1996}%
  \BibitemOpen
  \bibfield  {author} {\bibinfo {author} {\bibfnamefont {A.}~\bibnamefont
  {Barbu}}\ and\ \bibinfo {author} {\bibfnamefont {A.}~\bibnamefont
  {Lidiard}},\ }\bibfield  {title} {\enquote {\bibinfo {title} {Solute
  segregation in dilute bcc alloys under irradiation},}\ }\href {\doibase
  10.1080/01418619608243537} {\bibfield  {journal} {\bibinfo  {journal}
  {Philos. Mag. A}\ }\textbf {\bibinfo {volume} {74}},\ \bibinfo {pages} {709}
  (\bibinfo {year} {1996})}\BibitemShut {NoStop}%
\bibitem [{\citenamefont {Ghate}(1964)}]{Ghate1964}%
  \BibitemOpen
  \bibfield  {author} {\bibinfo {author} {\bibfnamefont {P.~B.}\ \bibnamefont
  {Ghate}},\ }\bibfield  {title} {\enquote {\bibinfo {title} {Screened
  interaction model for impurity diffusion in {Zinc}},}\ }\href {\doibase
  10.1103/PhysRev.133.A1167} {\bibfield  {journal} {\bibinfo  {journal} {Phys.
  Rev.}\ }\textbf {\bibinfo {volume} {133}},\ \bibinfo {pages} {A1167--A1175}
  (\bibinfo {year} {1964})}\BibitemShut {NoStop}%
\bibitem [{\citenamefont {Batra}(1967)}]{Batra1967}%
  \BibitemOpen
  \bibfield  {author} {\bibinfo {author} {\bibfnamefont {A.~P.}\ \bibnamefont
  {Batra}},\ }\bibfield  {title} {\enquote {\bibinfo {title} {Anisotropic
  isotope effect for diffusion of {Zinc} and {Cadmium} in {Zinc}},}\ }\href
  {\doibase 10.1103/PhysRev.159.487} {\bibfield  {journal} {\bibinfo  {journal}
  {Phys. Rev.}\ }\textbf {\bibinfo {volume} {159}},\ \bibinfo {pages}
  {487--499} (\bibinfo {year} {1967})}\BibitemShut {NoStop}%
\bibitem [{\citenamefont {Allnatt}, \citenamefont {Belova},\ and\ \citenamefont
  {Murch}(2014)}]{Allnatt2014}%
  \BibitemOpen
  \bibfield  {author} {\bibinfo {author} {\bibfnamefont {A.}~\bibnamefont
  {Allnatt}}, \bibinfo {author} {\bibfnamefont {I.}~\bibnamefont {Belova}}, \
  and\ \bibinfo {author} {\bibfnamefont {G.}~\bibnamefont {Murch}},\ }\bibfield
   {title} {\enquote {\bibinfo {title} {Diffusion kinetics in dilute binary
  alloys with the h.c.p. crystal structure},}\ }\href {\doibase
  10.1080/14786435.2014.916426} {\bibfield  {journal} {\bibinfo  {journal}
  {Philos. Mag.}\ }\textbf {\bibinfo {volume} {94}},\ \bibinfo {pages}
  {2487--2504} (\bibinfo {year} {2014})}\BibitemShut {NoStop}%
\bibitem [{\citenamefont {Murch}(1984)}]{Murch1984}%
  \BibitemOpen
  \bibfield  {author} {\bibinfo {author} {\bibfnamefont {G.~E.}\ \bibnamefont
  {Murch}},\ }\bibfield  {title} {\enquote {\bibinfo {title} {Simulation of
  diffusion kinetics with the monte carlo method,},}\ }in\ \href {\doibase
  10.1016/B978-0-12-522662-2.50012-1} {\emph {\bibinfo {booktitle} {Diffusion
  in Crystalline Solids}}},\ \bibinfo {editor} {edited by\ \bibinfo {editor}
  {\bibfnamefont {G.~E.}\ \bibnamefont {Murch}}\ and\ \bibinfo {editor}
  {\bibfnamefont {A.~S.}\ \bibnamefont {Nowick}}}\ (\bibinfo  {publisher}
  {Orlando, Florida: Academic Press},\ \bibinfo {year} {1984})\ Chap.~\bibinfo
  {chapter} {7}, pp.\ \bibinfo {pages} {379--427}\BibitemShut {NoStop}%
\bibitem [{\citenamefont {Belova}\ and\ \citenamefont
  {Murch}(2000)}]{Belova2000}%
  \BibitemOpen
  \bibfield  {author} {\bibinfo {author} {\bibfnamefont {I.~V.}\ \bibnamefont
  {Belova}}\ and\ \bibinfo {author} {\bibfnamefont {G.~E.}\ \bibnamefont
  {Murch}},\ }\bibfield  {title} {\enquote {\bibinfo {title} {Collective
  diffusion in the binary random alloy},}\ }\href {\doibase
  10.1080/01418610008212070} {\bibfield  {journal} {\bibinfo  {journal}
  {Philos. Mag. A}\ }\textbf {\bibinfo {volume} {80}},\ \bibinfo {pages}
  {599--607} (\bibinfo {year} {2000})}\BibitemShut {NoStop}%
\bibitem [{\citenamefont {Belova}\ and\ \citenamefont
  {Murch}(2001)}]{Belova2001}%
  \BibitemOpen
  \bibfield  {author} {\bibinfo {author} {\bibfnamefont {I.~V.}\ \bibnamefont
  {Belova}}\ and\ \bibinfo {author} {\bibfnamefont {G.~E.}\ \bibnamefont
  {Murch}},\ }\bibfield  {title} {\enquote {\bibinfo {title} {Behaviour of the
  diffusion vacancy-wind factors in the concentrated random alloy},}\ }\href
  {\doibase 10.1080/01418610108216635} {\bibfield  {journal} {\bibinfo
  {journal} {Philos. Mag. A}\ }\textbf {\bibinfo {volume} {81}},\ \bibinfo
  {pages} {1749--1758} (\bibinfo {year} {2001})}\BibitemShut {NoStop}%
\bibitem [{\citenamefont {Belova}\ and\ \citenamefont
  {Murch}(2003{\natexlab{a}})}]{Belova2003a}%
  \BibitemOpen
  \bibfield  {author} {\bibinfo {author} {\bibfnamefont {I.~V.}\ \bibnamefont
  {Belova}}\ and\ \bibinfo {author} {\bibfnamefont {G.~E.}\ \bibnamefont
  {Murch}},\ }\bibfield  {title} {\enquote {\bibinfo {title} {Computer
  simulation of solute-enhanced diffusion kinetics in dilute fcc alloys},}\
  }\href {\doibase 10.1080/0141861021000025810} {\bibfield  {journal} {\bibinfo
   {journal} {Philos. Mag.}\ }\textbf {\bibinfo {volume} {83}},\ \bibinfo
  {pages} {377--392} (\bibinfo {year} {2003}{\natexlab{a}})}\BibitemShut
  {NoStop}%
\bibitem [{\citenamefont {Belova}\ and\ \citenamefont
  {Murch}(2003{\natexlab{b}})}]{Belova2003b}%
  \BibitemOpen
  \bibfield  {author} {\bibinfo {author} {\bibfnamefont {I.~V.}\ \bibnamefont
  {Belova}}\ and\ \bibinfo {author} {\bibfnamefont {G.~E.}\ \bibnamefont
  {Murch}},\ }\bibfield  {title} {\enquote {\bibinfo {title} {Solvent diffusion
  kinetics in the dilute random alloy},}\ }\href {\doibase
  10.1080/0141861021000025838} {\bibfield  {journal} {\bibinfo  {journal}
  {Philos. Mag.}\ }\textbf {\bibinfo {volume} {83}},\ \bibinfo {pages}
  {393--399} (\bibinfo {year} {2003}{\natexlab{b}})}\BibitemShut {NoStop}%
\bibitem [{\citenamefont {Nastar}, \citenamefont {Dobretsov},\ and\
  \citenamefont {Martin}(2000)}]{Nastar2000}%
  \BibitemOpen
  \bibfield  {author} {\bibinfo {author} {\bibfnamefont {M.}~\bibnamefont
  {Nastar}}, \bibinfo {author} {\bibfnamefont {V.~Y.}\ \bibnamefont
  {Dobretsov}}, \ and\ \bibinfo {author} {\bibfnamefont {G.}~\bibnamefont
  {Martin}},\ }\bibfield  {title} {\enquote {\bibinfo {title} {Self-consistent
  formulation of configurational kinetics close to equilibrium: The
  phenomenological coefficients for diffusion in crystalline solids},}\ }\href
  {\doibase 10.1080/01418610008212047} {\bibfield  {journal} {\bibinfo
  {journal} {Philos. Mag. A}\ }\textbf {\bibinfo {volume} {80}},\ \bibinfo
  {pages} {155--184} (\bibinfo {year} {2000})}\BibitemShut {NoStop}%
\bibitem [{\citenamefont {Nastar}(2005)}]{Nastar2005}%
  \BibitemOpen
  \bibfield  {author} {\bibinfo {author} {\bibfnamefont {M.}~\bibnamefont
  {Nastar}},\ }\bibfield  {title} {\enquote {\bibinfo {title} {A mean field
  theory for diffusion in a dilute multi-component alloy: a new model for the
  effect of solutes on self-diffusion},}\ }\href {\doibase
  10.1080/14786430500228390} {\bibfield  {journal} {\bibinfo  {journal}
  {Philos. Mag.}\ }\textbf {\bibinfo {volume} {85}},\ \bibinfo {pages}
  {3767--3794} (\bibinfo {year} {2005})}\BibitemShut {NoStop}%
\bibitem [{\citenamefont {Belashchenko}\ and\ \citenamefont
  {Vaks}(1998)}]{Belashchenko1998}%
  \BibitemOpen
  \bibfield  {author} {\bibinfo {author} {\bibfnamefont {K.~D.}\ \bibnamefont
  {Belashchenko}}\ and\ \bibinfo {author} {\bibfnamefont {V.~G.}\ \bibnamefont
  {Vaks}},\ }\bibfield  {title} {\enquote {\bibinfo {title} {The master
  equation approach to configurational kinetics of alloys via the vacancy
  exchange mechanism: general relations and features of microstructural
  evolution},}\ }\href {\doibase 10.1088/0953-8984/10/9/004} {\bibfield
  {journal} {\bibinfo  {journal} {J. Phys. CM}\ }\textbf {\bibinfo {volume}
  {10}},\ \bibinfo {pages} {1965--1983} (\bibinfo {year} {1998})}\BibitemShut
  {NoStop}%
\bibitem [{\citenamefont {Vaks}\ \emph {et~al.}(2014)\citenamefont {Vaks},
  \citenamefont {Stroev}, \citenamefont {Pankratov},\ and\ \citenamefont
  {Zabolotskiy}}]{Vaks2014}%
  \BibitemOpen
  \bibfield  {author} {\bibinfo {author} {\bibfnamefont {V.~G.}\ \bibnamefont
  {Vaks}}, \bibinfo {author} {\bibfnamefont {A.~Y.}\ \bibnamefont {Stroev}},
  \bibinfo {author} {\bibfnamefont {I.~R.}\ \bibnamefont {Pankratov}}, \ and\
  \bibinfo {author} {\bibfnamefont {A.~D.}\ \bibnamefont {Zabolotskiy}},\
  }\bibfield  {title} {\enquote {\bibinfo {title} {Statistical theory of
  diffusion in concentrated alloys},}\ }\href {\doibase
  10.1134/S1063776114070103} {\bibfield  {journal} {\bibinfo  {journal} {J.
  Exper. Theo. Phys.}\ }\textbf {\bibinfo {volume} {119}},\ \bibinfo {pages}
  {272--299} (\bibinfo {year} {2014})}\BibitemShut {NoStop}%
\bibitem [{\citenamefont {Vaks}\ \emph {et~al.}(2016)\citenamefont {Vaks},
  \citenamefont {Khromov}, \citenamefont {Pankratov},\ and\ \citenamefont
  {Popov}}]{Vaks2016}%
  \BibitemOpen
  \bibfield  {author} {\bibinfo {author} {\bibfnamefont {V.~G.}\ \bibnamefont
  {Vaks}}, \bibinfo {author} {\bibfnamefont {K.~Y.}\ \bibnamefont {Khromov}},
  \bibinfo {author} {\bibfnamefont {I.~R.}\ \bibnamefont {Pankratov}}, \ and\
  \bibinfo {author} {\bibfnamefont {V.~V.}\ \bibnamefont {Popov}},\ }\bibfield
  {title} {\enquote {\bibinfo {title} {Statistical theory of diffusion in
  concentrated {BCC} and {FCC} alloys and concentration dependences of
  diffusion coefficients in {BCC} alloys {FeCu}, {FeMn}, {FeNi}, and {FeCr}},}\
  }\href {\doibase 10.7868/S0044451016070087} {\bibfield  {journal} {\bibinfo
  {journal} {J. Exper. Theo. Phys.}\ }\textbf {\bibinfo {volume} {150}},\
  \bibinfo {pages} {69--96} (\bibinfo {year} {2016})}\BibitemShut {NoStop}%
\bibitem [{\citenamefont {Novotny}(1995)}]{Novotny1995}%
  \BibitemOpen
  \bibfield  {author} {\bibinfo {author} {\bibfnamefont {M.~A.}\ \bibnamefont
  {Novotny}},\ }\bibfield  {title} {\enquote {\bibinfo {title} {{Monte Carlo}
  algorithms with absorbing {Markov} chains: {Fast} local algorithms for slow
  dynamics},}\ }\href {\doibase 10.1103/PhysRevLett.74.1} {\bibfield  {journal}
  {\bibinfo  {journal} {Phys. Rev. Lett.}\ }\textbf {\bibinfo {volume} {74}},\
  \bibinfo {pages} {1--5} (\bibinfo {year} {1995})}\BibitemShut {NoStop}%
\bibitem [{\citenamefont {Puchala}, \citenamefont {Falk},\ and\ \citenamefont
  {Garikipati}(2010)}]{Puchala2010}%
  \BibitemOpen
  \bibfield  {author} {\bibinfo {author} {\bibfnamefont {B.}~\bibnamefont
  {Puchala}}, \bibinfo {author} {\bibfnamefont {M.~L.}\ \bibnamefont {Falk}}, \
  and\ \bibinfo {author} {\bibfnamefont {K.}~\bibnamefont {Garikipati}},\
  }\bibfield  {title} {\enquote {\bibinfo {title} {An energy basin finding
  algorithm for kinetic {Monte Carlo} acceleration},}\ }\href {\doibase
  10.1063/1.3369627} {\bibfield  {journal} {\bibinfo  {journal} {J. Chem.
  Phys.}\ }\textbf {\bibinfo {volume} {132}},\ \bibinfo {pages} {134104}
  (\bibinfo {year} {2010})}\BibitemShut {NoStop}%
\bibitem [{\citenamefont {Nandipati}, \citenamefont {Shim},\ and\ \citenamefont
  {Amar}(2010)}]{Nandipati2010}%
  \BibitemOpen
  \bibfield  {author} {\bibinfo {author} {\bibfnamefont {G.}~\bibnamefont
  {Nandipati}}, \bibinfo {author} {\bibfnamefont {Y.}~\bibnamefont {Shim}}, \
  and\ \bibinfo {author} {\bibfnamefont {J.~G.}\ \bibnamefont {Amar}},\
  }\bibfield  {title} {\enquote {\bibinfo {title} {First-passage time approach
  to kinetic {Monte Carlo} simulations of metal (100) growth},}\ }\href
  {\doibase 10.1103/PhysRevB.81.235415} {\bibfield  {journal} {\bibinfo
  {journal} {Phys. Rev. B}\ }\textbf {\bibinfo {volume} {81}},\ \bibinfo
  {pages} {235415} (\bibinfo {year} {2010})}\BibitemShut {NoStop}%
\bibitem [{\citenamefont {Chatterjee}\ and\ \citenamefont
  {Voter}(2010)}]{Chatterjee2010}%
  \BibitemOpen
  \bibfield  {author} {\bibinfo {author} {\bibfnamefont {A.}~\bibnamefont
  {Chatterjee}}\ and\ \bibinfo {author} {\bibfnamefont {A.~F.}\ \bibnamefont
  {Voter}},\ }\bibfield  {title} {\enquote {\bibinfo {title} {Accurate
  acceleration of kinetic {Monte Carlo} simulations through the modification of
  rate constants},}\ }\href {\doibase 10.1063/1.3409606} {\bibfield  {journal}
  {\bibinfo  {journal} {J. Chem. Phys.}\ }\textbf {\bibinfo {volume} {132}},\
  \bibinfo {pages} {194101} (\bibinfo {year} {2010})}\BibitemShut {NoStop}%
\bibitem [{\citenamefont {Fichthorn}\ and\ \citenamefont
  {Lin}(2013)}]{Fichthorn2013}%
  \BibitemOpen
  \bibfield  {author} {\bibinfo {author} {\bibfnamefont {K.~A.}\ \bibnamefont
  {Fichthorn}}\ and\ \bibinfo {author} {\bibfnamefont {Y.}~\bibnamefont
  {Lin}},\ }\bibfield  {title} {\enquote {\bibinfo {title} {A local superbasin
  kinetic {Monte Carlo} method},}\ }\href {\doibase 10.1063/1.4801869}
  {\bibfield  {journal} {\bibinfo  {journal} {J. Chem. Phys.}\ }\textbf
  {\bibinfo {volume} {138}},\ \bibinfo {pages} {164104} (\bibinfo {year}
  {2013})}\BibitemShut {NoStop}%
\bibitem [{\citenamefont {Li}\ and\ \citenamefont {Trinkle}(2016)}]{Li2016}%
  \BibitemOpen
  \bibfield  {author} {\bibinfo {author} {\bibfnamefont {Z.}~\bibnamefont
  {Li}}\ and\ \bibinfo {author} {\bibfnamefont {D.~R.}\ \bibnamefont
  {Trinkle}},\ }\bibfield  {title} {\enquote {\bibinfo {title} {Kinetic monte
  carlo investigation of tetragonal strain on {Onsager} matrices},}\ }\href
  {\doibase 10.1103/PhysRevE.93.053305} {\bibfield  {journal} {\bibinfo
  {journal} {Phys. Rev. E}\ }\textbf {\bibinfo {volume} {93}},\ \bibinfo
  {pages} {053305} (\bibinfo {year} {2016})}\BibitemShut {NoStop}%
\bibitem [{\citenamefont {Nastar}(2014)}]{Nastar2014}%
  \BibitemOpen
  \bibfield  {author} {\bibinfo {author} {\bibfnamefont {M.}~\bibnamefont
  {Nastar}},\ }\bibfield  {title} {\enquote {\bibinfo {title} {Atomic diffusion
  theory challenging the {Cahn-Hilliard} method},}\ }\href {\doibase
  10.1103/PhysRevB.90.144101} {\bibfield  {journal} {\bibinfo  {journal} {Phys.
  Rev. B}\ }\textbf {\bibinfo {volume} {90}},\ \bibinfo {pages} {144101}
  (\bibinfo {year} {2014})}\BibitemShut {NoStop}%
\bibitem [{\citenamefont {Barbe}\ and\ \citenamefont
  {Nastar}(2007{\natexlab{a}})}]{Barbe2007a}%
  \BibitemOpen
  \bibfield  {author} {\bibinfo {author} {\bibfnamefont {V.}~\bibnamefont
  {Barbe}}\ and\ \bibinfo {author} {\bibfnamefont {M.}~\bibnamefont {Nastar}},\
  }\bibfield  {title} {\enquote {\bibinfo {title} {Split interstitials in an
  interacting bcc alloy. {I. Extension} of the cluster variation method},}\
  }\href {\doibase 10.1103/PhysRevB.76.054205} {\bibfield  {journal} {\bibinfo
  {journal} {Phys. Rev. B}\ }\textbf {\bibinfo {volume} {76}},\ \bibinfo
  {pages} {054205} (\bibinfo {year} {2007}{\natexlab{a}})}\BibitemShut
  {NoStop}%
\bibitem [{\citenamefont {Barbe}\ and\ \citenamefont
  {Nastar}(2007{\natexlab{b}})}]{Barbe2007b}%
  \BibitemOpen
  \bibfield  {author} {\bibinfo {author} {\bibfnamefont {V.}~\bibnamefont
  {Barbe}}\ and\ \bibinfo {author} {\bibfnamefont {M.}~\bibnamefont {Nastar}},\
  }\bibfield  {title} {\enquote {\bibinfo {title} {Split interstitials in an
  interacting bcc alloy. {II. Transport} coefficients},}\ }\href {\doibase
  10.1103/PhysRevB.76.054206} {\bibfield  {journal} {\bibinfo  {journal} {Phys.
  Rev. B}\ }\textbf {\bibinfo {volume} {76}},\ \bibinfo {pages} {054206}
  (\bibinfo {year} {2007}{\natexlab{b}})}\BibitemShut {NoStop}%
\bibitem [{\citenamefont {Garnier}\ \emph
  {et~al.}(2013{\natexlab{a}})\citenamefont {Garnier}, \citenamefont {Manga},
  \citenamefont {Trinkle}, \citenamefont {Nastar},\ and\ \citenamefont
  {Bellon}}]{Garnier2013a}%
  \BibitemOpen
  \bibfield  {author} {\bibinfo {author} {\bibfnamefont {T.}~\bibnamefont
  {Garnier}}, \bibinfo {author} {\bibfnamefont {V.~R.}\ \bibnamefont {Manga}},
  \bibinfo {author} {\bibfnamefont {D.~R.}\ \bibnamefont {Trinkle}}, \bibinfo
  {author} {\bibfnamefont {M.}~\bibnamefont {Nastar}}, \ and\ \bibinfo {author}
  {\bibfnamefont {P.}~\bibnamefont {Bellon}},\ }\bibfield  {title} {\enquote
  {\bibinfo {title} {Stress-induced anisotropic diffusion in alloys: Complex
  {Si} solute flow near a dislocation core in {Ni}},}\ }\href {\doibase
  10.1103/PhysRevB.88.134108} {\bibfield  {journal} {\bibinfo  {journal} {Phys.
  Rev. B}\ }\textbf {\bibinfo {volume} {88}},\ \bibinfo {pages} {134108}
  (\bibinfo {year} {2013}{\natexlab{a}})}\BibitemShut {NoStop}%
\bibitem [{\citenamefont {Garnier}\ \emph
  {et~al.}(2013{\natexlab{b}})\citenamefont {Garnier}, \citenamefont {Nastar},
  \citenamefont {Bellon},\ and\ \citenamefont {Trinkle}}]{Garnier2013c}%
  \BibitemOpen
  \bibfield  {author} {\bibinfo {author} {\bibfnamefont {T.}~\bibnamefont
  {Garnier}}, \bibinfo {author} {\bibfnamefont {M.}~\bibnamefont {Nastar}},
  \bibinfo {author} {\bibfnamefont {P.}~\bibnamefont {Bellon}}, \ and\ \bibinfo
  {author} {\bibfnamefont {D.~R.}\ \bibnamefont {Trinkle}},\ }\bibfield
  {title} {\enquote {\bibinfo {title} {Solute drag by vacancies in
  body-centered cubic alloys},}\ }\href {\doibase 10.1103/PhysRevB.88.134201}
  {\bibfield  {journal} {\bibinfo  {journal} {Phys. Rev. B}\ }\textbf {\bibinfo
  {volume} {88}},\ \bibinfo {pages} {134201} (\bibinfo {year}
  {2013}{\natexlab{b}})}\BibitemShut {NoStop}%
\bibitem [{\citenamefont {Garnier}\ \emph
  {et~al.}(2014{\natexlab{b}})\citenamefont {Garnier}, \citenamefont {Trinkle},
  \citenamefont {Nastar},\ and\ \citenamefont {Bellon}}]{Garnier2014a}%
  \BibitemOpen
  \bibfield  {author} {\bibinfo {author} {\bibfnamefont {T.}~\bibnamefont
  {Garnier}}, \bibinfo {author} {\bibfnamefont {D.~R.}\ \bibnamefont
  {Trinkle}}, \bibinfo {author} {\bibfnamefont {M.}~\bibnamefont {Nastar}}, \
  and\ \bibinfo {author} {\bibfnamefont {P.}~\bibnamefont {Bellon}},\
  }\bibfield  {title} {\enquote {\bibinfo {title} {Quantitative modeling of
  solute drag by vacancies in face-centered-cubic alloys},}\ }\href {\doibase
  10.1103/PhysRevB.89.144202} {\bibfield  {journal} {\bibinfo  {journal} {Phys.
  Rev. B}\ }\textbf {\bibinfo {volume} {89}},\ \bibinfo {pages} {144202}
  (\bibinfo {year} {2014}{\natexlab{b}})}\BibitemShut {NoStop}%
\bibitem [{\citenamefont {Kikuchi}(1966)}]{Kikuchi1966}%
  \BibitemOpen
  \bibfield  {author} {\bibinfo {author} {\bibfnamefont {R.}~\bibnamefont
  {Kikuchi}},\ }\bibfield  {title} {\enquote {\bibinfo {title} {The path
  probability method},}\ }\href {\doibase 10.1143/PTPS.35.1} {\bibfield
  {journal} {\bibinfo  {journal} {Prog. Theor. Phys. Suppl.}\ }\textbf
  {\bibinfo {volume} {35}},\ \bibinfo {pages} {1--64} (\bibinfo {year}
  {1966})}\BibitemShut {NoStop}%
\bibitem [{\citenamefont {Sato}\ and\ \citenamefont
  {Kikuchi}(1983)}]{Sato1983}%
  \BibitemOpen
  \bibfield  {author} {\bibinfo {author} {\bibfnamefont {H.}~\bibnamefont
  {Sato}}\ and\ \bibinfo {author} {\bibfnamefont {R.}~\bibnamefont {Kikuchi}},\
  }\bibfield  {title} {\enquote {\bibinfo {title} {Theory of many-body
  diffusion by the path-probability method: Conversion from ensemble averaging
  to time averaging},}\ }\href {\doibase 10.1103/PhysRevB.28.648} {\bibfield
  {journal} {\bibinfo  {journal} {Phys. Rev. B}\ }\textbf {\bibinfo {volume}
  {28}},\ \bibinfo {pages} {648--664} (\bibinfo {year} {1983})}\BibitemShut
  {NoStop}%
\bibitem [{\citenamefont {Sato}, \citenamefont {Ishikawa},\ and\ \citenamefont
  {Kikuchi}(1985)}]{Sato1985}%
  \BibitemOpen
  \bibfield  {author} {\bibinfo {author} {\bibfnamefont {H.}~\bibnamefont
  {Sato}}, \bibinfo {author} {\bibfnamefont {T.}~\bibnamefont {Ishikawa}}, \
  and\ \bibinfo {author} {\bibfnamefont {R.}~\bibnamefont {Kikuchi}},\
  }\bibfield  {title} {\enquote {\bibinfo {title} {Correlation factor in tracer
  diffusion for high tracer concentrations},}\ }\href {\doibase
  10.1016/0022-3697(85)90074-5} {\bibfield  {journal} {\bibinfo  {journal} {J.
  Phys. Chem. Solids}\ }\textbf {\bibinfo {volume} {46}},\ \bibinfo {pages}
  {1361--1370} (\bibinfo {year} {1985})}\BibitemShut {NoStop}%
\bibitem [{\citenamefont {Montroll}\ and\ \citenamefont
  {Weiss}(1965)}]{Montroll1965}%
  \BibitemOpen
  \bibfield  {author} {\bibinfo {author} {\bibfnamefont {E.~W.}\ \bibnamefont
  {Montroll}}\ and\ \bibinfo {author} {\bibfnamefont {G.~H.}\ \bibnamefont
  {Weiss}},\ }\bibfield  {title} {\enquote {\bibinfo {title} {Random walks on
  lattices. {II}},}\ }\href {\doibase 10.1063/1.1704269} {\bibfield  {journal}
  {\bibinfo  {journal} {J. Math. Phys.}\ }\textbf {\bibinfo {volume} {6}},\
  \bibinfo {pages} {167--181} (\bibinfo {year} {1965})}\BibitemShut {NoStop}%
\bibitem [{\citenamefont {Koiwa}\ and\ \citenamefont
  {Ishioka}(1983{\natexlab{a}})}]{Koiwa1983}%
  \BibitemOpen
  \bibfield  {author} {\bibinfo {author} {\bibfnamefont {M.}~\bibnamefont
  {Koiwa}}\ and\ \bibinfo {author} {\bibfnamefont {S.}~\bibnamefont
  {Ishioka}},\ }\bibfield  {title} {\enquote {\bibinfo {title} {Integral
  methods in the calculation of correlation factors for impurity diffusion},}\
  }\href {\doibase 10.1080/01418618308243130} {\bibfield  {journal} {\bibinfo
  {journal} {Philos. Mag. A}\ }\textbf {\bibinfo {volume} {47}},\ \bibinfo
  {pages} {927--938} (\bibinfo {year} {1983}{\natexlab{a}})}\BibitemShut
  {NoStop}%
\bibitem [{\citenamefont {Watson}(1939)}]{Watson1939}%
  \BibitemOpen
  \bibfield  {author} {\bibinfo {author} {\bibfnamefont {G.~N.}\ \bibnamefont
  {Watson}},\ }\bibfield  {title} {\enquote {\bibinfo {title} {Three triple
  integrals},}\ }\href {\doibase 10.1093/qmath/os-10.1.266} {\bibfield
  {journal} {\bibinfo  {journal} {Q. J. Math.}\ }\textbf {\bibinfo {volume}
  {10}},\ \bibinfo {pages} {266--276} (\bibinfo {year} {1939})}\BibitemShut
  {NoStop}%
\bibitem [{\citenamefont {Inoue}(1974)}]{Inoue1974}%
  \BibitemOpen
  \bibfield  {author} {\bibinfo {author} {\bibfnamefont {M.}~\bibnamefont
  {Inoue}},\ }\bibfield  {title} {\enquote {\bibinfo {title} {Lattice green's
  function for the face centered cubic lattice},}\ }\href {\doibase
  http://dx.doi.org/10.1063/1.1666714} {\bibfield  {journal} {\bibinfo
  {journal} {J. Math. Phys.}\ }\textbf {\bibinfo {volume} {15}},\ \bibinfo
  {pages} {704--707} (\bibinfo {year} {1974})}\BibitemShut {NoStop}%
\bibitem [{\citenamefont {Mano}(1974)}]{Mano1974}%
  \BibitemOpen
  \bibfield  {author} {\bibinfo {author} {\bibfnamefont {K.}~\bibnamefont
  {Mano}},\ }\bibfield  {title} {\enquote {\bibinfo {title} {A remark on the
  green's function for the face-centered cubic lattices},}\ }\href {\doibase
  http://dx.doi.org/10.1063/1.1666598} {\bibfield  {journal} {\bibinfo
  {journal} {J. Math. Phys.}\ }\textbf {\bibinfo {volume} {15}},\ \bibinfo
  {pages} {2175--2176} (\bibinfo {year} {1974})}\BibitemShut {NoStop}%
\bibitem [{\citenamefont {Morita}(1975)}]{Morita1975}%
  \BibitemOpen
  \bibfield  {author} {\bibinfo {author} {\bibfnamefont {T.}~\bibnamefont
  {Morita}},\ }\bibfield  {title} {\enquote {\bibinfo {title} {Use of a
  recurrence formula in computing the lattice {Green} function},}\ }\href
  {\doibase 10.1088/0305-4470/8/4/008} {\bibfield  {journal} {\bibinfo
  {journal} {J. Phys. A: Math. Gen.}\ }\textbf {\bibinfo {volume} {8}},\
  \bibinfo {pages} {478} (\bibinfo {year} {1975})}\BibitemShut {NoStop}%
\bibitem [{\citenamefont {Ishioka}\ and\ \citenamefont
  {Koiwa}(1978)}]{Koiwa1978}%
  \BibitemOpen
  \bibfield  {author} {\bibinfo {author} {\bibfnamefont {S.}~\bibnamefont
  {Ishioka}}\ and\ \bibinfo {author} {\bibfnamefont {M.}~\bibnamefont
  {Koiwa}},\ }\bibfield  {title} {\enquote {\bibinfo {title} {Random walks on
  diamond and hexagonal close packed lattices},}\ }\href {\doibase
  10.1080/01418617808239187} {\bibfield  {journal} {\bibinfo  {journal}
  {Philos. Mag. A}\ }\textbf {\bibinfo {volume} {37}},\ \bibinfo {pages}
  {517--533} (\bibinfo {year} {1978})}\BibitemShut {NoStop}%
\bibitem [{\citenamefont {Koiwa}\ and\ \citenamefont
  {Ishioka}(1979)}]{Koiwa1979}%
  \BibitemOpen
  \bibfield  {author} {\bibinfo {author} {\bibfnamefont {M.}~\bibnamefont
  {Koiwa}}\ and\ \bibinfo {author} {\bibfnamefont {S.}~\bibnamefont
  {Ishioka}},\ }\bibfield  {title} {\enquote {\bibinfo {title} {Random walks on
  three-dimensional lattices with the coordination number four},}\ }\href
  {\doibase 10.1080/01418617908234863} {\bibfield  {journal} {\bibinfo
  {journal} {Philos. Mag. A}\ }\textbf {\bibinfo {volume} {40}},\ \bibinfo
  {pages} {625--635} (\bibinfo {year} {1979})}\BibitemShut {NoStop}%
\bibitem [{\citenamefont {Trinkle}(2016{\natexlab{b}})}]{OnsagerCalc}%
  \BibitemOpen
  \bibfield  {author} {\bibinfo {author} {\bibfnamefont {D.~R.}\ \bibnamefont
  {Trinkle}},\ }\href {\doibase 10.5281/zenodo.59487} {\enquote {\bibinfo
  {title} {\textsc{Onsager}},}\ }\bibinfo {howpublished}
  {http://dallastrinkle.github.io/Onsager} (\bibinfo {year}
  {2016}{\natexlab{b}})\BibitemShut {NoStop}%
\bibitem [{\citenamefont {Ashcroft}\ and\ \citenamefont
  {Mermin}(1976)}]{Ashcroft76}%
  \BibitemOpen
  \bibfield  {author} {\bibinfo {author} {\bibfnamefont {N.~W.}\ \bibnamefont
  {Ashcroft}}\ and\ \bibinfo {author} {\bibfnamefont {N.~D.}\ \bibnamefont
  {Mermin}},\ }\href@noop {} {\emph {\bibinfo {title} {Solid State Physics}}}\
  (\bibinfo  {publisher} {Philadelphia: Saunders College},\ \bibinfo {year}
  {1976})\BibitemShut {NoStop}%
\bibitem [{\citenamefont {Senhaji}\ \emph {et~al.}(1992)\citenamefont
  {Senhaji}, \citenamefont {Treglia}, \citenamefont {Legrand}, \citenamefont
  {Barrett}, \citenamefont {Guillot},\ and\ \citenamefont
  {Villette}}]{Senhaji1992}%
  \BibitemOpen
  \bibfield  {author} {\bibinfo {author} {\bibfnamefont {A.}~\bibnamefont
  {Senhaji}}, \bibinfo {author} {\bibfnamefont {G.}~\bibnamefont {Treglia}},
  \bibinfo {author} {\bibfnamefont {B.}~\bibnamefont {Legrand}}, \bibinfo
  {author} {\bibfnamefont {N.}~\bibnamefont {Barrett}}, \bibinfo {author}
  {\bibfnamefont {C.}~\bibnamefont {Guillot}}, \ and\ \bibinfo {author}
  {\bibfnamefont {B.}~\bibnamefont {Villette}},\ }\bibfield  {title} {\enquote
  {\bibinfo {title} {Is the segregation-dissolution kinetics driven by a
  surface local equilibrium? {An} answer via the kinetic tight-binding {Ising}
  model},}\ }\href {\doibase 10.1016/0039-6028(92)90535-E} {\bibfield
  {journal} {\bibinfo  {journal} {Surface Science}\ }\textbf {\bibinfo {volume}
  {274}},\ \bibinfo {pages} {297 -- 305} (\bibinfo {year} {1992})}\BibitemShut
  {NoStop}%
\bibitem [{\citenamefont {{Van der Ven}}\ \emph {et~al.}(2001)\citenamefont
  {{Van der Ven}}, \citenamefont {Ceder}, \citenamefont {Asta},\ and\
  \citenamefont {Tepesch}}]{Vanderven2001}%
  \BibitemOpen
  \bibfield  {author} {\bibinfo {author} {\bibfnamefont {A.}~\bibnamefont {{Van
  der Ven}}}, \bibinfo {author} {\bibfnamefont {G.}~\bibnamefont {Ceder}},
  \bibinfo {author} {\bibfnamefont {M.}~\bibnamefont {Asta}}, \ and\ \bibinfo
  {author} {\bibfnamefont {P.~D.}\ \bibnamefont {Tepesch}},\ }\bibfield
  {title} {\enquote {\bibinfo {title} {First-principles theory of ionic
  diffusion with nondilute carriers},}\ }\href {\doibase
  10.1103/PhysRevB.64.184307} {\bibfield  {journal} {\bibinfo  {journal} {Phys.
  Rev. B}\ }\textbf {\bibinfo {volume} {64}},\ \bibinfo {pages} {184307}
  (\bibinfo {year} {2001})}\BibitemShut {NoStop}%
\bibitem [{\citenamefont {Yasi}\ and\ \citenamefont
  {Trinkle}(2012)}]{Yasi2012b}%
  \BibitemOpen
  \bibfield  {author} {\bibinfo {author} {\bibfnamefont {J.~A.}\ \bibnamefont
  {Yasi}}\ and\ \bibinfo {author} {\bibfnamefont {D.~R.}\ \bibnamefont
  {Trinkle}},\ }\bibfield  {title} {\enquote {\bibinfo {title} {Direct
  calculation of lattice green function with arbitrary interactions for general
  crystals},}\ }\href {\doibase 10.1103/PhysRevE.85.066706} {\bibfield
  {journal} {\bibinfo  {journal} {Phys. Rev. E}\ }\textbf {\bibinfo {volume}
  {85}},\ \bibinfo {pages} {066706} (\bibinfo {year} {2012})}\BibitemShut
  {NoStop}%
\bibitem [{Note1()}]{Note1}%
  \BibitemOpen
  \bibinfo {note} {Diffusive and relaxive modes are analogous to acoustic and
  optical modes for phonons in a crystal, and a similar separation is used by
  Yasi and Trinkle to solve for the force-constant lattice Green function.\cite
  {Yasi2012b}}\BibitemShut {NoStop}%
\bibitem [{\citenamefont {Tewary}\ and\ \citenamefont
  {Bullough}(1971)}]{Tewary1971}%
  \BibitemOpen
  \bibfield  {author} {\bibinfo {author} {\bibfnamefont {V.~K.}\ \bibnamefont
  {Tewary}}\ and\ \bibinfo {author} {\bibfnamefont {R.}~\bibnamefont
  {Bullough}},\ }\bibfield  {title} {\enquote {\bibinfo {title} {On a
  semi-continuum green function method for lattice dynamics with application to
  copper},}\ }\href@noop {} {\bibfield  {journal} {\bibinfo  {journal} {J.
  Phys. F}\ }\textbf {\bibinfo {volume} {1}},\ \bibinfo {pages} {554--569}
  (\bibinfo {year} {1971})}\BibitemShut {NoStop}%
\bibitem [{\citenamefont {Trinkle}(2008)}]{TrinkleLGF2008}%
  \BibitemOpen
  \bibfield  {author} {\bibinfo {author} {\bibfnamefont {D.~R.}\ \bibnamefont
  {Trinkle}},\ }\bibfield  {title} {\enquote {\bibinfo {title} {Lattice green
  function for extended defect calculations: Computation and error estimation
  with long-range forces},}\ }\href {\doibase 10.1103/PhysRevB.78.014110}
  {\bibfield  {journal} {\bibinfo  {journal} {Phys. Rev. B}\ }\textbf {\bibinfo
  {volume} {78}},\ \bibinfo {pages} {014110} (\bibinfo {year}
  {2008})}\BibitemShut {NoStop}%
\bibitem [{\citenamefont {Ghazisaeidi}\ and\ \citenamefont
  {Trinkle}(2009)}]{Ghazisaeidi2009}%
  \BibitemOpen
  \bibfield  {author} {\bibinfo {author} {\bibfnamefont {M.}~\bibnamefont
  {Ghazisaeidi}}\ and\ \bibinfo {author} {\bibfnamefont {D.~R.}\ \bibnamefont
  {Trinkle}},\ }\bibfield  {title} {\enquote {\bibinfo {title} {Convergence
  rate for numerical computation of the lattice {Green's} function},}\ }\href
  {\doibase 10.1103/PhysRevE.79.037701} {\bibfield  {journal} {\bibinfo
  {journal} {Phys. Rev. E}\ }\textbf {\bibinfo {volume} {79}},\ \bibinfo
  {pages} {037701} (\bibinfo {year} {2009})}\BibitemShut {NoStop}%
\bibitem [{\citenamefont {Gradshteyn}\ and\ \citenamefont
  {Ryzhik}(1994)}]{IntegralTables}%
  \BibitemOpen
  \bibfield  {author} {\bibinfo {author} {\bibfnamefont {I.~S.}\ \bibnamefont
  {Gradshteyn}}\ and\ \bibinfo {author} {\bibfnamefont {I.~M.}\ \bibnamefont
  {Ryzhik}},\ }\href@noop {} {\emph {\bibinfo {title} {Table of Integrals,
  Series, and Products}}},\ \bibinfo {edition} {5th}\ ed.,\ edited by\ \bibinfo
  {editor} {\bibfnamefont {A.}~\bibnamefont {Jeffrey}}\ (\bibinfo  {publisher}
  {Academic Press},\ \bibinfo {year} {1994})\BibitemShut {NoStop}%
\bibitem [{\citenamefont {Monkhorst}\ and\ \citenamefont
  {Pack}(1976)}]{Monkhorst76}%
  \BibitemOpen
  \bibfield  {author} {\bibinfo {author} {\bibfnamefont {H.~J.}\ \bibnamefont
  {Monkhorst}}\ and\ \bibinfo {author} {\bibfnamefont {J.~D.}\ \bibnamefont
  {Pack}},\ }\bibfield  {title} {\enquote {\bibinfo {title} {Special points for
  brillouin-zone integrations},}\ }\href {\doibase 10.1103/PhysRevB.13.5188}
  {\bibfield  {journal} {\bibinfo  {journal} {Phys. Rev. B}\ }\textbf {\bibinfo
  {volume} {13}},\ \bibinfo {pages} {5188--5192} (\bibinfo {year}
  {1976})}\BibitemShut {NoStop}%
\bibitem [{Note2()}]{Note2}%
  \BibitemOpen
  \bibinfo {note} {It should be noted that our use of ``kinetic shell'' is
  different than Nastar~\protect \textit {et~al.}\ Our kinetic shell is defined
  for states with changes in rates only---making it a finite space for a finite
  interaction range---while Nastar~\protect \textit {et~al.}\ uses the kinetic
  shell to identify the set of states for which the effective Hamiltonian will
  be non-zero, and hence it acts as a convergence parameter in SCMF
  calculations.}\BibitemShut {Stop}%
\bibitem [{\citenamefont {Glazer}\ and\ \citenamefont
  {Burns}(2013)}]{Glazer2013}%
  \BibitemOpen
  \bibfield  {author} {\bibinfo {author} {\bibfnamefont {M.}~\bibnamefont
  {Glazer}}\ and\ \bibinfo {author} {\bibfnamefont {G.}~\bibnamefont {Burns}},\
  }\href {\doibase ISBN:978-0-12-394400-9} {\emph {\bibinfo {title} {Space
  Groups for Solid State Scientists}}},\ \bibinfo {edition} {3rd}\ ed.\
  (\bibinfo  {publisher} {Elsevier},\ \bibinfo {year} {2013})\BibitemShut
  {NoStop}%
\bibitem [{\citenamefont {Garnier}\ \emph
  {et~al.}(2014{\natexlab{c}})\citenamefont {Garnier}, \citenamefont {Manga},
  \citenamefont {Bellon},\ and\ \citenamefont {Trinkle}}]{Garnier2014b}%
  \BibitemOpen
  \bibfield  {author} {\bibinfo {author} {\bibfnamefont {T.}~\bibnamefont
  {Garnier}}, \bibinfo {author} {\bibfnamefont {V.~R.}\ \bibnamefont {Manga}},
  \bibinfo {author} {\bibfnamefont {P.}~\bibnamefont {Bellon}}, \ and\ \bibinfo
  {author} {\bibfnamefont {D.~R.}\ \bibnamefont {Trinkle}},\ }\bibfield
  {title} {\enquote {\bibinfo {title} {Diffusion of {Si} impurities in {Ni}
  under stress: A first-principles study},}\ }\href {\doibase
  10.1103/PhysRevB.90.024306} {\bibfield  {journal} {\bibinfo  {journal} {Phys.
  Rev. B}\ }\textbf {\bibinfo {volume} {90}},\ \bibinfo {pages} {024306}
  (\bibinfo {year} {2014}{\natexlab{c}})}\BibitemShut {NoStop}%
\bibitem [{\citenamefont {Garnier}\ \emph
  {et~al.}(2014{\natexlab{d}})\citenamefont {Garnier}, \citenamefont {Manga},
  \citenamefont {Bellon},\ and\ \citenamefont {Trinkle}}]{Garnier2014bData}%
  \BibitemOpen
  \bibfield  {author} {\bibinfo {author} {\bibfnamefont {T.}~\bibnamefont
  {Garnier}}, \bibinfo {author} {\bibfnamefont {V.~R.}\ \bibnamefont {Manga}},
  \bibinfo {author} {\bibfnamefont {P.}~\bibnamefont {Bellon}}, \ and\ \bibinfo
  {author} {\bibfnamefont {D.~R.}\ \bibnamefont {Trinkle}},\ }\href
  {hdl.handle.net/11115/239} {\enquote {\bibinfo {title} {Data citation:
  Diffusion of {Si} impurities in {Ni} under stress: A first-principles
  study},}\ } (\bibinfo {year} {2014}{\natexlab{d}})\BibitemShut {NoStop}%
\bibitem [{\citenamefont {Agarwal}\ and\ \citenamefont
  {Trinkle}(2017)}]{Agarwal2017}%
  \BibitemOpen
  \bibfield  {author} {\bibinfo {author} {\bibfnamefont {R.}~\bibnamefont
  {Agarwal}}\ and\ \bibinfo {author} {\bibfnamefont {D.~R.}\ \bibnamefont
  {Trinkle}},\ }\bibfield  {title} {\enquote {\bibinfo {title} {Exact model of
  vacancy-mediated solute transport in magnesium},}\ }\href {\doibase
  10.1103/PhysRevLett.118.105901} {\bibfield  {journal} {\bibinfo  {journal}
  {Phys. Rev. Lett.}\ }\textbf {\bibinfo {volume} {118}},\ \bibinfo {pages}
  {105901} (\bibinfo {year} {2017})}\BibitemShut {NoStop}%
\bibitem [{\citenamefont {Hahn}(1996)}]{Wyckoff}%
  \BibitemOpen
  \bibinfo {editor} {\bibfnamefont {T.}~\bibnamefont {Hahn}},\ ed.,\ \href@noop
  {} {\emph {\bibinfo {title} {International Tables for Crystallography}}},\
  \bibinfo {edition} {4th}\ ed.,\ Vol.~\bibinfo {volume} {A}\ (\bibinfo
  {publisher} {Dordrecht: Kluwer Academic},\ \bibinfo {year}
  {1996})\BibitemShut {NoStop}%
\bibitem [{\citenamefont {Manning}(1964)}]{Manning1964}%
  \BibitemOpen
  \bibfield  {author} {\bibinfo {author} {\bibfnamefont {J.~R.}\ \bibnamefont
  {Manning}},\ }\bibfield  {title} {\enquote {\bibinfo {title} {Correlation
  factors for impurity diffusion. bcc, diamond, and fcc structures},}\ }\href
  {\doibase 10.1103/PhysRev.136.A1758} {\bibfield  {journal} {\bibinfo
  {journal} {Phys. Rev.}\ }\textbf {\bibinfo {volume} {136}},\ \bibinfo {pages}
  {A1758--A1766} (\bibinfo {year} {1964})}\BibitemShut {NoStop}%
\bibitem [{\citenamefont {Koiwa}\ and\ \citenamefont
  {Ishioka}(1983{\natexlab{b}})}]{Koiwa1983b}%
  \BibitemOpen
  \bibfield  {author} {\bibinfo {author} {\bibfnamefont {M.}~\bibnamefont
  {Koiwa}}\ and\ \bibinfo {author} {\bibfnamefont {S.}~\bibnamefont
  {Ishioka}},\ }\bibfield  {title} {\enquote {\bibinfo {title} {Random walks
  and correlation factor in diffusion in a three-dimensional lattice with
  coordination number 8},}\ }\href {\doibase 10.1080/01418618308234882}
  {\bibfield  {journal} {\bibinfo  {journal} {Philos. Mag. A}\ }\textbf
  {\bibinfo {volume} {48}},\ \bibinfo {pages} {1--9} (\bibinfo {year}
  {1983}{\natexlab{b}})}\BibitemShut {NoStop}%
\bibitem [{\citenamefont {Carlson}\ and\ \citenamefont
  {Wilson}(2016)}]{Carlson2016}%
  \BibitemOpen
  \bibfield  {author} {\bibinfo {author} {\bibfnamefont {W.~D.}\ \bibnamefont
  {Carlson}}\ and\ \bibinfo {author} {\bibfnamefont {C.~R.}\ \bibnamefont
  {Wilson}},\ }\bibfield  {title} {\enquote {\bibinfo {title} {Correlation
  factors for impurity diffusion on the sublattice of dodecahedral sites in
  garnet},}\ }\href {\doibase DOI 10.1007/s00269-016-0800-2} {\bibfield
  {journal} {\bibinfo  {journal} {Phys. Chem. Minerals}\ }\textbf {\bibinfo
  {volume} {43}},\ \bibinfo {pages} {363--369} (\bibinfo {year}
  {2016})}\BibitemShut {NoStop}%
\bibitem [{\citenamefont {Koiwa}\ and\ \citenamefont
  {Ishioka}(1983{\natexlab{c}})}]{Koiwa1983c}%
  \BibitemOpen
  \bibfield  {author} {\bibinfo {author} {\bibfnamefont {M.}~\bibnamefont
  {Koiwa}}\ and\ \bibinfo {author} {\bibfnamefont {S.}~\bibnamefont
  {Ishioka}},\ }\bibfield  {title} {\enquote {\bibinfo {title} {Random walks
  and correlation factors in diffusion via the vacancy mechanism in
  hexagonal-close-packed lattices},}\ }\href {\doibase
  10.1080/01418618308245263} {\bibfield  {journal} {\bibinfo  {journal}
  {Philos. Mag. A}\ }\textbf {\bibinfo {volume} {47}},\ \bibinfo {pages}
  {767--774} (\bibinfo {year} {1983}{\natexlab{c}})}\BibitemShut {NoStop}%
\bibitem [{\citenamefont {Gibbs}\ and\ \citenamefont
  {Smith}(1965)}]{Gibbs1965}%
  \BibitemOpen
  \bibfield  {author} {\bibinfo {author} {\bibfnamefont {G.~V.}\ \bibnamefont
  {Gibbs}}\ and\ \bibinfo {author} {\bibfnamefont {J.~V.}\ \bibnamefont
  {Smith}},\ }\bibfield  {title} {\enquote {\bibinfo {title} {Refinement of the
  crystal structure of synthetic pyrope},}\ }\href@noop {} {\bibfield
  {journal} {\bibinfo  {journal} {Am. Mineral.}\ }\textbf {\bibinfo {volume}
  {50}},\ \bibinfo {pages} {2023--2039} (\bibinfo {year} {1965})}\BibitemShut
  {NoStop}%
\bibitem [{\citenamefont {Henkelman}, \citenamefont {Uberuaga},\ and\
  \citenamefont {J{\'o}nsson}(2000)}]{Henkelman2000}%
  \BibitemOpen
  \bibfield  {author} {\bibinfo {author} {\bibfnamefont {G.}~\bibnamefont
  {Henkelman}}, \bibinfo {author} {\bibfnamefont {B.~P.}\ \bibnamefont
  {Uberuaga}}, \ and\ \bibinfo {author} {\bibfnamefont {H.}~\bibnamefont
  {J{\'o}nsson}},\ }\bibfield  {title} {\enquote {\bibinfo {title} {A climbing
  image nudged elastic band method for finding saddle points and minimum energy
  paths},}\ }\href@noop {} {\bibfield  {journal} {\bibinfo  {journal} {J. Chem.
  Phys.}\ }\textbf {\bibinfo {volume} {113}},\ \bibinfo {pages} {9901--9904}
  (\bibinfo {year} {2000})}\BibitemShut {NoStop}%
\bibitem [{\citenamefont {Shang}\ \emph {et~al.}(2011)\citenamefont {Shang},
  \citenamefont {{Hector, Jr.}}, \citenamefont {Wang},\ and\ \citenamefont
  {Liu}}]{Shang2011}%
  \BibitemOpen
  \bibfield  {author} {\bibinfo {author} {\bibfnamefont {S.~L.}\ \bibnamefont
  {Shang}}, \bibinfo {author} {\bibfnamefont {L.~G.}\ \bibnamefont {{Hector,
  Jr.}}}, \bibinfo {author} {\bibfnamefont {Y.}~\bibnamefont {Wang}}, \ and\
  \bibinfo {author} {\bibfnamefont {Z.~K.}\ \bibnamefont {Liu}},\ }\bibfield
  {title} {\enquote {\bibinfo {title} {Anomalous energy pathway of vacancy
  migration and self-diffusion in hcp {Ti}},}\ }\href {\doibase
  10.1103/PhysRevB.83.224104} {\bibfield  {journal} {\bibinfo  {journal} {Phys.
  Rev. B}\ }\textbf {\bibinfo {volume} {83}},\ \bibinfo {pages} {224104}
  (\bibinfo {year} {2011})}\BibitemShut {NoStop}%
\end{thebibliography}
%
\end{document}